%% file: main.tex
\documentclass[sigconf, nonacm]{acmart}

\usepackage{subfigure}
\usepackage[linesnumbered,ruled,vlined]{algorithm2e}
\usepackage{multirow}
\usepackage{enumitem}
\usepackage{balance}
\usepackage{xcolor}
\usepackage{color}
\usepackage{array}
\usepackage{makecell}
\usepackage{marvosym}

\newcommand\vldbdoi{XX.XX/XXX.XX}
\newcommand\vldbpages{XXX-XXX}
\newcommand\vldbvolume{18}
\newcommand\vldbissue{2}
\newcommand\vldbyear{2024}
\newcommand\vldbauthors{\authors}
\newcommand\vldbtitle{\shorttitle} 
\newcommand\vldbavailabilityurl{URL_TO_YOUR_ARTIFACTS}
\newcommand\vldbpagestyle{empty} 

\renewcommand{\thefootnote}{\fnsymbol{footnote}}

\begin{document}
\title{Less is More: Efficient Time Series Dataset Condensation via Two-fold Modal Matching--Extended Version}

\author{Hao Miao}
\affiliation{Aalborg University, Denmark}
\email{haom@cs.aau.dk}

\author{Ziqiao Liu}
\affiliation{University of Electronic Science and Technology of China, China}
\email{liuziqiao@std.uestc.edu.cn}

\author{Yan Zhao}
\affiliation{Aalborg University, Denmark \Letter}
\email{yanz@cs.aau.dk}

\author{Chenjuan Guo}
\author{Bin Yang}
\affiliation{East China Normal \\University, China \Letter}
\email{{cjguo, byang}@dase.ecnu.edu.cn}

\author{Kai Zheng}
\affiliation{University of Electronic Science and Technology of China, China}
\email{zhengkai@uestc.edu.cn}

\author{Christian S. Jensen}
\affiliation{Aalborg University, Denmark}
\email{csj@cs.aau.dk}
%

\begin{abstract}
The expanding instrumentation of processes throughout society with sensors yields a proliferation of time series data that may in turn enable important applications, e.g., related to transportation infrastructures or power grids. Machine-learning based methods are increasingly being used to extract value from such data. We provide means of reducing the resulting considerable computational and data storage costs. We achieve this by providing means of condensing large time series datasets such that models trained on the condensed data achieve performance comparable to those trained on the original, large data. Specifically, we propose a \underline{time} series \underline{d}ataset \underline{c}ondensation framework, TimeDC, that employs two-fold modal matching, encompassing frequency matching and training trajectory matching. Thus, TimeDC performs time series feature extraction and decomposition-driven frequency matching to preserve complex temporal dependencies in the reduced time series. Further, TimeDC employs curriculum training trajectory matching to ensure effective and generalized time series dataset condensation. To avoid memory overflow and to reduce the cost of dataset condensation, the framework includes an expert buffer storing pre-computed expert trajectories. Extensive experiments on real data offer insight into the effectiveness and efficiency of the proposed solutions.
\end{abstract}

\maketitle
\footnotetext{Yan Zhao and Chenjuan Guo are corresponding authors.}

\pagestyle{\vldbpagestyle}
\begingroup\small\noindent\raggedright\textbf{PVLDB Reference Format:}\\
\vldbauthors. \vldbtitle. PVLDB, \vldbvolume(\vldbissue): \vldbpages, \vldbyear. \\
\href{https://doi.org/\vldbdoi}{doi:\vldbdoi}
\endgroup
\begingroup
\renewcommand\thefootnote{}\footnote{\noindent
This work is licensed under the Creative Commons BY-NC-ND 4.0 International License. Visit \url{https://creativecommons.org/licenses/by-nc-nd/4.0/} to view a copy of this license. For any use beyond those covered by this license, obtain permission by emailing \href{mailto:info@vldb.org}{info@vldb.org}. Copyright is held by the owner/author(s). Publication rights licensed to the VLDB Endowment. \\
\raggedright Proceedings of the VLDB Endowment, Vol. \vldbvolume, No. \vldbissue\ %
ISSN 2150-8097. \\
\href{https://doi.org/\vldbdoi}{doi:\vldbdoi} \\
}\addtocounter{footnote}{-1}\endgroup

\ifdefempty{\vldbavailabilityurl}{}{
\vspace{.3cm}
\begingroup\small\noindent\raggedright\textbf{PVLDB Artifact Availability:}\\
The source code, data, and/or other artifacts have been made available at \url{https://github.com/uestc-liuzq/STdistillation}.
\endgroup
}
\input{chapter/01intro}
\input{chapter/02problem}
\input{chapter/03method}
\input{chapter/04experiment}
\input{chapter/05relatedwork}
\input{chapter/06conclusion}


\bibliographystyle{ACM-Reference-Format}
\bibliography{sample}
\input{chapter/07Appendix}


\end{document}

%% file: chapter/01intro.tex
\section{Introduction}
\label{introduction}
With the proliferation of edge computing and mobile sensing, massive volumes of time series data, comprising millions of observations, are being collected and stored into time series database systems (TSMSs)~\cite{jensen2018modelardb, wang2020apache, lai2023lightcts}, enabling various real-world applications~\cite{wu2023autocts+, xu2024pefad}. We are seeing impressive advances in machine learning~\cite{DBLP:conf/icde/Liu00L0024, yanwww2022}, especially in deep learning~\cite{wu2023autocts+, zhou2022fedformer, cheng2024memfromer}, that are successful at extracting information and creating value from large time series datasets. However, the proposed methods are also resource-intensive. Thus, storing and preprocessing datasets is costly, and model training often calls for the use of specialized equipment and infrastructure~\cite{li2022camel, zhao2021DC, yu2020two, DBLP:conf/iclr/ChenZ0SWW0G24}, limiting the application on edge devices~\cite{wang2020apache, xiao2022time}.

An effective way to reduce costs associated with the use of large data is coreset construction~\cite{agarwal2004approximating, chai2023goodcore, li2022camel, hasani2018efficient}, which often involves clustering~\cite{feldman2011unified}, Gaussian mixture models~\cite{lucic2018training}, and streaming learning~\cite{li2022camel}. 
Coreset construction methods~\cite{lucic2018training, feldman2011unified} often define heuristic criteria to select the most representative subsets from full datasets to form small coresets, such that models trained on the coresets are competitive with those built on the full datasets. Unfortunately, heuristic coreset construction methods guarantee neither optimal solutions nor promise the presence of representative observations in relation to downstream tasks~\cite{zhao2021DC}. Given these limitations, a recent alternative approach, dataset condensation ~\cite{zhao2023improved} aims to directly synthesize small condensed, optimized datasets, not relying on representative subset-selection. 

\begin{figure}[t]
    \centering
    \includegraphics[scale=0.48]{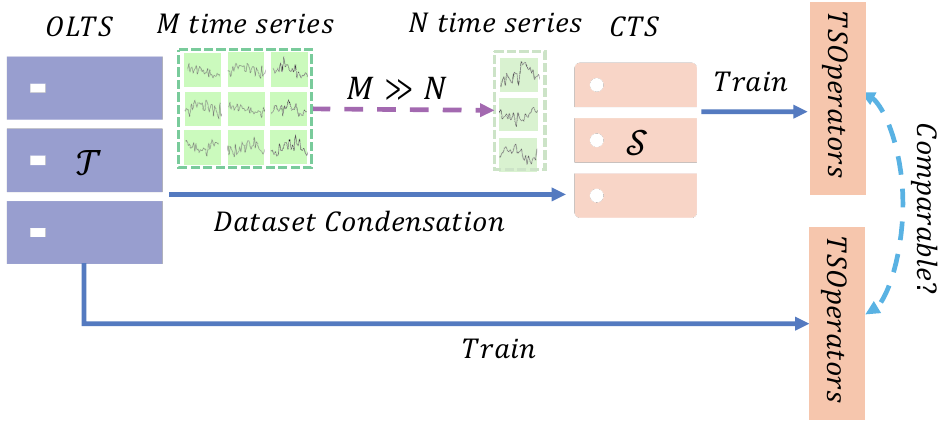}
    \vspace{-0.35cm}
    \caption{Time Series Condensation}
    \vspace{-0.48cm}
    \label{motivation}
\end{figure} 

We consider the novel problem of time series dataset condensation, where the goal is to synthesize a small but informative condensed time series dataset $\mathcal{S}$ derived from an original large dataset $\mathcal{T}$. A model, composed of stacked time series operators (\emph{TSOperators}) that perform feature extraction,
is trained on the condensed time series (CTS) dataset with the objective of achieving performance comparable to that of a model trained on the original large time series (OLTS) dataset for downstream tasks (see Figure~\ref{motivation}). 
The stacked \emph{TSOperators} are multiple \emph{TSOperators} arranged sequentially so that the output of one \emph{TSOperator} is fed into the next \emph{TSOperator}. Stacked \emph{TSOperators} enable more complex feature extraction. 
Time series are usually stored in TSMSs over edge devices~\cite{jensen2018modelardb, xiao2022time}. In the setting of continuously produced time series, time series dataset condensation aims to alleviate the storage burden of TSMSs considering the limited processing capability of edge devices, which further contributes to the \emph{database community}~\cite{aggarwal2008static, angiulli2007fast}. Additionally, due to the small-scaled CTS, time series dataset condensation is expected to bring significant efficiency improvements in repeated training scenarios in TSMSs, e.g., streaming learning~\cite{DBLP:conf/icde/00010GY0HXJ24} and model selection~\cite{DBLP:journals/vldb/WuWYZGQHSJ24}. 
Although substantial research has been devoted to inventing effective dataset condensation methods~\cite{zhao2021DC, cazenavette2022dataset, zhao2023improved}, 
such as gradient matching~\cite{zhao2021DC}, distribution matching~\cite{zhao2023improved}, and multi-step parameters matching~\cite{cazenavette2022dataset}, 
existing methods target image data and cannot be applied to time series data directly as they are not built to contend with the unique temporal dependencies present in time series~\cite{zhou2022fedformer, MiaoSCXW23}. These methods thus fail to capture inherent temporal dependencies such as trends and seasonalities~\cite{wen2021robustperiod}. New methods are needed to enable time series condensation. However, developing such methods is non-trivial, due to the following challenges.

\emph{Challenge I: Effectiveness and Generalization Ability.} It is challenging to guarantee the effectiveness and generalization ability of the condensed time series dataset~\cite{aggarwal2008static, cazenavette2022dataset}. 
First, typical dataset condensation methods require a bi-level (even triple-level) optimization to jointly learn a minimum of two objectives: model parameters and a condensed time series dataset. Such complex non-convex optimization cannot guarantee optimal solutions, thus significantly limiting its effectiveness as the representative of the large original time series dataset. Second, the condensed time series datasets should be generalized to train different networks. Nonetheless, the condensed data may suffer from various types of overfitting~\cite{cazenavette2022dataset}, e.g., overfitting to a certain network architecture. Also, the downstream models might overfit the condensed data during training.

\emph{Challenge II: Complex Temporal Dependencies.} Existing popular dataset condensation methods mainly focus on computer vision without specific modules to capture the unique time series characteristics~\cite{bonifati2022time2feat, wu2023autocts+}. Existing dataset condensation methods are ill-equipped for modeling the complex temporal dependencies of time series. For example, traffic conditions during afternoon rush hour may be similar on consecutive workdays. 
Moreover, morning rush hours may gradually start later when winter arrives as people get up later and later due to the gradual decrease in temperature and later sunrise. 
It is important to learn complex temporal dependencies in a time series dataset so that a condensed version of the dataset exhibits similar temporal patterns to the original dataset.
In addition, time series datasets are often multivariate, encompassing correlated variables, or features (channels). Such features are often coupled to perform feature extraction simultaneously. We argue that analysis in independent channels enables more effective feature extraction. For example, traffic flow is negatively correlated with vehicle speed, but positively correlated with road occupancy~\cite{wang2022multivariate}. 
Coupling features with varying and complex interactions might confuse a model, leading to decreased model performance~\cite{wang2022multivariate}. 

\emph{Challenge III: Scalability.} Existing dataset condensation methods often suffer from poor scalability, as bi-level optimization is generally time-consuming~\cite{zhou2022dataset}. For example, gradient-matching methods~\cite{zhao2021DC} set different hyper-parameters of the outer and the inner loop optimization for different learning settings, which requires extensive cross-validation, rendering condensed dataset synthesis costly. In addition, these methods may incur memory overflow: they require an entire model to reside in memory during training, which is infeasible in many resource-constrained environments~\cite{tai2018sketching}. It is highly desirable, but also non-trivial, to develop an efficient time series dataset condensation method.

This study addresses the above challenges by providing an efficient \underline{time} series \underline{d}ataset \underline{c}ondensation (TimeDC) framework, which features a two-fold matching mechanism: time series frequency matching and training trajectory matching. TimeDC synthesizes a small time series dataset that summarizes a large dataset such that models trained on the small dataset achieve comparable performance to models trained on the large dataset. It encompasses three major modules: a time series feature extraction module, a decomposition-driven frequency matching module, and a curriculum training trajectory matching module. Time series of stock prices are among the most common time series in real-world applications. We thus use such time series in a running example to to explain and convey the intuition underlying definitions and the proposed modules throughout the paper. 

To achieve effective and generalized TS dataset condensation (\emph{Challenge I}), we propose a curriculum training trajectory matching (CT$^2$M) module (see Figure~\ref{CTQM}), where we consider an original time series dataset as the gold standard and seek to imitate the long-term training dynamics of models trained on it. The parameter trajectories trained on the original dataset are called expert trajectories.
CT$^2$M first trains a set of expert trajectories of a model, composed of a set of stacked \emph{TSOperators}, on the large dataset. These are computed offline and are stored in an expert buffer and then serve as guidance for the condensed dataset optimization. Next, we sample trajectories from the expert buffer with curriculum trajectory queries and conduct long-term trajectory matching to align the segments of trajectories between the condensed time series dataset and the large dataset, enabling comprehensive and general knowledge transfer to the condensed dataset. 

To support the capture of complex temporal dependencies (\emph{Challenge II}), we propose a time series feature extraction (TSFE) module. In particular, TSFE contains a channel-independent mechanism and a set of stacked \emph{TSOperators}, where a \emph{TSOperator} includes a set of self-attention and fully connected layers. A decomposition block separates and refines the frequency, i.e., trend and seasonality, progressively from the intermediate results extracted by each \emph{TSOperator}. We design a decomposition-driven frequency matching (DDFM) module to facilitate consistent temporal pattern preservation between the condensed time series dataset and the original one. It maps the decomposed frequencies when optimizing the condensed dataset. 

To address the issue of high computation costs and poor scalability (\emph{Challenge III}), we propose an expert buffer to save expert trajectories pre-computed on original time series data to avoid memory overflow and reduce the overall training time. Further, we employ a patching mechanism that splits the time series into patches, thereby accelerating time series feature extraction in the TSFE module and also enabling effective local semantics modeling for time series.

The main contributions are summarized as follows.
\begin{itemize}[leftmargin=18pt, topsep=0pt]
    \item To the best of our knowledge, this is the first study to learn dataset condensation over time series. We propose a framework called TimeDC that aims to condense large time series datasets into small synthetic time series datasets while retaining the expressiveness of the large datasets.
    \item We design a TSFE module to capture temporal dependencies of time series effectively. We further propose a DDFM module to reduce the discrepancy of frequencies between original and condensed time series. 
    \item A novel curriculum trajectory query and matching module is proposed to penalize condensed time series data based on how far the synthetically trained model deviates from expert trajectories.
    \item We report on experiments using real data,  offering evidence of the effectiveness and efficiency of the proposals.
\end{itemize}

The remainder of this paper is organized as follows. Section~\ref{problem} covers preliminary concepts and formalizes the problem of time series dataset condensation. We detail the TimeDC framework in Section~\ref{method}, followed by the experimental study in Section~\ref{experiment}. Section~\ref{relatedwork} surveys related work, and Section~\ref{conclusion} concludes the paper.

%% file: chapter/02problem.tex
\section{Preliminaries}
\label{problem}
We proceed to present the necessary preliminaries and then define the problem addressed. 

\begin{definition}[Time Series]
    A time series $T = \langle t_1, t_2, \cdots, t_n \rangle$ is a time ordered sequence of $n$ observations, where each observation $t_i \in \mathbb{R}^C$ is a $C$-dimensional vector. If $C=1$, $T$ is univariate; otherwise, $T$ is multivariate. For example, a series of stock prices with multiple features, e.g., the opening price, closing price, highest price, and lowest price, is an example of multivariate time series.
\end{definition}

\begin{definition}[Time Series Dataset]
    A time series dataset $\mathcal{T}$ is a set of time series $\mathcal{T} = \left\{T_1, T_2, \cdots, T_M\right\}$, where $M$ is the cardinality.
\end{definition}

\subsection{Dataset Condensation over Time Series}

\begin{definition}[Condensed Time Series Dataset]
    Given a time series dataset $\mathcal{T}$, a time series dataset $\mathcal{S} = \left\{\widetilde{T}_1, \widetilde{T}_2, \cdots, \widetilde{T}_N\right\}$ is condensed if $N$ is much smaller than $M$, where $\widetilde{T}$ is the condensed time series, $M$ is the cardinality, and $\mathcal{S}$ is derived from $\mathcal{T}$.
\end{definition}
Note that we can initialize each item in $\mathcal{S}$ with a random time series in the original dataset or a Gaussian noise. After dataset condensation, we expect to learn an optimized $\mathcal{S}$ to replace $\mathcal{T}$.

Given a large time series dataset $\mathcal{T} = \left\{T_1, T_2, \cdots, T_M\right\}$, we aim to learn an optimal differentiable function $f$ parameterized by $\theta$ that correctly performs time series tasks, e.g., time series forecasting. 
We use $\mathcal{T}$ as a training set to learn the optimal parameters $\theta^\mathcal{T}$ by minimizing an empirical loss term as follows.
\begin{equation}
\label{thetaoptim}
    \theta^\mathcal{T} = \mathop{\arg\min}_{\theta}\mathcal{L}^\mathcal{T}(\theta),
\end{equation}
where $\mathcal{L}^\mathcal{T}(\theta) = \frac{1}{M}\sum_{T_i\in\mathcal{T}}\ell(f_\theta, T_i)$, and $\ell(\cdot)$ is a task specific loss (e.g., Mean Squared Error). We denote the generalization performance of the obtained model $f_{\theta^\mathcal{T}}$ by $\mathbb{E}_{T_i \sim P_{\mathcal{T}}}[\ell(f_{\theta^\mathcal{T}}, T_i)]$, where $P_{\mathcal{T}}$ is the data distribution of $\mathcal{T}$.

\begin{figure*}[!h]
    \centering
    \vspace{-0.4cm}
    \includegraphics[scale=0.58]{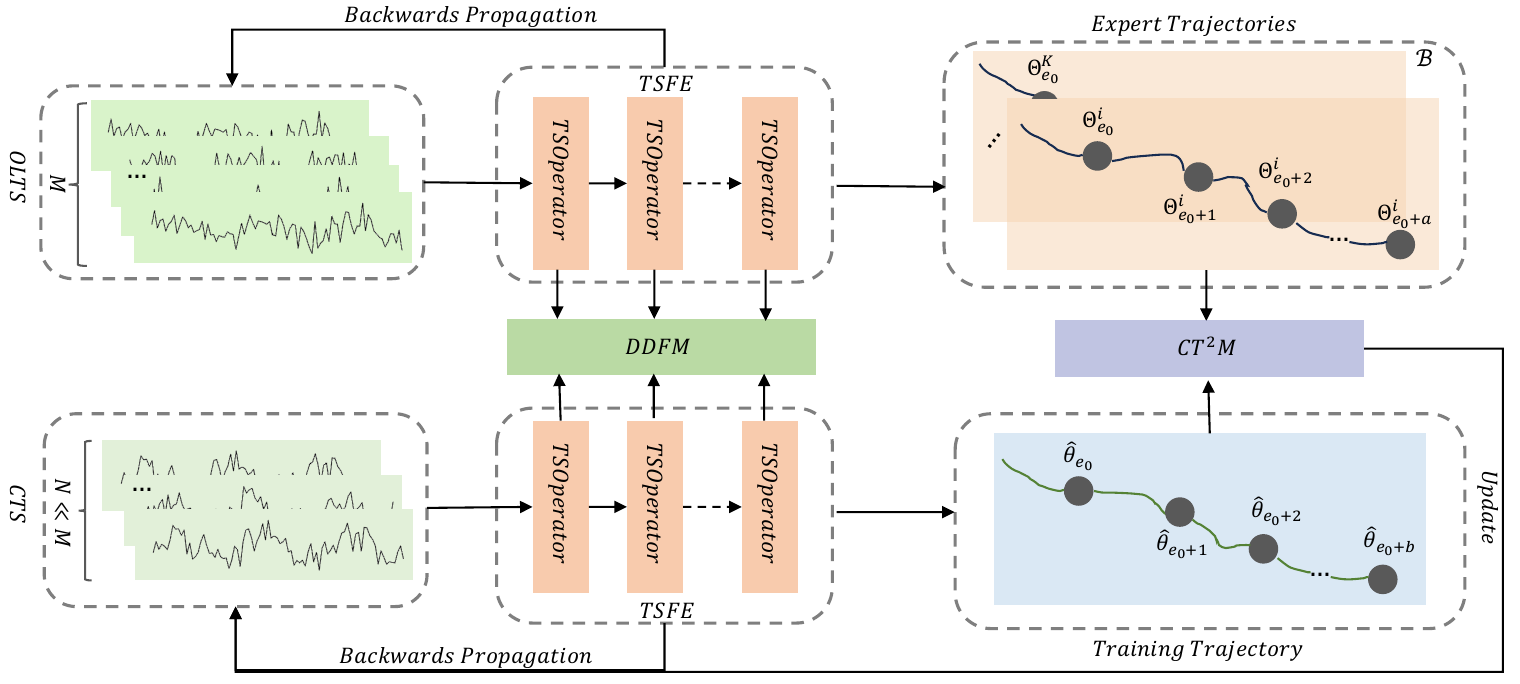}
    \vspace{-0.4cm}
    \caption{TimeDC Framework Overview}
    \vspace{-0.4cm}
    \label{overall_framework}
\end{figure*}

\textbf{Dataset Condensation over Time Series. } 
Given a large time series dataset $\mathcal{T}$, our goal is to synthesize a small time series dataset $\mathcal{S} = \left\{\widetilde{T}_1, \widetilde{T}_2, \cdots, \widetilde{T}_N\right\}$ that preserves most of the information in $\mathcal{T}$ and $N \ll M$. Once the condensed time series set $\mathcal{S}$ is learned, we can train $f$ on it by replacing $\mathcal{T}$ by $\mathcal{S}$ in Equation~\ref{thetaoptim}.
As the condensed set $\mathcal{S}$ is significantly smaller than $\mathcal{T}$, we expect the optimization to be considerably faster. In addition, we expect the generalization performance of $f_{\theta^\mathcal{S}}$ to be close to $f_{\theta^\mathcal{T}}$, i.e., $\mathbb{E}_{T_i \sim P_{\mathcal{T}}}[\ell(f_{\theta^\mathcal{T}}, T_i)] \simeq \mathbb{E}_{T_i \sim P_{\mathcal{S}}}[\ell(f_{\theta^\mathcal{S}}, T_i)]$, on the real data distribution $P_{\mathcal{T}}$. 

Thus, given a learnable function $f$ parameterized by $\theta$, our problem, \emph{dataset condensation over time series}, can be formulated as a bi-level optimization problem~\cite{zhao2021DC} as follows.
\begin{equation}
    \overbrace{\min_{\mathcal{S}} \mathcal{L}(f_{\theta^\mathcal{S}}, \mathcal{T})}^{\mathit{outer\, level}}, \, s.t. \, \theta^\mathcal{S} = \overbrace{\mathop{\arg\min}_{\theta}\mathcal{L}(f_{\theta}, \mathcal{S})}^{\mathit{inner\, level}},
\end{equation}
where the outer loop optimizes the condensed time series set $\mathcal{S}$, while the inner loop aims to learn an optimal function $f_{\theta^\mathcal{S}}$ on $\mathcal{S}$.

\textbf{Offline Expert Trajectories.} We denote the time sequence of parameters $\{\theta_e^k\}_{e=1}^E$ as expert trajectories, that are obtained during the training of the proposed TSFE module on the full, original time series dataset, where $E$ is the training epoch intervals. To generate these expert trajectories, we train TSFE several times on original datasets and save their snapshot parameters at every epoch. We call these parameter sequences as expert trajectories since they represent the theoretical upper bound for the downstream applications.

%% file: chapter/03method.tex
\section{Methodology}
\label{method}
We proceed to detail the efficient time series dataset condensation framework, TimeDC. We first give an overview of the framework and then provide specifics on each module in the framework.

\subsection{Framework Overview}
As illustrated in Figure~\ref{overall_framework}, given a model composed of stacked \emph{TSOperators}, we first pre-train a set of training trajectories on the original large time series (OLTS) dataset and store these pre-trained trajectories as expert trajectories in an expert buffer. As the original dataset is used to guide the network training, we denote the parameters trained on the original dataset as expert trajectory. Then, we train the same model on the condensed time series (CTS) dataset and perform two-fold modal matching: trajectory matching and frequency matching. Specifically, we optimize the condensed dataset with respect to the distance between the synthetically training trajectories and the trajectories trained on the original dataset. Further, we align the frequencies (i.e., trend and seasonality) between the original dataset and the condensed dataset to preserve temporal correlations when synthesizing the condensed dataset.
The framework encompasses three major modules: time series feature extraction (TSFE), decomposition-driven frequency matching (DDFM), and curriculum training trajectory matching (CT$^2$M). 

\begin{itemize}[leftmargin=12pt, topsep=0pt]
    \item \emph{Time Series Feature Extraction.} This module aims to extract effective high-dimensional features from the input time series. A channel-independent mechanism is adopted to decouple the time series $T_{\mathit{input}} \in \mathbb{R}^{\mathcal{B} \times n\times C} (C \geq 1)$ into $C$ univariate time series $\{T_{\mathit{input}}^c\}_{c=1}^C \in \mathbb{R}^{\mathcal{B} \times n\times 1}$ for subsequent feature extraction. Next, the module splits each decoupled univariate (i.e., channel-independent) time series into $P$ patches to learn local semantic information employing the patching mechanism. Then, the module stacks several \emph{TSOperators}, composed of self-attention and fully connected layers, for channel-independent feature extraction. 
    Finally, the channel-independent features are concatenated to enable prediction.
    \item \emph{Decomposition-Driven Frequency Matching.} This module adopts a decomposition-driven frequency matching mechanism to reduce the discrepancy of decomposed frequencies between the original time series dataset and the condensed time series dataset, aiming to preserve temporal dependencies from the original dataset when synthesizing the condensed dataset. Specifically, it incorporates a moving-average based~\cite{cleveland1990stl, wen2021robustperiod} series decomposition block into each \emph{TSOperator}, which separates the trend and seasonal information progressively from the learned features. This decomposition block enables a \emph{TSOperator} to alternately decompose and refine the intermediate results during the time series feature extraction.
    \item \emph{Curriculum Training Trajectory Matching.} This module first trains a set of training trajectories (i.e., parameters) of stacked \emph{TSOperators} on the large original time series dataset, which are stored in an expert buffer. In addition, we design a curriculum trajectory query and matching mechanism to organize training trajectories matching from similar (easy) to dissimilar (hard). In particular, we sample trajectories from the expert buffer with minimally modified parameters to those with highly modified parameters, iteratively.
\end{itemize}

We train TimeDC on the original dataset to extract frequencies and compute original parameters denoted as expert trajectories, aiming at guiding the effective condensed dataset synthesis. Thus, extracting frequencies and expert trajectories from the original dataset supports the purpose of TimeDC, which enables models trained on the small dataset to achieve \emph{comparable performance} to models trained on the large dataset. Next, we provide the technical details of each module.

\subsection{Time Series Feature Extraction}
\begin{figure}[t]
    \centering
    \includegraphics[scale=0.65]{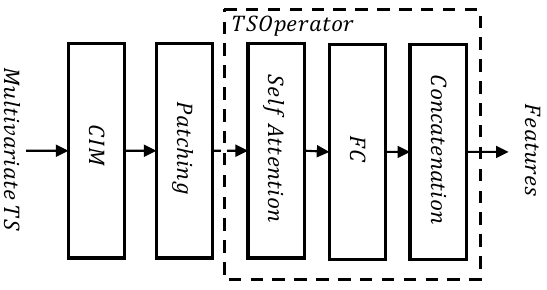}
    \vspace{-0.3cm}
    \caption{Feature Extraction Module}
    \vspace{-0.5cm}
    \label{featureextraction}
\end{figure}

We first consider the time series feature extraction (TSFE) module. As shown in Figure~\ref{featureextraction}, TSFE is composed of a channel-independent mechanism, a patching mechanism, and stacked \emph{TSOperators}. 

\textbf{Channel Independent Mechanism.} Most existing time series modeling methods~\cite{wu2023autocts+, wu2021autoformer} perform time series feature extraction by coupling all features and projecting them together into an embedding space. However, simply coupling different features may influence model performance very uncontrollably~\cite{wang2022multivariate}. To avoid this problem, we employ the channel-independent mechanism $\mathit{CIM(\cdot)}$ to model each time series feature independently.

We consider a time series with $C$ features as a sequence of multi-channel variables. Specifically, given a time series $T_{input} = \langle t_1, t_2, \cdots, t_n \rangle \in \mathbb{R}^{\mathcal{B}\times n \times C} (C \geq 1)$, the channel independent mechanism separates $T_{input}$ into $C$ univariate time series along the feature (i.e., channel) dimension, where $\mathcal{B}$ is the batch size. 
For example, in a series of stock prices with multiple features, e.g., opening prices and closing prices, the channel-independent mechanism separates these features by forming univariate time series. This means that the opening and closing prices are considered as two different time series in the subsequent feature extraction. 
Formally, the channel-independent mechanism works as follows.
\begin{equation}
    CIM(T_{\mathit{input}}) = T_{\mathit{input}}^1, \cdots, T_{\mathit{input}}^c, \cdots, T_{\mathit{input}}^C,
    \label{CIM}
\end{equation}
where $ T_{\mathit{input}}^c \in \mathbb{R}^{\mathcal{B} \times n \times 1}$. The separated univariate time series are then fed independently into the stacked \emph{TSOperators}. The \emph{TSOperators} share the same architecture, but their forward processes are independent.  


\textbf{Patching Mechanism.} Time series feature extraction aims to model the correlations between observations at different time steps. We aggregate observations over several time steps of the separated univariate time series $T_{input}^c$ into subseries-level patches~\cite{Yuqietal-2023-PatchTST} to enable local semantics modeling, which allows the model to see the longer historical sequences to improve feature extraction. For instance, in the context of stock prices, if the original time series represents hourly prices over a week, the patching mechanism may create patches where each patch contains the stock prices for each day, enhancing the capture of local semantics capturing.

We feed the univariate time series obtained from the channel independent mechanism into the patching mechanism that then divides them into patches that can be set to be overlapping or non-overlapping. Given a patch length $L$ and a stride (i.e., the non-overlapping region between two consecutive patches) $S$, a sequence of patches $T_{p}^c \in \mathbb{R}^{\mathcal{B}\times L \times \mathcal{P}}$ is generated, where $\mathcal{P} = \lfloor \frac{n-L}{S} + 2 \rfloor$ is the number of patches. The last value $t_n$ repeated $S-1$ times is padded to the end of the input time series before patching. 

The patching decreases the memory usage and computational complexity of the following \emph{TSOperator}, containing self-attention and fully connected (FC) layers quadratically, by a factor of $S$ since the input time series length reduced from $n$ to approximately $n/S$. 

\textbf{\emph{TSOperator}.} We then feed the patches $T_{p}^c$ into the stacked \emph{TSOperators} to learn latent features for the following frequency matching and trajectory matching. As illustrated in Figure~\ref{featureextraction}, a \emph{TSOperator} includes a multi-head self-attention layer followed by a fully-connected (FC) layer. In the context of stock prices, the multi-head self-attention mechanism learns the temporal correlations between different time steps. This may include how the opening prices at the beginning of a day influence the prices in the next few days. The FC layer summarizes these temporal correlations for future prediction. 
The $j$-$th$ \emph{TSOperator} can be formulated as follows.
\begin{equation}
h^j = \mathit{TSOperator}(h^{j-1}) = \mathit{Norm(FC(MultiHead}(h^{j-1}))),
\label{TSop}
\end{equation}
where $h^0 = T_p^c$, and $Norm(\cdot)$ is the normalization layer (i.e., BatchNorm).
The multi-head self-attention layer has three components: \emph{Query}, \emph{Key}, and \emph{Value}. The multi-head self-attention is computed by concatenating the output matrix of each attention head. 
For $i$-$th$ attention head, the input latent features $h^j$ are transformed as follows.
\begin{equation}
\small
\begin{split}
    &A_i^j = \mathit{Attention}(h^jW_i^Q, h^jW_i^K, h^jW_i^V)\\
    &\mathit{Attention}(Q, K, V) = \mathit{softmax}(\frac{Q\cdot K^T}{\sqrt{d_k}})\cdot V\\
    &h^{j+1} = \mathit{Norm}(h^j + \mathit{FC}_i(A_i^jh^j)),
\end{split}
\label{attention}
\end{equation}
where $d_k$ is a scaling factor, and $W_i^Q, W_i^K$, and $ W_i^V$ are the linear transformation parameter of the \emph{Query}, \emph{Key}, and \emph{Value}, respectively.

The time series feature extraction is shown in Algorithm~\ref{TTFMAlgorithm}. Specifically, we first separate the batch of input time series $T_{\mathit{input}}$ into $C$ univariate time series (line~\ref{chanel}) and apply the patching mechanism to split the univariate time series into patches (line~\ref{split}). Then, we input the patches obtained in line~\ref{split} into the stacked \emph{TSOperators} for feature extraction (lines~\ref{f1}--\ref{f2}). Finally, $C$ extracted features are concatenated to get the overall hidden features $h$ (line~\ref{op}). The space and time complexities of the TSFE module are both $\mathcal{O}(\frac{n^2}{S^2})$.

\begin{algorithm}[t]
    \caption{Time Series Feature Extraction}
    \label{TTFMAlgorithm}
    \SetKwInput{Parameters}{Input}
    \SetKwInput{Output}{Output}
    \Parameters{a batch of time series: $T_{input} \in \mathcal{B}\times n \times C$; the number of \emph{TSOperators}: $N_{op}$.}
    \Output{extracted features: $h$.}

    $T_{\mathit{input}}^1, \cdots, T_{\mathit{input}}^c, \cdots, T_{\mathit{input}}^C \leftarrow$ Separate $T_{input}$ into $C$ univariate channel-independent time series with Equation~\ref{CIM};
    \label{chanel}

    Split each univariate time series into patches;
    \label{split}

    \For{$1\leq c\leq C$}{
    \label{f1}
    $h_{c}^0 \leftarrow T_{input}^c$;
    
    \For{$0< j< N_{op}$}{
    Feature extraction with Equation~\ref{TSop};
    
    $h_{c}^j \leftarrow \mathit{TSOperator}(h_{c}^{j-1})$;
    }
    $h_{c}\leftarrow h_{c}^{N_{op}-1}$;
    }
    \label{f2}

    $h \leftarrow$ Concatenate $\{h_{c}\}_{c=1}^{C}$ along the channel;\\
    \label{op}
    \Return $h$.
\end{algorithm}

\subsection{Decomposition-Driven Frequency Matching}
Time series data usually exhibit distinct frequencies, such as trend and seasonality, separating from other modalities (e.g., images). For instance, the trend in stock prices indicates the prolonged direction of prices over time, which can be upward or downward. Seasonality refers to the regular, periodic fluctuations in stock prices, often influenced by seasonal business cycles or holidays. 
We argue that it would give more guidance on time series condensation if the condensed and original time series share similar frequencies. Thus, to preserve temporal patterns in the condensed time series dataset, we propose a novel decomposition-driven frequency matching (DDFM) mechanism that incorporates a decomposition block into the stacked \emph{TSOperators}. 

As shown in Figure~\ref{overall_framework}, for each \emph{TSOperator}, DDFM decomposes the intermediate hidden features, learned from the original dataset $\mathcal{T}$ and the condensed dataset $\mathcal{S}$, into their frequencies (i.e., trend and seasonality) progressively. It then aligns the frequencies to reduce the discrepancy of the temporal patterns between the original dataset and the condensed dataset based on the moving average. 
Concretely, for the hidden features $h^j$ of the $j$-$th$ \emph{TSOperator}, we adopt the moving average by means of average pooling $\mathit{AvgPool}(\cdot)$ to smooth periodic fluctuations and highlight the trend and seasonality. In particular, the process of series decomposition is as follows.
\begin{equation}
        h_{TRE}^j = \mathit{AvgPool(Padding}(h^j)), \, h_{\mathit{SEA}}^j =  h^j - h_{\mathit{TRE}}^j,
\end{equation}
where $h_{TRE}^j$ and $h_{SEA}^j$ are the decomposed trend and seasonality, respectively. We employ the average pooling $\mathit{AvgPool}(\cdot)$ with the padding operation $\mathit{Padding}(\cdot)$ to keep the time series length unchanged. We denote the decomposition process as follows.
\begin{equation}
    h_{\mathit{TRE}}^j, h_{\mathit{SEA}}^j = \mathit{SeriesDecomposition}(h^j)
    \label{decompose}
\end{equation}

Given the extracted features $h_{\mathcal{T}}^j$ and $h_{\mathcal{S}}^j$ of the original dataset $\mathcal{T}$ and the condensed dataset $\mathcal{S}$, we decompose the corresponding trend and seasonality according to Equation~\ref{decompose},
We use cosine similarity $\mathit{cos}(\cdot, \cdot)$ to measure similarities between the trends $h_{\mathit{TRE}_\mathcal{T}}^j$, $h_{\mathit{TRE}_\mathcal{S}}^j$ and seasonalities $h_{\mathit{SEA}_\mathcal{T}}^j$, $h_{\mathit{SEA}_\mathcal{S}}^j$, formulated as follows.
\begin{equation}
\begin{split}
    cos(h_{\mathit{TRE}_\mathcal{T}}^j, h_{\mathit{TRE}_\mathcal{S}}^j) = \frac{h_{\mathit{TRE}_\mathcal{T}}^j}{||h_{\mathit{TRE}_\mathcal{T}}^j||_2} \cdot \frac{h_{\mathit{TRE}_\mathcal{S}}^j}{||h_{\mathit{TRE}_\mathcal{S}}^j||_2}\\
    cos(h_{\mathit{SEA}_\mathcal{T}}^j, h_{\mathit{SEA}_\mathcal{S}}^j) = \frac{h_{\mathit{SEA}_\mathcal{T}}^j}{||h_{\mathit{SEA}_\mathcal{T}}^j||_2} \cdot \frac{h_{\mathit{SEA}_\mathcal{S}}^j}{||h_{\mathit{SEA}_\mathcal{S}}^j||_2},
\end{split}
\end{equation}
where $||\cdot||_2$ is \emph{$l_2$-$\mathit{norm}$}.

We aim to maximize the cosine similarity to reduce the discrepancy of frequencies between the condensed data and the original data. The objective function of DDFM is formulated as follows.
\begin{equation}
    L_{\mathit{Fre}} =  -\frac{1}{N_{op}}\sum_j^{N_{op}}(\mathit{cos}(h_{\mathit{TRE}_\mathcal{T}}^j, h_{\mathit{TRE}_\mathcal{S}}^j) + cos(h_{\mathit{SEA}_\mathcal{T}}^j, h_{\mathit{SEA}_\mathcal{S}}^j)),
    \label{fre}
\end{equation}
where $N_{op}$ is the number of \emph{TSOperators}.

\subsection{Curriculum Training Trajectory Matching}
To enable effective dataset condensation, we propose a curriculum training trajectory matching mechanism. Existing popular dataset condensation methods, which are often based on gradient matching~\cite{zhao2021DC}, conduct online gradient matching step by step. These methods have drawbacks when used for short-range matching, causing short-sightedness issues, thus failing to imitate holistic learning behaviors, and reducing the quality of condensed time series data. 

Motivated by an existing study~\cite{cazenavette2022dataset}, we match the long-term training trajectories of the TTFM module with the offline guidance of expert trajectories, which are pre-trained and stored in the expert buffer. We consider the expert trajectories (i.e., the performance of stacked \emph{TSOperators} trained on the original dataset $\mathcal{T}$) as the theoretical upper bound for the downstream application tasks. Our goal is to obtain a condensed dataset $\mathcal{S}$ that will induce a similar trajectory as that induced by the real training data $\mathcal{T}$ such that models trained on $\mathcal{T}$ and $\mathcal{S}$ achieve similar performance. In addition, we propose a curriculum training trajectory query to perform trajectory matching from similar to dissimilar expert trajectories, which further accelerates the model training. For example, expert trajectories derived from historical stock prices will then guide the alignment of training trajectories extracted from a condensed dataset. This alignment ensures that the condensed dataset's training trajectories capture similar dynamics (such as relationships across different time steps) to those learned from historical data. 

In particular, we first train $K$ time series models with the same architecture, i.e., stacked \emph{TSOperators}, denoted as $f_{\mathcal{T}}$, on the original large time series dataset. Then, we can obtain $K$ numbers of expert training trajectories that have a holistic knowledge of the original time series dataset in terms of $f_{\mathcal{T}}$'s training process. We save these model parameters $\{\Theta_{\mathcal{T}}^k\}_{k=1}^K = \{\theta_e^k\}_{e=1}^E$ at certain epoch intervals $E$ in the expert buffer $\mathcal{B}$. Finally, we design a curriculum trajectory query mechanism to sample training trajectories of the expert buffer from similar to dissimilar for trajectory matching. Here, the pre-training of $f_\mathcal{T}$ is offline and can be separated from the end-to-end time series condensation, reducing the online computation costs.

\begin{figure}[t]
    \centering
    \includegraphics[scale=0.3]{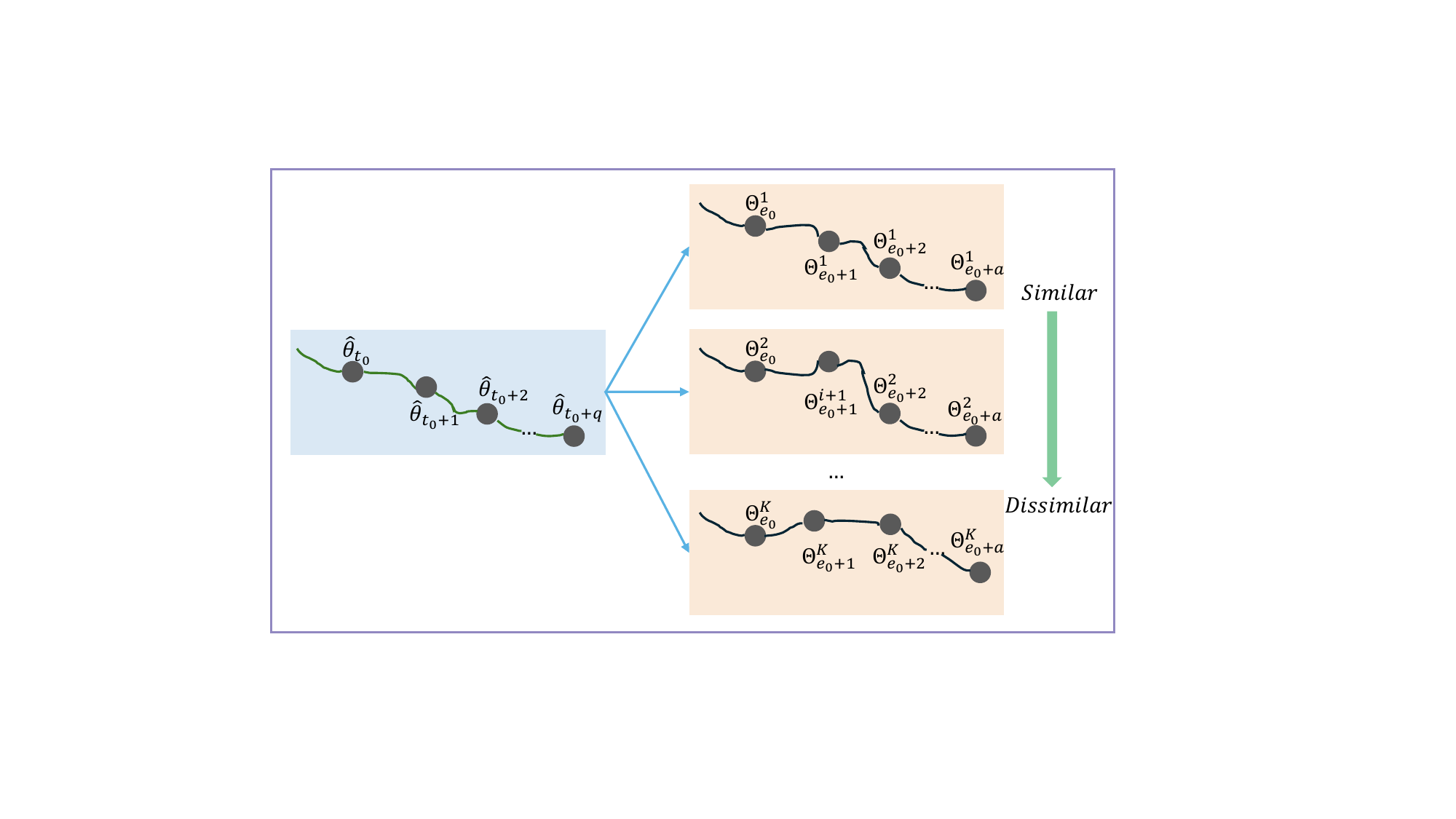}
    \vspace{-0.25cm}
    \caption{Curriculum Trajectory Query and Matching}
    \vspace{-0.3cm}
    \label{CTQM}
\end{figure}

\textbf{Training Trajectory Matching.} We use the original dataset to guide the network training, and the parameters trained on this dataset are called an expert trajectory. If a condensed time series dataset is capable of forcing network training dynamics to follow expert trajectories, the idea is that the synthetically trained network will be located close to the model trained on the original time series dataset and will achieve similar test performance for downstream applications.

When sampling a pre-computed parameter trajectory $\theta_e$ from the expert buffer $\mathcal{B}$, we aim to minimize the distance between the parameters $\widetilde{\theta_e}$ trained on the condensed dataset and $\theta_e$:
\begin{equation}
    \arg \min_{\mathcal{S}} \mathbb{E}_{\theta_e \sim \mathcal{B}}[L_{\mathit{tmm}}(\theta_e|_{e=e_0}^a, \widetilde{\theta_e}|_{e=e_0}^b)],
\end{equation}
where $\mathbb{E}$ is the expectation, $\theta_e|_{e=e_0}^a, \widetilde{\theta_e}|_{e=e_0}^b$ are the parameters with range $(e_0, e_0+a)$ and $(e_0, e_0+b)$, where $e_0<e_0+a<E$. More specifically, we update the condensed TS data according to the trajectory matching loss $L_{tmm}$.
\begin{equation}
    L_{\mathit{tmm}} = \frac{||\widetilde{\theta}_{e_0+b}- \theta_{e_0+a}||_2^2}{||\widetilde{\theta}_{e_0}- \theta_{e_0+a}||_2^2}
    \label{ltmm}
\end{equation}
Note that we initialize the parameters of $f_{\theta^\mathcal{S}}$ with those of $f_\mathcal{T}$ at $e_0$ training step, i.e., $\widetilde{\theta}_{e_0} = \theta_{e_0}$, for more focused training. This way, we align the learning behavior of $f_\mathcal{T}$ with $a$-steps optimization to $f_{\theta^\mathcal{S}}$ with $b$-steps optimization, imitating the long-term learning behavior of time series modeling.

For the inner loop, we train $f_{\theta^\mathcal{S}}$ on the condensed time series data for optimization until the optimal $\widetilde{\theta}^{*}$. Thus, the objective function of the CT$^2$M is defined as follows.
\begin{equation}
    \min_{\mathcal{S}} \mathbb{E}_{\theta_e^* \sim \mathcal{B}}[L_{tmm}(\theta_e|_{e=e_0}^a, \widetilde{\theta_e}|_{e=e_0}^b)] \quad
    s.t.\, \widetilde{\theta}^{*, \mathcal{S}} = \arg \min_{\widetilde{\theta}}\mathcal{L}(f_{\widetilde{\theta}}, \mathcal{S}),
\end{equation}
where $\mathcal{L}(\cdot)$ is the task-specific loss (e.g., Mean Square Error (MSE)). 

Utilizing training trajectory matching, we can reduce computation and memory costs during the condensation process by sampling pre-trained expert trajectories offline. Moreover, long-term trajectory matching provides a more holistic and comprehensive method to imitate the learning behaviors over the original dataset by avoiding that the condensed dataset fits to certain optimization steps short-sightedly.

\begin{algorithm}[t]
    \caption{Curriculum Training Trajectory Query and Matching}
    \label{curriculum}
    \SetKwInput{Parameters}{Input}
    \SetKwInput{Output}{Output}
    \Parameters{A buffer $\mathcal{B}$ with a set of trajectories pre-trained on the original TS dataset $\mathcal{T}$ parameterized by $\{\Theta_{\mathcal{T}}^k\}_{k=1}^K$; current model parameters $\widetilde{\theta}^{\mathcal{S}}$ on $\mathcal{S}$.}
    \Output{Trajectory matching loss $L_{tmm}$.}

    Distance list $DT \leftarrow []$;
    \label{DT0}

    Pre-update $\widetilde{\theta}^{\mathcal{S}}$ for $a$-steps with Equation~\ref{gradient};

    $\widetilde{\theta}_{e_0+a} \leftarrow \widetilde{\theta}_{e_0} - \sum_{s=1}^a(\alpha\nabla\mathcal{L}(f_{\theta^\mathcal{S}}, \mathcal{S}))$;

    \For{$\Theta_{\mathcal{T}}^k \in \{\Theta_{\mathcal{T}}^k\}_{k=1}^K$}{
    Compute the distance $dis_k$ between $\widetilde{\theta}|_{e_0}^a$ and $\Theta_{\mathcal{T}}^k$ with Equation~\ref{distance};

    $dis_k \leftarrow - D(\widetilde{\theta}|_{e=e_0}^a, \theta^k|_{e=e_0}^a)$;

    $DT \leftarrow (k, dis_k) $;
    }
    Rank $DT$ in a descending order;
    \label{DT}

    $\beta \leftarrow 0$
    \label{DT1}

    \While{$\beta<K$}{

    $k \leftarrow DT[\beta][0]$
    
    $L_{tmm} \leftarrow$ Sample trajectory $\Theta_{\mathcal{T}}^k$ and match the training trajectory according to Equation~\ref{ltmm};

    $\beta \leftarrow \beta + 1$;
    \label{DT2}
    }
    \Return $L_{tmm}$

\end{algorithm}

\textbf{Curriculum Training Trajectory Query.}
As shown in Figure~\ref{CTQM}, we propose a curriculum training trajectory query method to further enhance the generalization of time series condensation and accelerate the model convergence~\cite{zheng2021pace, wang2022iedeal}. Curriculum learning aims to train a model with easy samples first, and then gradually increases the difficulty levels. In our setting, we first query similar training trajectories $\theta_e|_{e=e_0}^a$ and $\widetilde{\theta_e}|_{e=e_0}^b$, learned on the original and condensed datasets, respectively, to perform trajectory matching. Then, we gradually increase the dissimilarity of the original data training trajectory. 

To quantify the similarity between training trajectories of the original dataset $\mathcal{T}$ and the condensed dataset $\mathcal{S}$, we design a minimally interfered retrieval sampling strategy. 
Specifically, given current parameters $\widetilde{\theta}_{e_0}$ of $f_{\theta^\mathcal{S}}$ learned on the condensed dataset, trajectories $\{\Theta_\mathcal{T}^k\}_{k=1}^K$ in buffer $\mathcal{B}$, and a standard task-specific objective function $\min_{\theta^\mathcal{S}}\mathcal{L}(f_{\theta^\mathcal{S}}, \mathcal{S})$, we retrieve trajectories that will be close to $\widetilde{\theta}$ by the update of the foreseen (i.e., future) parameters to select more similar trajectories from $\mathcal{B}$. We update the parameter $\widetilde{\theta}_{e_0}$ for $a$-steps by gradient matching, as shown in Equation~\ref{gradient}.
\begin{equation}
    \widetilde{\theta}_{e_0+a} = \widetilde{\theta}_{e_0} - \sum_{s=1}^a(\alpha\nabla\mathcal{L}(f_{\theta^\mathcal{S}}, \mathcal{S})),
    \label{gradient}
\end{equation}
where $\alpha$ is the learning rate. 
Then, we compute the distance $D(\cdot, \cdot)$ between the foreseen trajectory $\widetilde{\theta}|_{e=e_0}^a$ of $\mathcal{S}$ and sub-trajectories $\theta^k|_{e=e_0}^a$ of $\{\Theta_\mathcal{T}^k\}_{k=1}^K$ in $\mathcal{B}$ as follows.
\begin{equation}
    dis_k = -D(\widetilde{\theta}|_{e=e_0}^a, \theta^k|_{e=e_0}^a)
\end{equation}
We use the cosine similarity to measure the distance $D(\cdot)$.
\begin{equation}
    D(\widetilde{\theta}|_{e=e_0}^a, \theta^k|_{e=e_0}^a) = \frac{\widetilde{\theta}|_{e=e_0}^a}{||\widetilde{\theta}|_{e=e_0}^a||_2}\cdot \frac{\theta^k|_{e=e_0}^a}{||\theta^k|_{e=e_0}^a||_2}
    \label{distance}
\end{equation}

Finally, we match current model trajectories of $\mathcal{S}$ with the pre-trained trajectories in descending order based on the similarity $\{dis_k\}_{k=1}^K$.

Algorithm~\ref{curriculum} shows the process of the proposed curriculum training trajectory matching. Lines~\ref{DT0}--\ref{DT} concern the curriculum trajectory query, and lines~\ref{DT1}--\ref{DT2} concern the trajectory matching. The space and time complexities of Algorithm~\ref{curriculum} are $\mathcal{O}(K\cdot n)$ and $\mathcal{O}((a+K)\cdot n + KlogK)$, respectively.

\subsection{Overall Objective Function}
The final loss contains three parts: a task-specific loss $\mathcal{L}$, a frequency matching loss $L_{Fre}$, and a trajectory matching loss $L_{tmm}$. We combine them together and the overall loss is as follows.
\begin{equation}
    L_{all} = \mathcal{L} + L_{\mathit{Fre}} + L_{\mathit{tmm}}.
    \label{OOF}
\end{equation}
The task-specific loss $\mathcal{L}$ is specific to the particular downstream tasks to achieve a better condensed dataset tailored for the intended use. For example, we adopt the cross-entropy loss for task-specific optimization for time series classification tasks to optimize accuracy, while we adopt the mean square error (MSE) loss to minimize the discrepancy between ground truth and predicted values for time series forecasting tasks. 

%% file: chapter/04experiment.tex
\section{Experimental Evaluation}
\label{experiment}

\subsection{Experimental Setup}

\input{chapter/table/baseline}

\subsubsection{Datasets}
Various time series analytics are on edge devices where the storage is limited. We aim at getting a really small condensed dataset. 
The experiments are carried out on six widely-used time series datasets, covering four application domains: weather, traffic, economics, and energy. 
\begin{itemize}[leftmargin=12pt]
    \item \textbf{Weather.} The Weather dataset contains 21 indicators of weather (e.g., air temperature and humidity), which are collected in Germany. The data is recorded every 10 minutes. 
    \item \textbf{Traffic.} The Traffic dataset contains hourly road occupancy rates obtained from sensors located at San Francisco freeways from 2015 to 2016. 
    \item \textbf{Electricity.} The Electricity dataset contains the hourly electricity consumption of 321 clients from 2012 to 2014. 
    \item \textbf{ETT.} The ETT dataset includes two hourly-level datasets (ETTh1 and ETTh2) and two 15-minute-level datasets (ETTm1 and ETTm2). Each dataset includes 7 oil and load features of electricity transformers between July 2016 and July 2018. 
\end{itemize}

We choose time series forecasting as a representative downstream task, as it is a popular analytics task.  In Section~\ref{TSclassification}, we also present the performance comparison on the task of time series classification. The numbers of condensed time series are set to 500 and 50 as default for forecasting and classification tasks, respectively. We employ the proposed stacked \emph{TSOperators} as the forecasting and classification models.

\subsubsection{Baselines}
We compare TimeDC with the following existing methods that include coreset construction methods (i.e., Random, Herding~\cite{welling2009herding}, and K-Center~\cite{farahani2009facility}), 
and dataset condensation methods (i.e., DC~\cite{zhao2021DC} and MTT~\cite{cazenavette2022dataset} ).

\begin{itemize}[leftmargin=12pt]
    \item \textbf{Random.} The Random method randomly selects certain numbers of time series as a coreset.
    \item \textbf{Herding.} The Herding method adds time series observations to coresets greedily~\cite{welling2009herding}.
    \item \textbf{K-Center.} The K-Center method first performs K-Center clustering on the original datasets and then chooses observations from each cluster~\cite{farahani2009facility}.
    \item \textbf{DC.} The DC method employs gradient matching to perform dataset condensation~\cite{zhao2021DC}.
    \item \textbf{MTT.} The MTT method matches the multi-step training parameters of the condensed data and the original data~\cite{cazenavette2022dataset}.
\end{itemize}


\subsubsection{Evaluation Metrics}
Mean Absolute Error (MAE) and Root Mean Square Error (RMSE) are adopted as the evaluation metrics, which are defined as follows.
\begin{equation}
\small
        \mathit{MAE} = \frac{1}{M}\sum^{M}_{m=1}|\hat{\mathcal{Y}^m}-\mathcal{Y}^m|,\,\mathit{MSE} = \frac{1}{M}\sum^{M}_{m=1}||\hat{\mathcal{Y}^m}-\mathcal{Y}^m||^2,
\end{equation}
where $M$ is the testing data size, $\hat{y}^t$ is the prediction and $y^t$ is the ground truth. 
The smaller the MAE and the RMSE are, the more accurate method is.
We also evaluate the efficiency of the models, including the training and dynamic tensor memory cost.


\subsubsection{Implementation Details}
We implement our model using the Pytorch framework on an NVIDIA GTX 3090 GPU. The hyper-parameters in the model are set as follows. The patch length and stride are set to 16 and 8, respectively. The initial learning rate is 0.0001. The number of \emph{TSOperator} layers is 3. The number of heads in the self-attention layer is set to 16. The number of expert trajectories in the expert buffer is set to 10 by default. ETT datasets and other datasets are split into the training data, validation data, and test data by the ratio of $6:2:2$ and $7:1:2$, respectively. The parameters of the baseline methods are set according to their original papers and any accompanying code. All of the models follow the same experimental setup with prediction length $PL \in \{96, 192, 336\}$ on all datasets. 

\subsection{Experimental Results}

\subsubsection{Overall Performance Comparison}
We report the MAE and RMSE values of the methods in Table~\ref{baseline}. The best performance by an existing method (Random, Herding, K-Center, DC, and MTT) is underlined, and the overall best performance is marked in bold. Whole Dataset indicates training on the original dataset and serves as an approximate upper-bound performance; is marked in italics. We use a set of stacked TSFE modules as the basic forecasting model for each baseline. The following observations are made.
\begin{itemize}[leftmargin=12pt]
    \item TimeDC achieves the best results on all datasets across all prediction lengths ($PL \in \{96, 192, 336\}$). TimeDC performs better than the best among the baselines by up to $13.49\%$ and $26.59\%$ in terms of MAE and RMSE, respectively. We observe that the performance improvements obtained by TimeDC on the Weather dataset exceed those on the Traffic, Electricity, ETTh1, and ETTh2 in most cases. This is because the Weather has much more training data than the other datasets. TimeDC trained with more training data results in better results. Though the ETTm1 and ETTm2 datasets have substantial training data, TimeDC performs slightly better than baselines on these two datasets. This is because the temporal patterns in ETTm1 and ETTm2 are highly regular and can be captured easily by existing methods.
    \item The coreset methods (i.e., Random, K-Center, and Herding), perform worse than the condensation methods (i.e., DC, MTT, and TimeDC), where Random has the worst performance in most cases. This is because coreset methods are based on heuristic metrics, that makes it hard to select representative observations and guarantee the optimal solutions.
    \item A popular condensation method, MTT performs the best among the existing methods except when $PL = 96$ on ETTh1, due to its powerful condensation capability that involves matching multi-step parameters.
\end{itemize}

The experiment offers evidence that TimeDC is more effective than existing coreset and dataset condensation methods.

\begin{figure}[!htbp] \centering
	 \vspace{-0.45cm}
	\subfigure[Weather] {
		\includegraphics[scale=0.24]{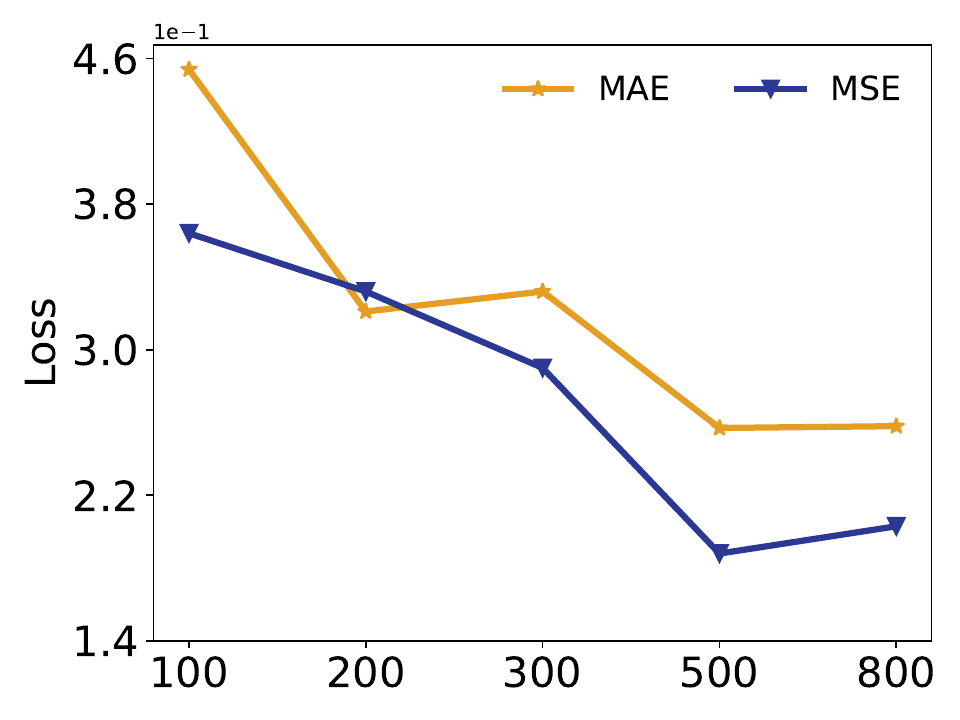} \label{weather_sample}
	}     
	\subfigure[Traffic] { 
		\includegraphics[scale=0.24]{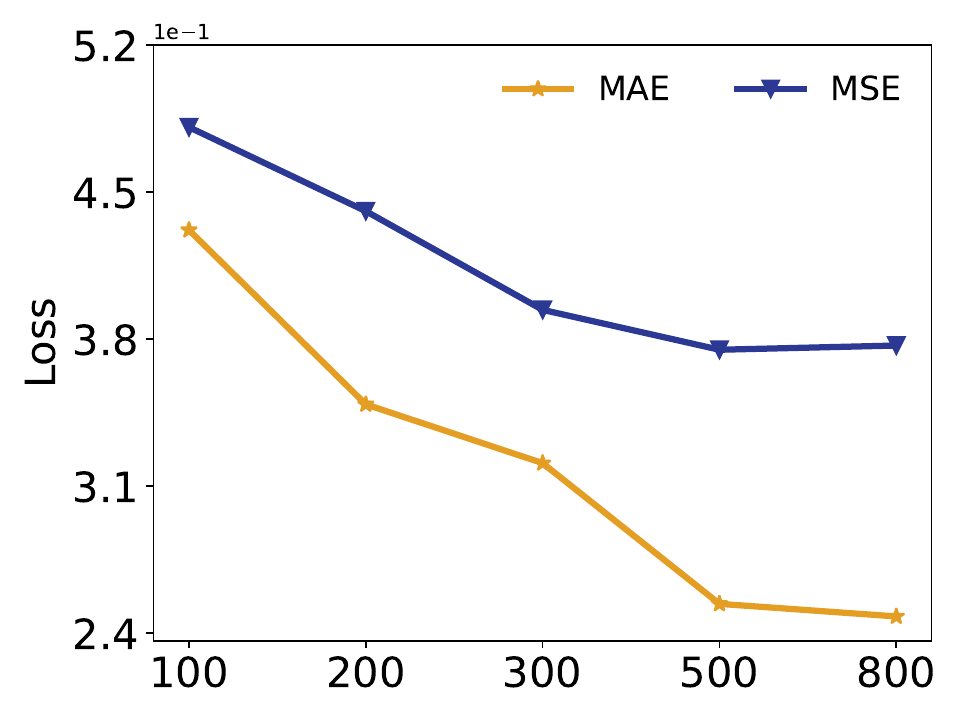}   \label{traffic_sample}  
	}  
        \subfigure[Electricity] {
		\includegraphics[scale=0.24]{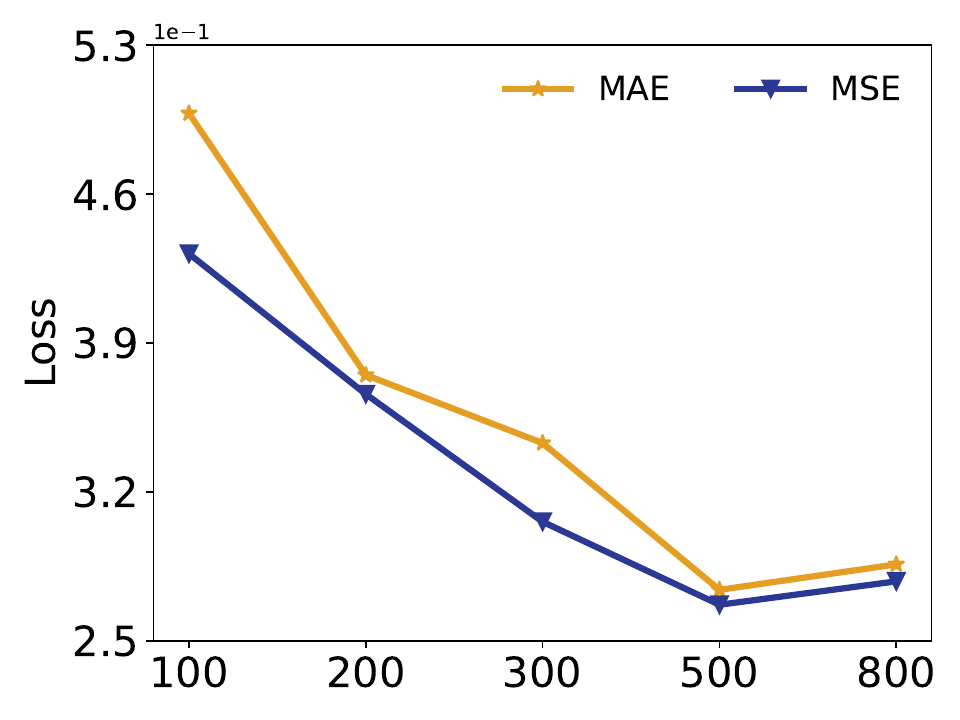} \label{electyri_sample}
	}     
	\subfigure[ETTh1] { 
		\includegraphics[scale=0.24]{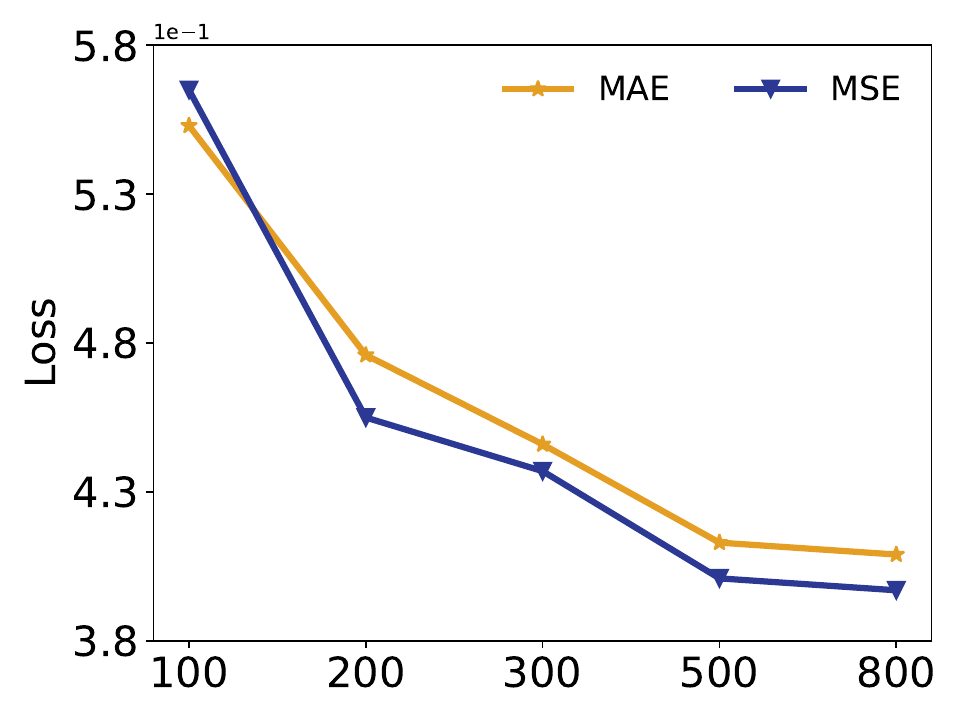}   \label{etth_sample}  
	}  
	\vspace{-0.4cm}
	\caption{Effect of the Size of Condensed TS Dataset on Four Datasets ($PL = 96$)}
	\label{sampleno} 
	\vspace{-0.34cm}
\end{figure}

\subsubsection{Effect of the Size of Condensed Time Series Dataset}
To study the effect of the size of a condensed time series dataset, we conduct experiments with 100, 200, 300, 500, and 800 condensed time series. The results are shown in Figure~\ref{sampleno}. We observe that the curves first drop significantly and then increase slightly (Figures~\ref{weather_sample} and~\ref{electyri_sample}) or remain almost the same (Figures~\ref{traffic_sample} and~\ref{etth_sample}). Generally, the results demonstrate that the model performance improves with an increase in the condensed time series data, as more condensed data yields more training data. In addition, it shows that using more condensed time series data for training is more likely to lead to better performance because more useful knowledge is learned from more data. One can also observe that TimeDC with 800 condensed time series performs slightly worse than TimeDC with 500 condensed time series data on the Weather and Electricity. This may be because the patterns in these datasets are relatively simple. Additional condensed time series data might introduce recurring patterns, thereby making the model overfit to these patterns and degrading performance on other data. 

\input{chapter/table/crossarchitecture}

\begin{figure}[!htbp] \centering
	\subfigure[Weather] {
		\includegraphics[scale=0.248]{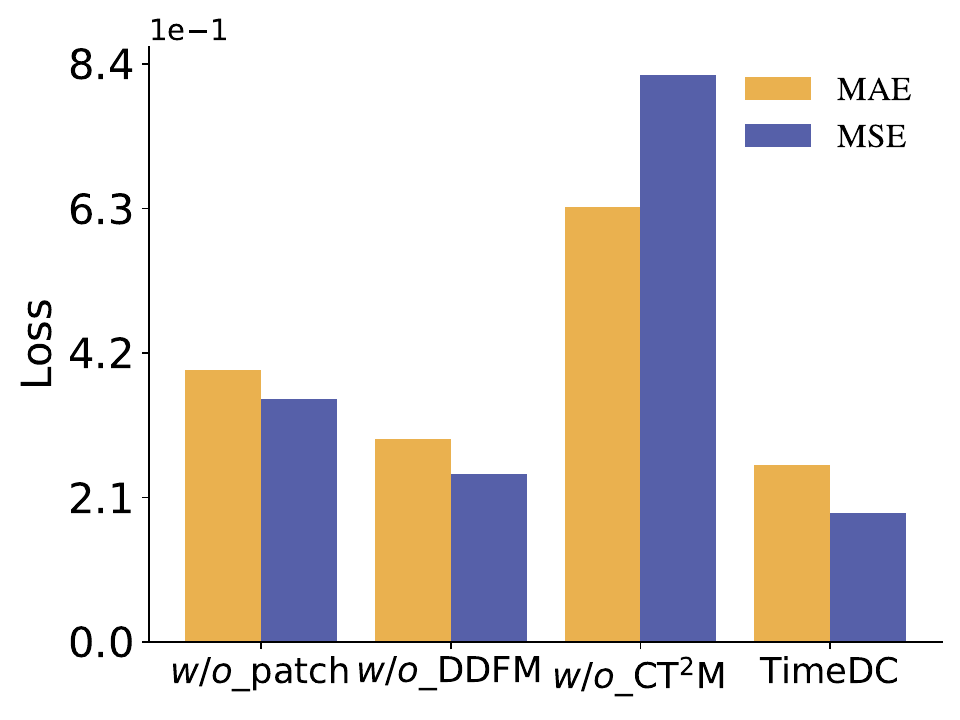} \label{weather_ablation}
	}     
	\subfigure[Traffic] { 
		\includegraphics[scale=0.248]{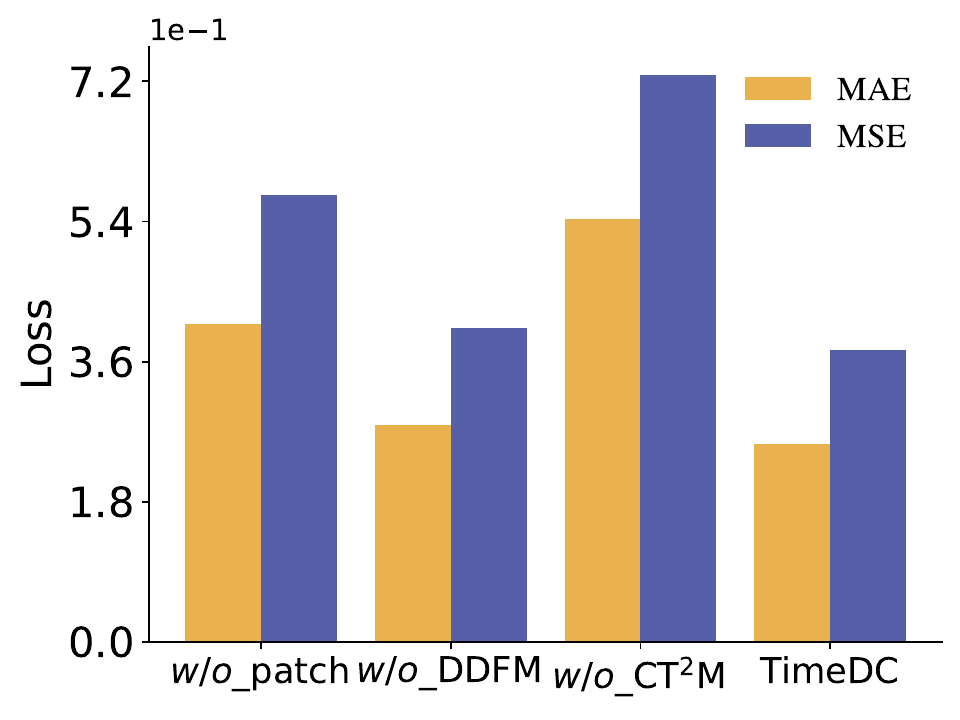}   \label{traffic_ablation}  
	}  
        \subfigure[Electricity] {
		\includegraphics[scale=0.248]{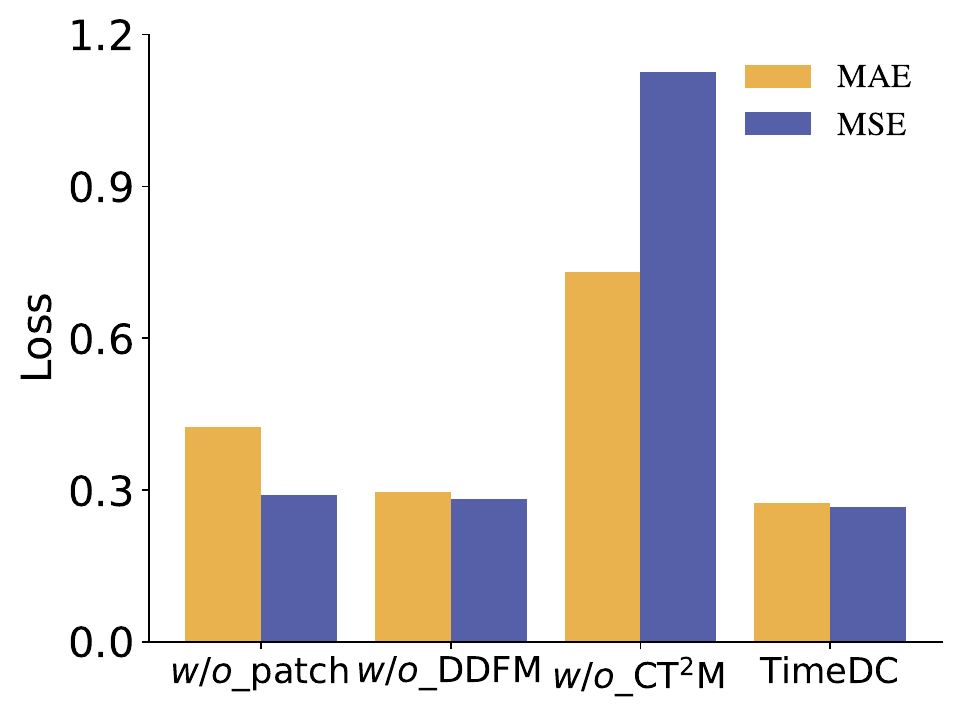} \label{elec_ablation}
	}     
	\subfigure[ETTh1] { 
		\includegraphics[scale=0.248]{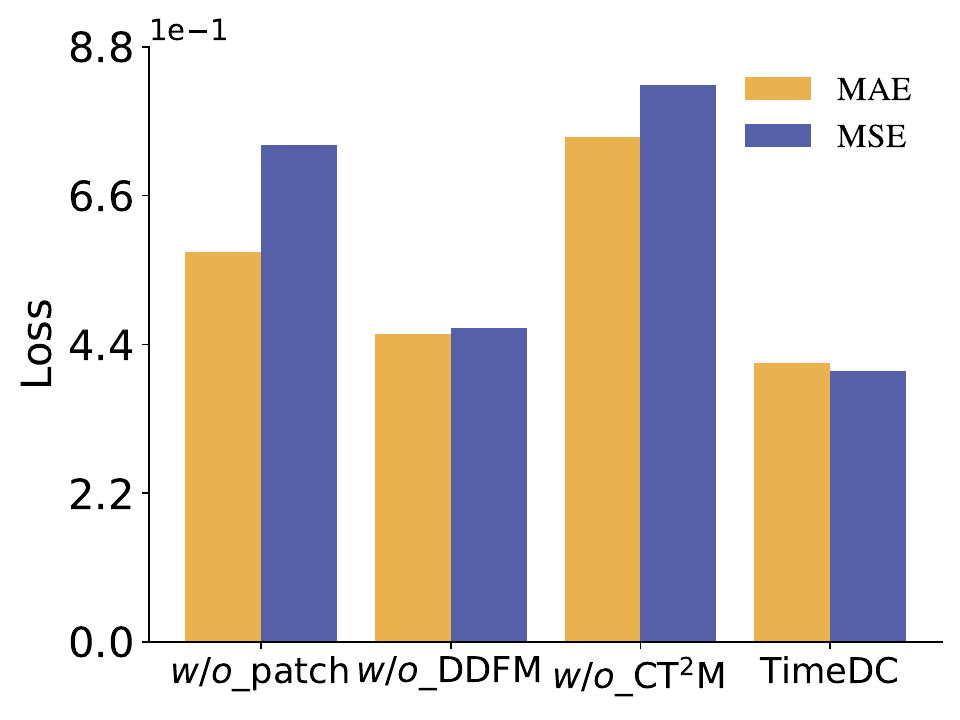}   \label{etth_ablation}  
	}  
	\vspace{-0.45cm}
	\caption{Performance of TimeDC and Its Variants on Four Datasets ($PL = 96$)}
	\label{ablation} 
	\vspace{-0.45cm}
\end{figure}

\subsubsection{Ablation Study}
To gain insight into the effects of the different components of TimeDC, including the patching mechanism (patch), decomposition-driven frequency matching (DDFM), and curriculum training trajectory matching (CT$^2$M), we evaluate three variants:
\begin{itemize}[leftmargin=12pt, topsep=0pt]
    \item \textbf{\emph{w/o\_Patch}}. TimeDC without the patching mechanism.
    \item \textbf{\emph{w/o\_DDFM}}. TimeDC without the DDFM module.
    \item \textbf{\emph{w/o\_CT$^2$M}}. TimeDC without the CT$^2$M module.
\end{itemize}

Figure~\ref{ablation} shows results on Weather, Traffic, Electricity, and ETTh1. Regardless of the datasets, TimeDC outperforms its counterparts without the patching mechanism, the DDFM module, and the CT$^2$M module. This shows that these three components are all useful for effective time series dataset condensation. TimeDC obtains MAE and RMSE reductions by up to $12.88\%$ and $22.95\%$, respectively, compared with w/o\_DDFM. Further, on all datasets, w/o\_CT$^2$M performs worst among all variants. TimeDC performs better than w/o\_CT$^2$M by at least 44.64\% and 48.49\% in terms of MAE and MSE, respectively, which indicates the effectiveness of the CT$^2$M module. 

\subsubsection{Cross-Architecture Performance}
Next, we consider the cross-architecture performance of TimeDC. It is important to determine whether the condensed time series data generated by TimeDC can be used to train an unseen network. To assess such cross-architecture performance comprehensively, we consider three representative state-of-the-art network architectures for time series forecasting, including \emph{Autoformer}~\cite{wu2021autoformer}, \emph{Informer}~\cite{zhou2021informer}, and \emph{Transformer}~\cite{vaswani2017attention}. We first synthesize condensed time series data with TimeDC and then train these networks with the condensed time series data. The hyper parameters of these network architectures are set based on their original papers and any accompanying code. For TimeDC, we use a set of stacked TSFE modules as the forecasting network.

The prediction results are given in Table~\ref{crossarchitecture}. Overall, TimeDC has the best performance in most cases, indicating its stable and superior performance, especially on Traffic and Electricity. It also illustrates the effectiveness of the TSFE module. We observe that Autoformer performs worse than TimeDC but better than Informer and Transformer, while the performances of Informer and Transformer are comparable, especially on ETTm1 and ETTm2. These observations are in line with their performances when trained on the original dataset~\cite{wu2021autoformer}, indicating that TimeDC learns a generalized condensed time series dataset that works across different network architectures. This is because of the powerful feature extraction capabilities of the proposed stacked \emph{TSOperators}, as well as the trajectory matching that imitates the long-term training dynamics of models trained on the original dataset. In addition, training a model on the condensed dataset consumes much less time than training that on the original dataset (see Table~\ref{costreduction_1}). This makes it possible to train different candidate models using the condensed data set when performing model selection in TSMSs, thus saving considerable training time.

\begin{table}[t]
\vspace{-0.05cm}
\caption{Dynamic Tensor Memory Cost on Four Datasets}
\vspace{-0.35cm}
\label{memory}
\setlength{\tabcolsep}{4mm}{
\begin{tabular}{c|cccc}
\hline
Dataset       & DC & MTT & TimeDC\\ \hline
Weather & 10.0 GB   & 8.9 GB & 3.3 GB\\
Traffic &17.8 GB&13.7 GB&10.9 GB\\
Electricity &8.5 GB&932.5 MB&516.0 MB\\
ETTh1  & 1.9 GB  & 845.5 MB   & 280.9 MB\\
\hline
\end{tabular}}
\vspace{-0.15cm}
\end{table}

\begin{figure}[h] \centering
	\vspace{-0.25cm}
        \subfigure[Weather] {
		\includegraphics[scale=0.24]{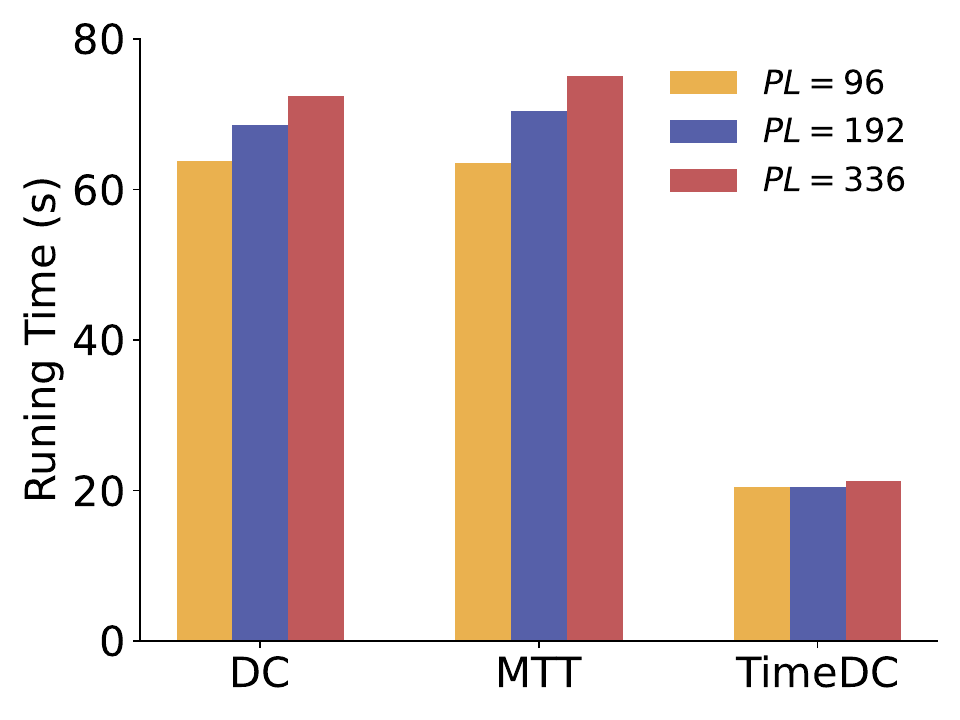} \label{weather}
	}     
	\subfigure[ETTh1] { 
		\includegraphics[scale=0.24]{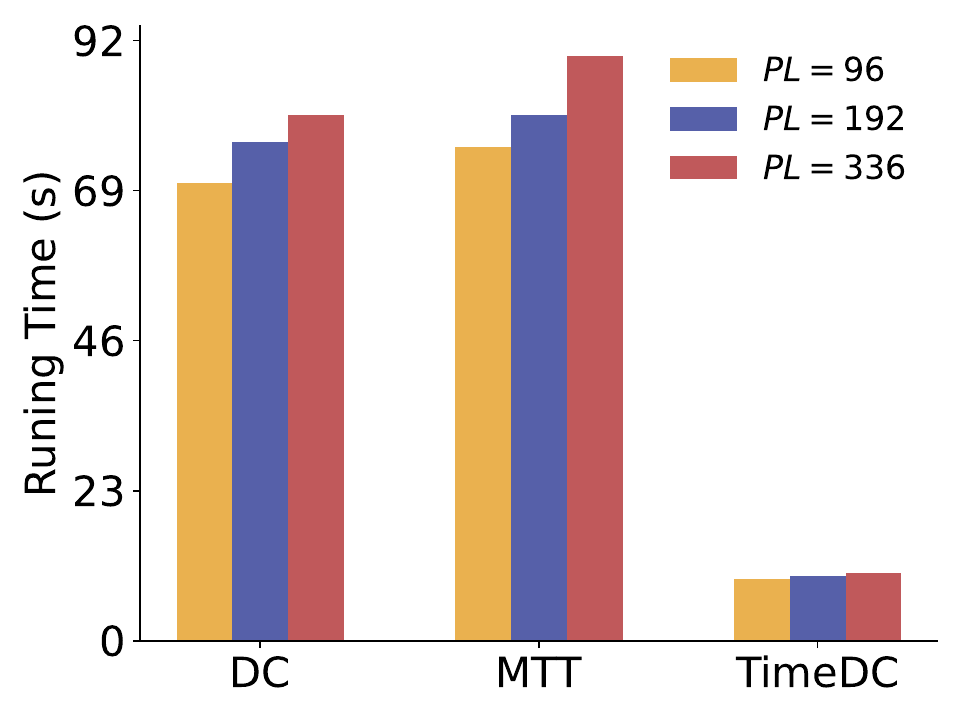}   \label{etth1}  
	} 
	\subfigure[Weather] {
		\includegraphics[scale=0.24]{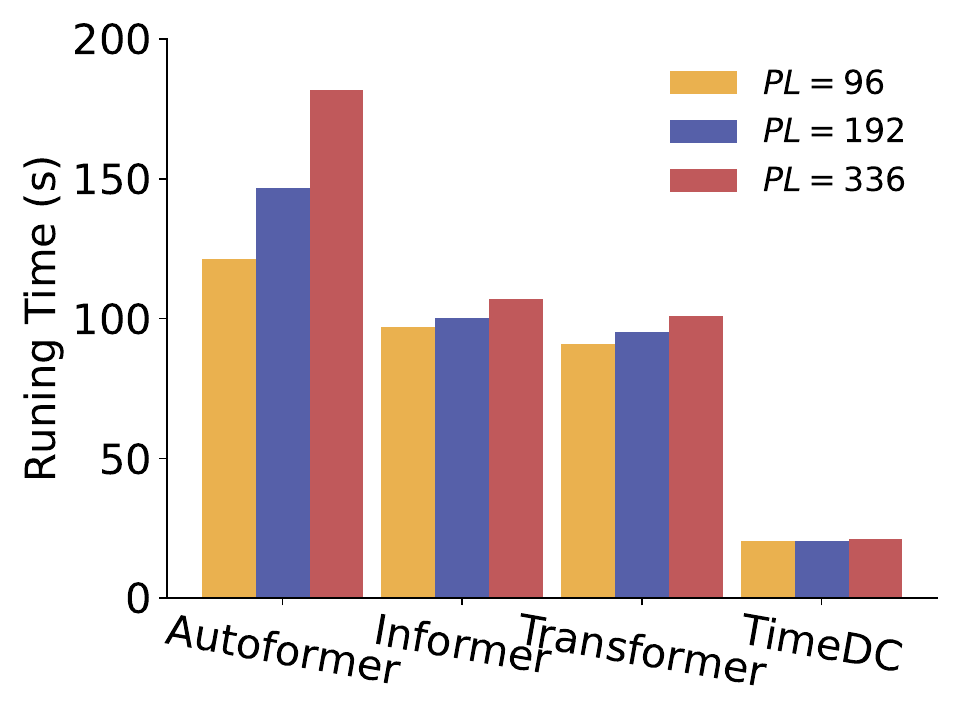} \label{weather_}
	}     
	\subfigure[ETTh1] { 
		\includegraphics[scale=0.24]{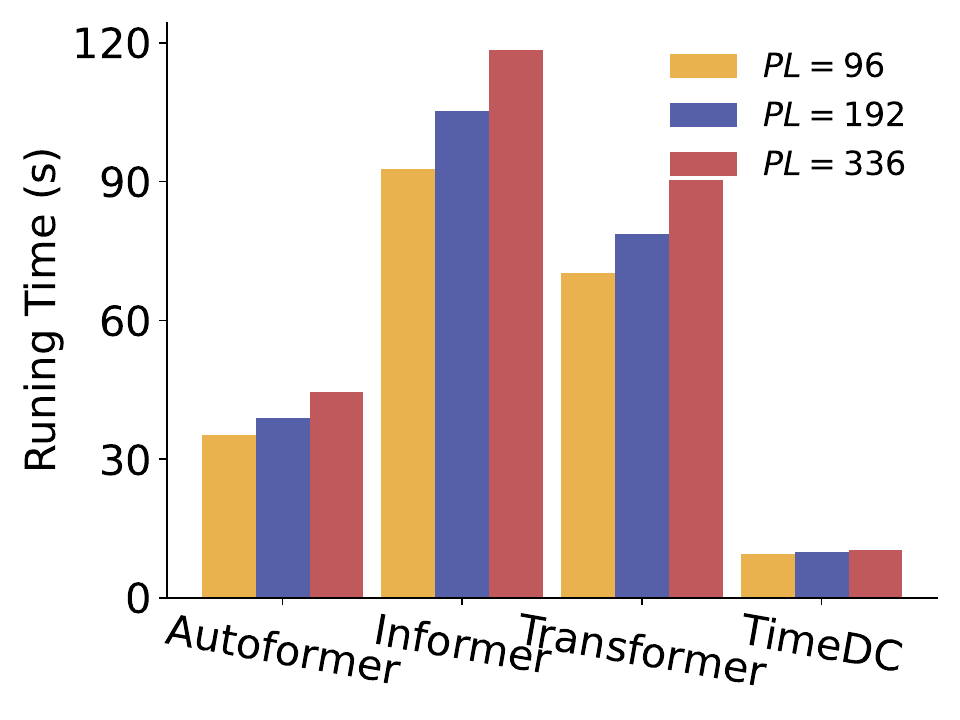}   \label{etth1_}  
	}   
	\vspace{-0.4cm}
	\caption{Training Time Comparison}
	\label{efficiency} 
	\vspace{-0.5cm}
\end{figure}

\subsubsection{Training Time}
As resource efficiency is important in dataset condensation to enable scalability, especially on resource-constrained edge computing devices, we study training time (of an epoch) for the condensation methods. Figures~\ref{weather} and~\ref{etth1} report the training time on Weather and ETTh1. We see that the training time of TimeDC is much lower than those of DC and MTT, which is largely because of the expert buffer in the CT$^2$M module that stores the pre-computed trajectories. This indicates the feasibility of TimeDC for model deployment in large time series dataset reduction scenarios. We also compare the training time of TimeDC and coreset methods, which are included in the code repository, showing TimeDC achieves training times comparable to those of coreset methods. But coreset methods, especially Herding, need more time to construct coresets while having worse performance.
Moreover, we compare the memory used by the dynamic (online) tensor across DC, MTT, and TimeDC in Table~\ref{memory}. TimeDC is able to reduce markedly the online memory and computation costs thanks to the training trajectories precomputed offline.

\begin{table}[h]
    \centering
    \vspace{-0.15cm}
    \caption{Training Time of TimeDC and Training Time on Condensed and Original Datasets (s/epoch)}
    \vspace{-0.35cm}
    \setlength{\tabcolsep}{2mm}
    \begin{tabular}{cccc}
    \hline
       Dataset  & TimeDC &Condensed Dataset & Original Dataset\\ \hline
       Weather & 22.39 & 4.31& 35.26\\
       Traffic & 232.34& 61.94&346.76\\
       Electricity &314.56& 41.14&522.85\\ 
       ETTh1 & 14.38& 4.93&20.43\\ \hline
    \end{tabular}
    \vspace{-0.1cm}
    \label{costreduction_1}
\end{table}

We also study training time across different network architectures on Weather and ETTh1---see Figures~\ref{weather_} and~\ref{etth1_}. It is clear that TimeDC consumes the least training time. TimeDC is faster than the other methods by at least 73.0\%, due to its patching mechanism, which reduces the complexity of the self-attention mechanism through input data simplification. Thus, TimeDC has lower training time across different networks, which still offering better performance. Finally, we compare the training time of TimeDC and the training time on original time series datasets based on the stacked \emph{TSOperators}, as shown in Table~\ref{costreduction_1}. One can see that the training time of TimeDC is notably lower than when using original datasets, showing the efficiency and practicality of time series dataset condensation. For example, the training time of TimeDC for dataset condensation is reduced by 39.84\% compared to using the original dataset on Electricity.

\begin{figure}[h] \centering
	 \vspace{-0.4cm}
        \subfigure[Weather] {
		\includegraphics[scale=0.245]{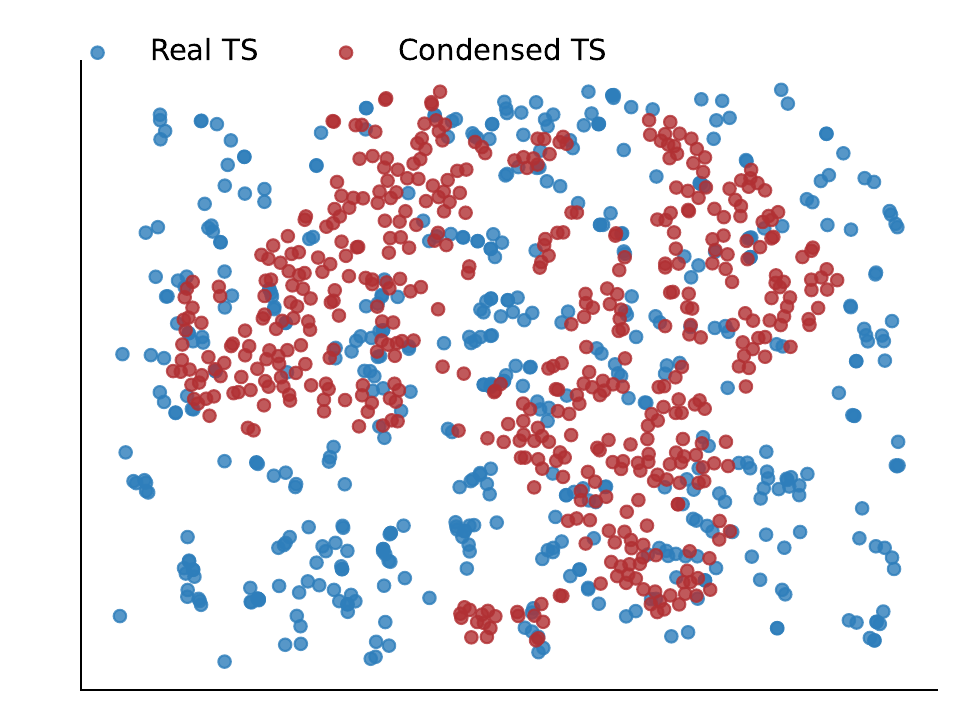} \label{tsne_weather}
	}     
	\subfigure[ETTh1] { 
		\includegraphics[scale=0.245]{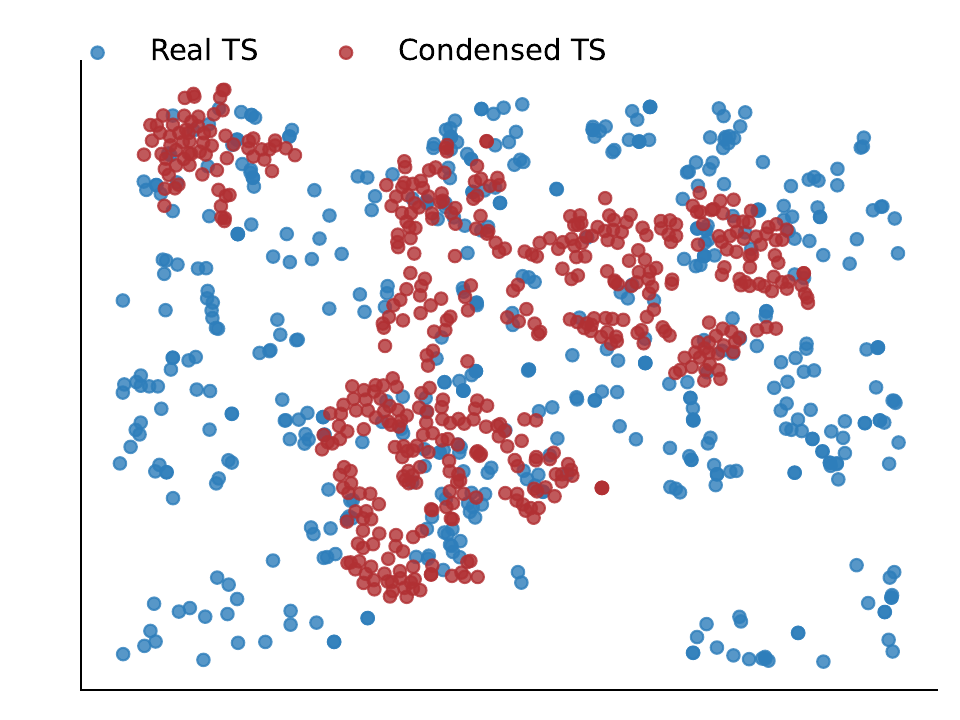}   \label{tsne_etth1}  
	} 
	\vspace{-0.4cm}
	\caption{Dataset Distribution Comparison on Two Datasets}
	\label{tsne} 
	\vspace{-0.35cm}
\end{figure}

\subsubsection{Case Study on Dataset Distribution}
To observe the effectiveness of TimeDC on synthesizing condensed time series datasets that cover the original time series distribution well, we visualize the t-SNE~\cite{van2008visualizing} graphs of the original time series dataset and the condensed time series dataset on Weather and ETTh1. Figure~\ref{tsne} compares the dataset distributions. Blue and red dots represent the original (i.e., real) time series dataset and the condensed time series dataset, respectively. We sample 500 time series from the original time series dataset for the visualization of the original time series. We observe that the red dots are integrated with the blue dots, indicating that the original time series and condensed time series datasets exhibit similar distributions. This indicates that the condensed dataset is of high quality and that the proposed method is effective. Moreover, the condensed dataset can cover the original dataset evenly on Weather, showing that TimeDC achieves a more diverse and robust condensation result on Weather that exhibits less complicated temporal patterns.

\begin{table}[h]
\vspace{-0.25cm}
\caption{Performance on Streaming Learning ($PL=96$)}
\vspace{-0.34cm}
\label{streaming}
\setlength{\tabcolsep}{0.65mm}{
\begin{tabular}{cccccc}
\hline
\multirow{2}{*}{Dataset}     & Metric & \multicolumn{2}{c}{MAE} & \multicolumn{2}{c}{RMSE} \\\cline{2-6} 
                             & Method & Autoformer${_f}$   & TimeDC   & Autoformer${_f}$    & TimeDC   \\ \hline
\multirow{3}{*}{Weather}     & $\mathcal{B}_0$     & 0.603        & 0.456    & 0.625         & 0.477    \\ \cline{2-6}
                             & $\mathcal{B}_1$     & 0.694        & 0.512    & 0.873         & 0.533    \\
                             & $\mathcal{I}$      & 0.823        & 0.656    & 1.014         & 0.512    \\ \hline
\multirow{3}{*}{Traffic}     & $\mathcal{B}_0$     & 0.541        & 0.465    & 0.657         & 0.489    \\ \cline{2-6}
                             & $\mathcal{B}_1$     & 0.635        & 0.502    & 0.829         & 0.611    \\
                             & $\mathcal{I}$      & 0.735        & 0.624    & 0.929         & 0.784    \\ \hline
\multirow{3}{*}{Electricity} & $\mathcal{B}_0$     & 0.557        & 0.437    & 0.601         & 0.542    \\ \cline{2-6}
                             & $\mathcal{B}_1$    & 0.628        & 0.504    & 0.875         & 0.599    \\
                             & $\mathcal{I}$      & 0.844        & 0.689    & 1.046         & 0.702    \\ \hline
\multirow{3}{*}{ETTh1}       & $\mathcal{B}_0$     & 0.655        & 0.573    & 0.812         & 0.705    \\\cline{2-6}
                             & $\mathcal{B}_1$     & 0.712        & 0.604    & 0.933         & 0.742    \\
                             & $\mathcal{I}$     & 1.297        & 0.762    & 2.015         & 0.979    \\ \hline
\end{tabular}}
\vspace{-0.35cm}
\end{table}

\subsection{Application on Streaming Time Series}
In real-world scenarios, time series data is often generated incrementally at edge devices that are distributed geographically. It may be preferable to process such streaming time series data on the edge directly to enable compliance with data access restrictions and exploit the potential for more efficient processing and lower latencies. However, time series data keeps growing while the capacities of edge devices are limited. We perform a performance, storage, and parameter comparison to determine whether our condensed time series data can be a key ingredient for streaming time series learning, thereby addressing also the catastrophic forgetting problem~\cite{chu2023continual, li2022camel}, due to its condensed nature. 


\subsubsection{Performance Comparison}
A naive solution to streaming time series learning is to use only newly arrived data, not all the data collected so far, to update model parameters continuously, called fine-tuning. We compare the performance of TimeDC and fine-tuning to test the effectiveness of TimeDC at streaming learning. For the fine-tuning, we use Autoformer as the basic model, entitled Autoformer$_f$. We split the original dataset into a base set $\mathcal{B}$ and an incremental set $\mathcal{I}$ with the ratio $7:3$. $\mathcal{B}$ and $\mathcal{I}$ are further split into training data, validation data, and test data with the ratios $7:1:2$. We first train and test the model on $\mathcal{B}$. The result is denoted as $\mathcal{B}_0$. Then, we update the learned model with $\mathcal{I}$ and test on $\mathcal{B}$ and $\mathcal{I}$ simultaneously, with the results denoted as $\mathcal{B}_1$ and $\mathcal{I}$, respectively. For the model update stage of TimeDC, we condense the base set and incorporate the condensed data into $\mathcal{I}$ for subsequent model training to address concept drift in evolving time series datasets. The results in Table~\ref{streaming} show that TimeDC has the best performance in terms of MAE and RMSE when compared with Autoformer$_f$ on four datasets. For example, TimeDC outperforms Autoformer$_f$ by $18.36\% \sim 21.54\%$ and $9.82\% \sim 32.89\%$ in terms of MAE and RMSE on Electricity, respectively. 
Autoformer$_f$ provides acceptable MAE and RMSE results on the base set, but their performance deteriorates on incremental set, especially on ETTh1. This indicates that a simple fine-tuning method is insufficient, due to forgetting problems caused by the possible concept drift.
TimeDC achieves relatively stable performance on all base sets and incremental sets, demonstrating its superiority.

\begin{table}[t]
\caption{Storage Comparison on Four Datasets}
 \vspace{-0.35cm}
\label{storage}
\begin{tabular}{c|cccc}
\hline
Storage       & Weather & Traffic & Electricity & ETTh1 \\ \hline
Whole Dataset & 2.9 GB   &20.0 GB & 11.2 GB     & 313.1 MB \\
Condensed TS  & 38.5 MB  & 827.5 MB   & 308.2 MB      & 12.8 MB \\
\hline
\end{tabular}
\vspace{-0.3cm}
\end{table}

\begin{table}[h]
\vspace{-0.25cm}
\caption{Parameter Comparison on Four Architectures}
\vspace{-0.35cm}
\label{parameter}
\begin{tabular}{c|cccc}
\hline
Dataset & Autoformer & Informer & Transformer & TimeDC \\ \hline
Weather & 10.6 M   &11.4 M & 10.6 M     & 1.8 M \\
Etth1  & 10.5 M  & 11.3 M   & 10.5 M      & 0.2 M \\
\hline
\end{tabular}
\vspace{-0.5cm}
\end{table}

\subsubsection{Storage and Parameter Comparison}
As storage size is a main concern in time series streaming learning on edge devices, we compare the storage cost of the pre-processed whole dataset and the condensed time series dataset using four datasets, as shown in Table~\ref{storage}. Generally, the condensed time series data of TimeDC reduces the storage space substantially compared with the whole dataset. The condensed time series data is only 4.09\% of the whole dataset on ETTh1. Thus, TimeDC enables much lower storage costs while achieving more promising performance on time series streaming learning.

Another characteristic of edge devices is limited computational capabilities. We thus compare the number of parameters of different network architectures including TimeDC (i.e., TSFE), Autoformer, Informer, and Transformer---see Table~\ref{parameter}. TimeDC achieves much fewer parameters, indicating that the TSFE module of TimeDC (mainly the patching mechanism) can significantly reduce computational costs. Tables~\ref{streaming},~\ref{storage}, and~\ref{parameter} indicate the feasibility and scalability of TimeDC for streaming learning on edge devices.

\begin{figure}[!htbp] \centering
	\vspace{-0.45cm}
	\subfigure[Accuracy] {
		\includegraphics[scale=0.24]{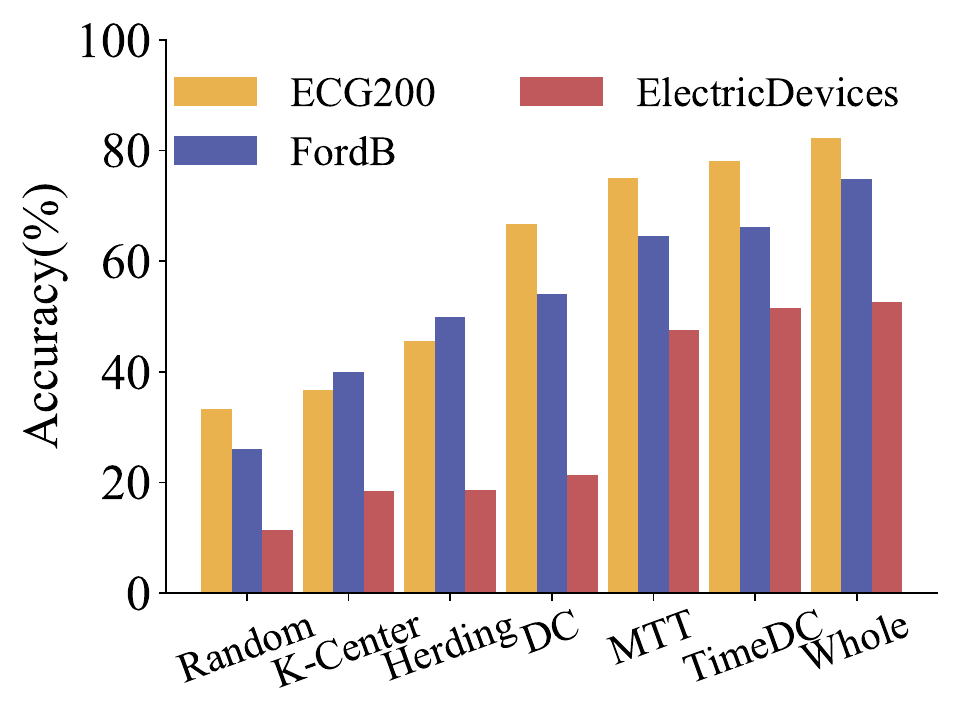} \label{cls_acc}
	}     
 \subfigure[Precision]{
 \includegraphics[scale=0.24]{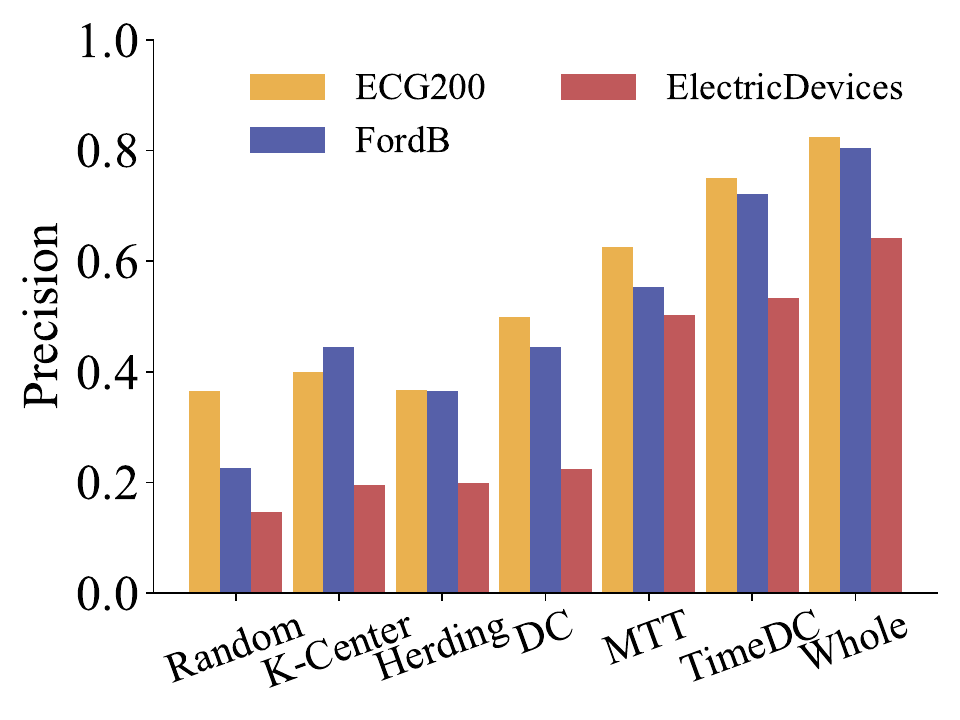} \label{cls_precision}
 }
	\vspace{-0.4cm}
	\caption{Time Series Classification Performance}
	\label{PCTSC} 
	\vspace{-0.5cm}
\end{figure}

\subsection{Performance on Time Series Classification}
\label{TSclassification}
It is important to show that the proposed TimeDC generalizes and can be used in other time series tasks. We conduct experiments on time series classification, considering three datasets (i.e., ECG200, ElectricDevices, and FordB) of the UCR time series archive\footnote[2]{\url{https://www.cs.ucr.edu/~eamonn/time_series_data_2018/}}. 
A set of stacked TSFE modules with an FC is used as the basic classification model. CrossEntropy Loss is used as the objective of time series classification. \emph{Accuracy} and \emph{Recall} are adopted as evaluation metrics. The number of condensed time series is set to 50 for each dataset. 
The overall performance results are provided in Figure~\ref{PCTSC}. The whole (dataset) indicates training on the whole original dataset, serving as an approximate upper-bound performance. TimeDC performs better than the best among the baselines by up to $10.40\%$ and $67.31\%$ in terms of \emph{Accuracy} and \emph{Precision}, respectively. All coreset construction methods perform worse than dataset condensation methods. Herding has the best performance in most cases among the coreset methods. Overall, TimeDC achieves the best results on three time series classification datasets among the baselines, which shows that TimeDC can be extended to other time series tasks.

%% file: chapter/table/baseline.tex
\begin{table*}[t]
\caption{Overall Performance Comparison on Seven Datasets}
\vspace{-0.4cm}
\label{baseline}
\setlength{\tabcolsep}{1.8mm}{
\begin{tabular}{cc|cccccccccccccc}
\hline
\multicolumn{2}{c|}{Baseline}         & \multicolumn{2}{c}{Random} & \multicolumn{2}{c}{K-Center} & \multicolumn{2}{c}{Herding} & \multicolumn{2}{c}{DC} & \multicolumn{2}{c}{MTT} & \multicolumn{2}{c}{TimeDC} & \multicolumn{2}{c}{Whole Dataset} \\ \hline
Dataset & $PL$         & MAE          & MSE         & MAE           & MSE          & MAE          & MSE          & MAE        & MSE       & MAE        & MSE        & MAE            & MSE       & MAE            & MSE    \\ \hline
\multirow{3}{*}{Weather}     & 96  & 0.731        & 1.256       & 0.452         & 0.687        &  0.478      &0.677        & 0.361      & 0.514     & \underline{0.295}      & \underline{0.244}      & \textbf{0.257}          & \textbf{0.188} & \textit{0.239}  & \textit{0.182}       \\
                             & 192 & 0.786        & 1.302       & 0.487         & 0.723        &   0.512      &0.688         & 0.413      & 0.527     & \underline{0.344}      & \underline{0.301}      & \textbf{0.285}          & \textbf{0.247}   & \textit{0.261} & \textit{0.195}      \\
                             & 336 & 0.794        & 1.311       & 0.524         & 0.756        &  0.554       &    0.712     & 0.444      & 0.567     & \underline{0.368}      & \underline{0.328}      & \textbf{0.330}           & \textbf{0.287}  & \textit{0.282} & \textit{0.241}       \\ \hline
\multirow{3}{*}{Traffic}     & 96  & 0.675        & 1.125       & 0.503         & 0.576        &    0.483     &0.554         & 0.375      & 0.603     & \underline{0.279}      & \underline{0.403}      & \textbf{0.254}          & \textbf{0.375} & \textit{0.247} &\textit{0.337}        \\
                             & 192 & 0.712        & 1.144       & 0.514         & 0.604        &   0.517      &0.606         & 0.432      & 0.633     & \underline{0.336}      & \underline{0.442}      & \textbf{0.297}          & \textbf{0.405}   & \textit{0.265} & \textit{0.338}      \\
                             & 336 & 0.729        & 1.117       & 0.523         & 0.611        &     0.553    &0.654         & 0.449      & 0.676     & \underline{0.355}      & \underline{0.471}      & \textbf{0.312}          & \textbf{0.423}     & \textit{0.297} &\textit{0.360}    \\ \hline
\multirow{3}{*}{Electricity} & 96  & 0.421        & 0.669       & 0.448         & 0.583        &  0.501      &0.592         & 0.376      & 0.513     & \underline{0.296}      & \underline{0.283}      & \textbf{0.274}          & \textbf{0.267} & \textit{0.252} & \textit{0.268}        \\
                             & 192 & 0.450         & 0.743       & 0.476         & 0.601        &    0.534    &0.628         & 0.419      & 0.532     & \underline{0.315}      & \underline{0.337}      & \textbf{0.285}          & \textbf{0.294}  & \textit{0.239} & \textit{0.255}       \\
                             & 336 & 0.491        & 0.853       & 0.506         & 0.622        &     0.569    &0.477          & 0.436      & 0.544     & \underline{0.339}      & \underline{0.356}      & \textbf{0.304}          & \textbf{0.322}    & \textit{0.271} & \textit{0.285}     \\ \hline
\multirow{3}{*}{ETTh1}       & 96  & 0.523        & 0.745       & 0.554         & 0.698        &    0.536     &0.656         & 0.503      & \underline{0.442}     & \underline{0.456}      & 0.464      & \textbf{0.413}          & \textbf{0.401} & \textit{0.354} &\textit{0.386}        \\
                             & 192 & 0.557        & 0.786       & 0.578         & 0.722        &    0.589     &0.698         & 0.552      & 0.508     & \underline{0.504}      & \underline{0.471}      & \textbf{0.436}          & \textbf{0.428}   &\textit{0.362} & \textit{0.355}      \\
                             & 336 & 0.588        & 0.802       & 0.604         & 0.745        &    0.603      & 0.723        & 0.556      & 0.513     & \underline{0.498}      & \underline{0.464}      & \textbf{0.447}          & \textbf{0.431}   &\textit{0.409} &\textit{0.387}      \\ \hline
\multirow{3}{*}{ETTh2}       & 96  & 0.487        & 0.655       & 0.589         & 0.711        &   0.521      &0.589         & 0.463      & 0.524     & \underline{0.388}      & \underline{0.342}      & \textbf{0.368}          & \textbf{0.271} & \textit{0.324} & \textit{0.255}        \\
                             & 192 & 0.509        & 0.673       & 0.605         & 0.732        &   0.553      &0.621         & 0.488      & 0.536     & \underline{0.416}      & \underline{0.384}      & \textbf{0.389}          & \textbf{0.302}   & \textit{0.332} &\textit{0.257}      \\
                             & 336 & 0.524        & 0.689       & 0.623         & 0.744        &  0.564       &0.640         & 0.505      & 0.540      & \underline{0.435}      & \underline{0.455}      & \textbf{0.411}          & \textbf{0.334}   & \textit{0.376} & \textit{0.296}      \\ \hline
\multirow{3}{*}{ETTm1}       & 96  & 0.743        & 1.124       & 0.525         & 0.492        &    0.607     &0.554         & 0.603      & 0.665     & \underline{0.512}      & \underline{0.453}      & \textbf{0.503}          & \textbf{0.442} &\textit{0.453} & \textit{0.403}         \\
                             & 192 & 0.764        & 1.245       & 0.566         & 0.510         &    0.628     &0.571         & 0.597      & 0.647     & \underline{0.563}      & \underline{0.501}      & \textbf{0.512}          & \textbf{0.465}    & \textit{0.464} & \textit{0.432}     \\
                             & 336 & 0.801        & 1.128       & 0.571         & 0.523        &    0.644     &0.582         & 0.624      & 0.668     & \underline{0.552}      & \underline{0.488}      & \textbf{0.500}            & \textbf{0.483}    & \textit{0.477} & \textit{0.455}     \\ \hline
\multirow{3}{*}{ETTm2}       & 96  & 0.664        & 0.795       & 0.486         & 0.623        &    0.524     &0.558         & 0.472      & 0.535     & \underline{0.376}      & \underline{0.421}      & \textbf{0.354}          & \textbf{0.391} & \textit{0.347} & \textit{0.381}         \\
                             & 192 & 0.687        & 0.804       & 0.512         & 0.643        &      0.549   &0.583         & 0.488      & 0.567     & \underline{0.453}      & \underline{0.479}      & \textbf{0.401}          & \textbf{0.421}  & \textit{0.358} & \textit{0.403}       \\
                             & 336 & 0.702        & 0.823       & 0.558         & 0.661        &   0.598     &0.624        & 0.493      & 0.556     & \underline{0.473}      & \underline{0.523}      & \textbf{0.453}          & \textbf{0.474} &\textit{0.406} &\textit{0.435}  \\ \hline     
\end{tabular}}
\end{table*}

%% file: chapter/table/crossarchitecture.tex
\begin{table*}[t]
\centering

\caption{Cross-Architecture Performance Comparison}
\vspace{-0.3cm}
\setlength{\tabcolsep}{3mm}{
\begin{tabular}{cccccccccc}
\hline
\multirow{2}{*}{Method}     & \multirow{2}{*}{Metric} & \multirow{2}{*}{PL} & \multirow{2}{*}{Weather} & \multirow{2}{*}{Traffic} & \multirow{2}{*}{Electricity} & \multicolumn{4}{c}{ETT}                                                                                           \\ \cline{7-10}
                             &                         &                     &                          &                          &                              & \multicolumn{1}{c}{ETTh1} & \multicolumn{1}{c}{ETTh2} & \multicolumn{1}{c}{ETTm1} & \multicolumn{1}{c}{ETTm2}                                  \\ \hline
\multirow{4}{*}{TimeDC}      & \multirow{2}{*}{MAE}    & 96                  & \textbf{0.257}                    & \textbf{0.254}                    & \textbf{0.274}                        & \multicolumn{1}{c}{\textbf{0.413}} & \multicolumn{1}{c}{\textbf{0.368}} & \multicolumn{1}{c}{\textbf{0.503}} & \textbf{0.354}                                                  \\ 
                             &                         & 192                 & \textbf{0.285}                    & \textbf{0.297}                    & \textbf{0.285}                        & \multicolumn{1}{c}{\textbf{0.436}} & \multicolumn{1}{c}{\textbf{0.389}} & \multicolumn{1}{c}{\textbf{0.512}} & \textbf{0.401}                                                   \\ 
                             \cline{2-10} 
                             & \multirow{2}{*}{MSE}    & 96                  & \textbf{0.188}                    & \textbf{0.375}                    & \textbf{0.267}                        & \multicolumn{1}{c}{\textbf{0.401}} & \multicolumn{1}{c}{\textbf{0.271}} & \multicolumn{1}{c}{\textbf{0.442}} & \textbf{0.391}                                                   \\ 
                             &                         & 192                 & \textbf{0.247}                    & \textbf{0.405}                    & \textbf{0.294}                        & \multicolumn{1}{c}{\textbf{0.428}} & \multicolumn{1}{c}{\textbf{0.302}} & \multicolumn{1}{c}{\textbf{0.465}} & \textbf{0.421}                                                  \\ 
                             \hline
\multirow{4}{*}{Autoformer}  & \multirow{2}{*}{MAE}    & 96                  & \underline{0.312}                    & \underline{0.370}                    & \underline{0.343}                        & \multicolumn{1}{c}{\underline{0.453}} & \multicolumn{1}{c}{\underline{0.473}} & \multicolumn{1}{c}{\underline{0.548}} & \underline{0.342}                                                  \\ 
                             &                         & 192                 & \underline{0.381}                    & \underline{0.385}                    & \underline{0.355}                        & \multicolumn{1}{c}{\underline{0.478}} & \multicolumn{1}{c}{\underline{0.491}} & \multicolumn{1}{c}{\underline{0.550}} & \underline{0.334}                                                   \\ 
                             \cline{2-10} 
                             & \multirow{2}{*}{MSE}    & 96                  & \underline{0.255}                    & 0.597                    & \underline{0.236}                        & \multicolumn{1}{c}{\underline{0.465}} & \multicolumn{1}{c}{\underline{0.412}} & \multicolumn{1}{c}{\underline{0.542}} & \underline{0.265}                                                   \\ 
                             &                         & 192                 & \underline{0.334}                    & 0.613                    & \underline{0.264}                        & \multicolumn{1}{c}{\underline{0.493}} & \multicolumn{1}{c}{\underline{0.488}} & \multicolumn{1}{c}{\underline{0.532}} & \underline{0.287}                                                  \\ 
                             \hline
\multirow{4}{*}{Informer}    & \multirow{2}{*}{MAE}    & 96                  & 0.423                    & 0.430                    & 0.428                        & \multicolumn{1}{c}{0.773} & \multicolumn{1}{c}{0.842} & \multicolumn{1}{c}{0.576} & 0.552                                                  \\ 
                             &                         & 192                 & 0.482                    & 0.476                    & 0.446                        & \multicolumn{1}{c}{0.788} & \multicolumn{1}{c}{0.954} & \multicolumn{1}{c}{0.597} & 0.532                                                  \\ 
                             \cline{2-10} 
                             & \multirow{2}{*}{MSE}    & 96                  & 0.354                    & 0.643                    & 0.253                        & \multicolumn{1}{c}{0.992} & \multicolumn{1}{c}{1.032} & \multicolumn{1}{c}{0.624} & 0.402                                                   \\ 
                             &                         & 192                 & 0.478                    & 0.710                    & 0.271                        & \multicolumn{1}{c}{0.987} & \multicolumn{1}{c}{1.055} & \multicolumn{1}{c}{0.653} & 0.432                                                  \\ 
                             \hline
\multirow{4}{*}{Transformer} & \multirow{2}{*}{MAE}    & 96                  & 0.389                    & 0.412                    & 0.398                        & \multicolumn{1}{c}{0.632} & \multicolumn{1}{c}{0.506} & \multicolumn{1}{c}{0.563} & 0.555                                                  \\ 
                             &                         & 192                 & 0.588                    & 0.431                    & 0.422                        & \multicolumn{1}{c}{0.612} & \multicolumn{1}{c}{0.513} & \multicolumn{1}{c}{0.588} & 0.576                                                   \\ 
                             \cline{2-10} 
                             & \multirow{2}{*}{MSE}    & 96                  & 0.344                    & \underline{0.578}                    & 0.267                        & \multicolumn{1}{c}{0.785} & \multicolumn{1}{c}{0.579} & \multicolumn{1}{c}{0.615} & 0.479                                                   \\ 
                             &                         & 192                 & 0.524                    & \underline{0.567}                    & 0.288                        & \multicolumn{1}{c}{0.732} & \multicolumn{1}{c}{0.542} & \multicolumn{1}{c}{0.643} & 0.455                                                  \\ 
                             \hline
\end{tabular}
\label{crossarchitecture}}
\end{table*}

%% file: chapter/05relatedwork.tex
\section{Related work}
\label{relatedwork}

\subsection{Time Series Modeling}
Time series modeling attracts increasing interest due to the growing availability of time series data and rich downstream applications~\cite{DBLP:conf/icde/KieuYGJZHZ22, kieu2024Team, DBLP:journals/pvldb/ZhaoGCHZY23, fang2024temporal, yanwww2022}, such as traffic prediction~\cite{wu2023autocts+, shekhar2007adaptive}, electricity prediction~\cite{zhou2021informer, Yuqietal-2023-PatchTST}, and anomaly detection~\cite{xu2024pefad, sylligardos2023choose}. Traditional time series analytic models are mostly based on statistical models~\cite{shekhar2007adaptive}. However, the statistical models cannot capture complex temporal correlations of time series data effectively due to their limited learning capacity. Recent advances in deep learning techniques have sparked a surge of interest in applying neural network architectures for time series modeling~\cite{wu2023autocts+, zhou2021informer, bonifati2022time2feat, xinle2024FACTS} outperforming traditional statistical models, including temporal convolutional network (TCN) based methods~\cite{liu2021modeling, cheng2023weakly}, recurrent neural network (RNN) based methods~\cite{salinas2020deepar}, and transformer based methods~\cite{zhou2021informer, zhou2022fedformer}. 
However, these methods are mostly supervised, and large training data is required resulting in high computational cost. At such scales, it becomes burdensome to store and preprocess the data and calls for specialized equipment and infrastructure to train machine learning models on them.

\subsection{Coreset and Dataset Condensation}
Coreset construction~\cite{agarwal2004approximating, li2022camel, chai2023goodcore, feldman2020turning, David2024Qcore} is the traditional dataset reduction approach that works by identifying the most representative training samples in an original dataset, aiming at achieving models trained on the coreset that are provably competitive with those built on the full dataset. Typically, coreset construction methods choose samples that are representative for training based on heuristic criteria~\cite{li2022camel, agarwal2004approximating, chai2023goodcore}, e.g., minimizing distance between coreset and whole-dataset centers~\cite{agarwal2004approximating}, 
Nonetheless, these heuristic methods cannot guarantee optimal solutions and the presence of representative samples for the downstream task. 
A recent approach, dataset condensation (or distillation)~\cite{zhao2021DC, cazenavette2022dataset}, is proposed to address these limitations by learning a small typical dataset that distills the most important knowledge from a given large dataset, such that a model trained on it can obtain comparable testing accuracy to that trained on the original training set. 
Existing dataset distillation methods have demonstrated superior performance, which can be categorized into matching-based methods~\cite{zhao2021DC, cazenavette2022dataset, zhao2023improved} and kernel-based methods~\cite{nguyen2020dataset, zhou2022dataset}. Matching-based methods generate synthetic datasets by matching gradients~\cite{zhao2021DC}, multi-step parameters~\cite{cazenavette2022dataset}, and distributions~\cite{zhao2023improved} between two surrogate models trained on the synthetic dataset and the original dataset. Kernel-based methods~\cite{nguyen2020dataset, zhou2022dataset} treat the synthetic dataset as the parameters to be optimized inspired by kernel functions e.g., neural tangent kernel. 
However, most of the above methods are designed for computer vision, and cannot be applied to time series directly due to complex temporal correlations such as seasonality and trend. In addition, these methods require substantial storage costs.

%% file: chapter/06conclusion.tex
\section{Conclusion}
\label{conclusion}
We present TimeDC, a new efficient time series dataset condensation framework that aims to synthesize a small but informative condensed time series dataset summarizing an original large time series dataset. To capture complex temporal dependencies, we design a time series feature extraction module with stacked \emph{TSOperators}. In addition, decomposition-driven frequency matching is proposed to ensure similar temporal patterns between the condensed and original time series datasets. To enable effective and generalized dataset condensation, we propose a curriculum training trajectory matching module with an expert buffer that aims to decrease the training cost.
Comprehensive experiments on original datasets offer evidence that TimeDC achieves state-of-the-art accuracy and requires fewer computational and storage resources.

\section{Acknowledgments}
This work was supported in part by the Independent Research Fund
Denmark under agreement 8048-00038B, the Villum Fonden under agreement 40567, the Innovation Fund Denmark project DIREC (9142-00001B), the National Natural Science Foundation of China (62372179, 62472068), Shenzhen Municipal Science and Technology R\&D Funding Basic Research Program (JCYJ20210324133607021), Municipal Government of Quzhou (2023D044), and Key Laboratory of Data Intelligence and Cognitive Computing, Longhua District, Shenzhen.

%% file: chapter/07Appendix.tex
\newpage
\appendix
\section{Appendix}

\begin{table}[h]
\renewcommand\arraystretch{1.1}
\footnotesize
\centering
\caption{\textbf{Summary of Notation}}
\vspace{-0.4cm}
\setlength\tabcolsep{17pt}
\scalebox{1}{
\begin{tabular}{cl}
    \hline
    \textbf{Symbol} &\textbf{Definition} \\
    \hline
    $w$ & Sliding window size\\

    $\mathit{TRE}$ & Trend\\

    $\mathit{SEA}$ & Seasonality\\
    
    $T$ & Time series\\

    $t_i$ & Observation in $T$ at the $i$-$th$ timestamp\\

    $\mathcal{T}$ & Time series dataset\\

    $M$ & Length of $\mathcal{T}$\\

    $\mathcal{S}$ & Condensed time series dataset\\

    $N$ & Length of $\mathcal{S}$ \\

    $\theta$ & Parameters learned on $\mathcal{T}$\\

    $\theta^\mathcal{S}$ & Parameters learned on $\mathcal{S}$\\

    $T_{\mathit{input}}^c$ & Channel-independent time series\\

    $\mathcal{B}$ & Expert buffer\\
    \hline
    
\end{tabular}}
\label{notation}
\end{table}
\subsection{Preliminaries}
\subsubsection{Notation}
Table~\ref{notation} lists notation used throughout the paper.

\subsection{Methodology}

\subsubsection{Algorithm}
\begin{algorithm}[h]
    \caption{The TimeDC Framework}
    \label{wholeprocess1}
    \SetKwInput{Parameters}{Input}
    \SetKwInput{Output}{Output}
    \Parameters{A buffer $\mathcal{B}$ with a set of pre-trained trajectories on the original time series dataset $\mathcal{T}$ parameterized by $\{\Theta_{\mathcal{T}}^k\}_{k=1}^K$; numbers of two-fold matching steps: $T_0$; $f_{\theta^\mathcal{S}}$ training steps: $T_1$; initialized condensed time series dataset: $\mathcal{S}$.}
    \Output{Optimized condensed time series dataset: $\mathcal{S}$.}
    \While{$\eta < T_0$}{
    Sample a pre-trained trajectory in $\mathcal{B}$ based on the curriculum trajectory query (see Algorithm~\ref{curriculum});
    \label{l_curr}
    
    calculate the $L_{tmm}$ according to Equation~\ref{fre});
    \label{l_trajm}

    \For{$\gamma < T_1$}{
    \label{l_fr}
    Train $f_{\theta^\mathcal{S}}$ via gradient matching and frequency matching (see Equation~\ref{ltmm});

    $\widetilde{\theta}_{\gamma+1} \leftarrow \widetilde{\theta}_{\gamma} - \alpha\nabla (\mathcal{L}(f_{\theta^\mathcal{S}}, \mathcal{S})+L_{Fre})$, where $\alpha$ is the learning rate;
    }
    \label{l_frend}
    Update the condensed time series dataset $\mathcal{S}$ according to Equation~\ref{OOF};
    \label{l_final}
    }

    \Return $\mathcal{S}$

\end{algorithm}
The whole process of TimeDC is shown in Algorithm~\ref{wholeprocess1}, where lines~\ref{l_curr}--\ref{l_trajm} cover the curriculum trajectory query and matching, lines~\ref{l_fr}--\ref{l_frend} cover the training process of $f_{\theta^\mathcal{S}}$ and frequency matching, and line~\ref{l_final} covers the optimization of $\mathcal{S}$.

\subsection{Experiment}
\subsubsection{Dataset Statistics}
\begin{table}[t]
    \centering
    \caption{Statistics of Forecasting Datasets}
    \vspace{-0.4cm}
    \begin{tabular}{cccc}
    \hline
       Dataset  & Feature & Time step & Granularity \\ \hline
    Weather  & 21 & 52696 & 10 minutes\\ 
    Traffic & 862 & 17544 & 1 hour\\ 
    Electricity& 321 & 26304 &1 hour \\ 
       ETTh1 \& ETTh2  &  7& 17420 & 1 hour\\ 
     ETTm1 \& ETTm2    &  7& 69680 & 15 minutes\\ \hline
    \end{tabular}
    \label{Dataset}
\end{table}

\begin{table}[h]
    \centering
    \caption{Statistics of Classification Datasets}
    \vspace{-0.4cm}
    \begin{tabular}{cccc}
    \hline
     Dataset  & Class & Length & Type \\ \hline
     ECG200 & 2& 96 & Electrocardiography\\ 
     ElectricDevices & 7& 96 &Device\\ 
     FordB & 2& 500 &Sensor\\ \hline
    \end{tabular}
    \label{Dataset_cls}
\end{table}
The forecasting and classification dataset statistics are provided in Tables~\ref{Dataset} and~\ref{Dataset_cls}, respectively. 

\begin{figure}[!htbp] \centering
	\subfigure[Performance on Weather] {
		\includegraphics[scale=0.22]{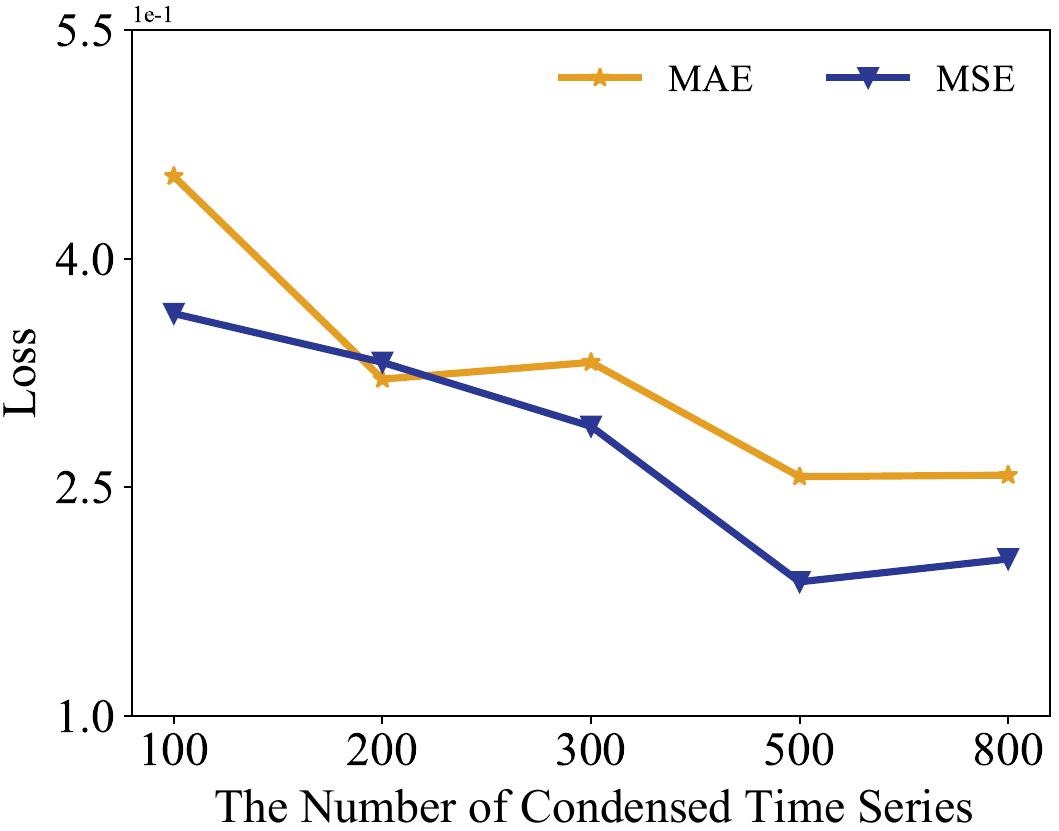} \label{weather_trade}
	}     
	\subfigure[Running Time on Weather] { 
		\includegraphics[scale=0.22]{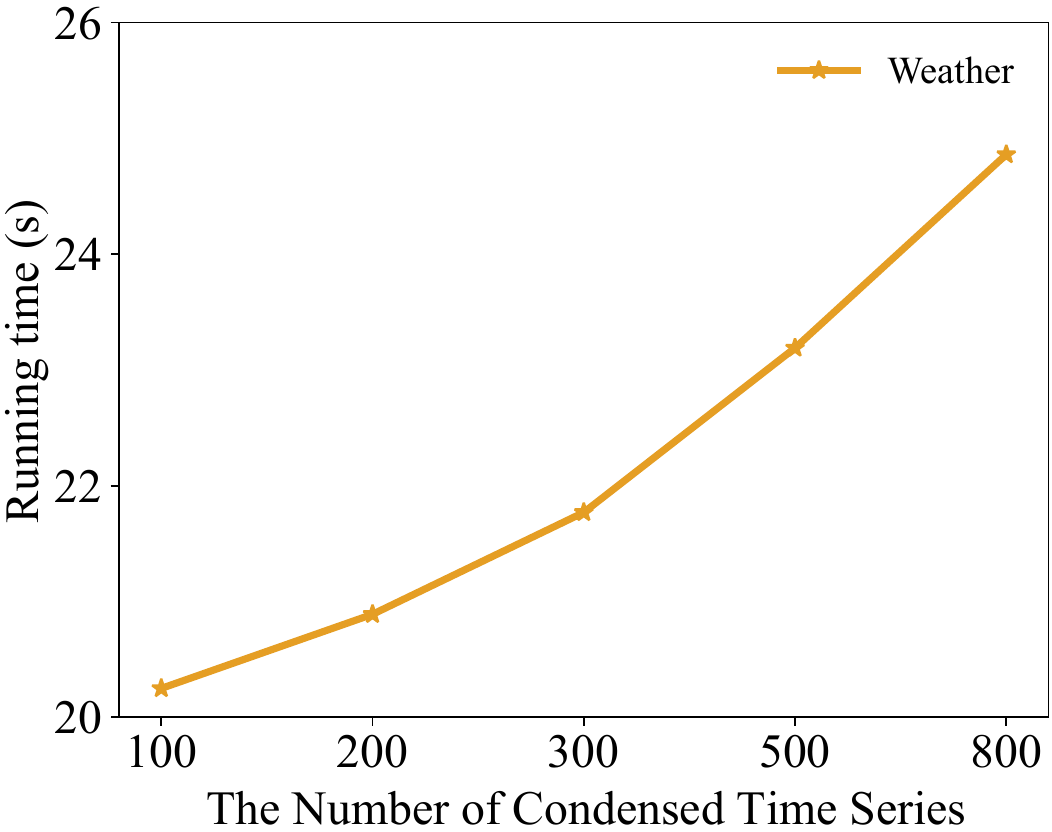}   \label{weather_efff}  
	}  
 \subfigure[Performance on Traffic] {
		\includegraphics[scale=0.22]{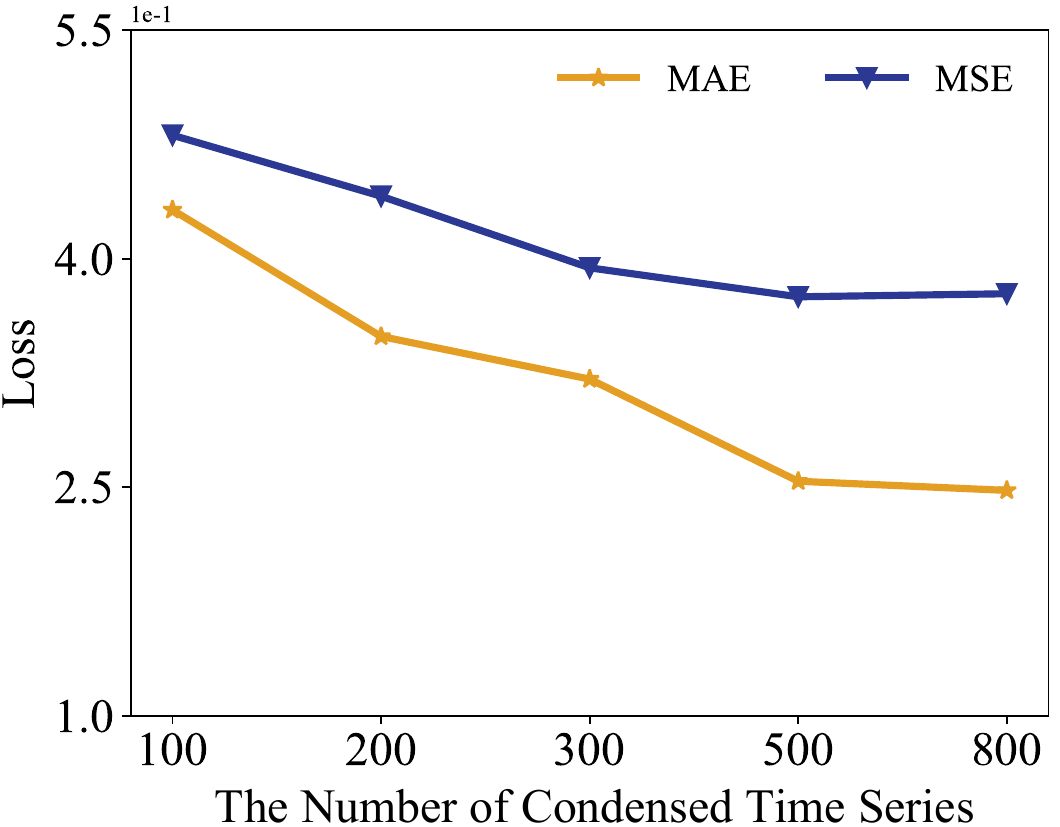} \label{traffic_trade}
	}     
	\subfigure[Running Time on Traffic] { 
		\includegraphics[scale=0.22]{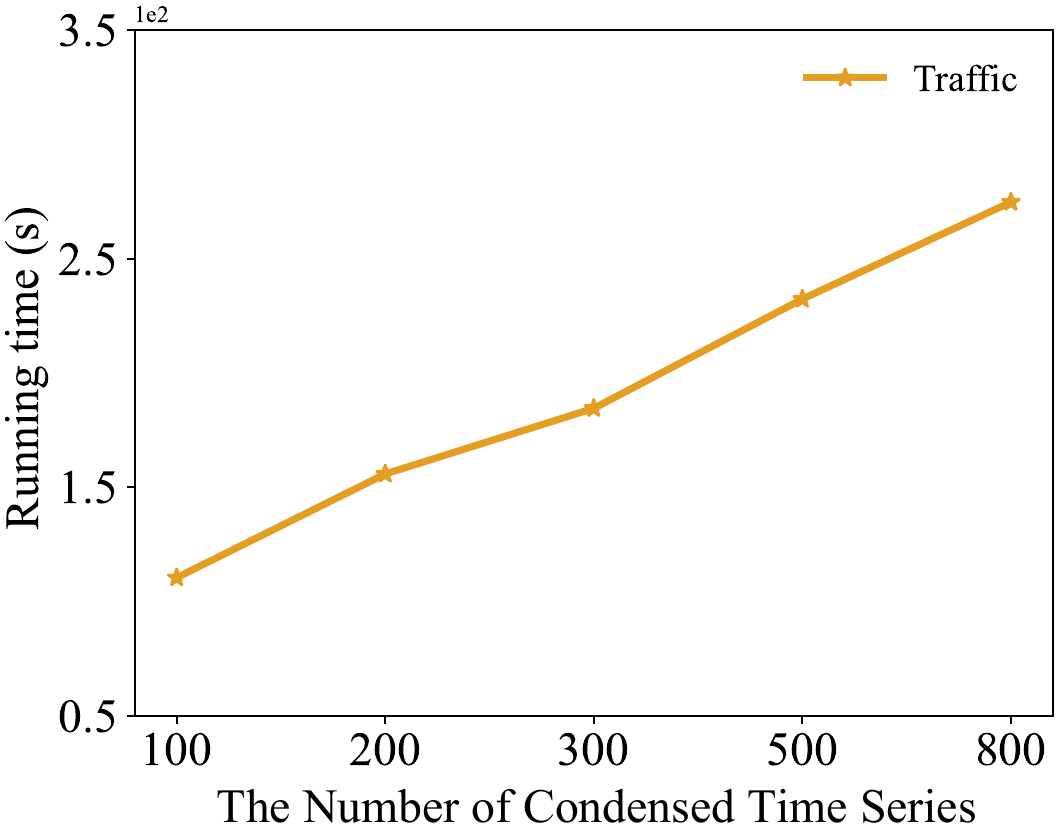}   \label{traffic_efff}  
	} 
	\vspace{-0.45cm}
	\caption{Trade-off between Performance and Efficiency on Two Datasets}
	\label{trade} 
	\vspace{-0.45cm}
\end{figure}

\subsubsection{Trade-off between Performance and Efficiency} 
To study the effect of the number of condensed time series on performance and running time, we conduct experiments with 100, 200, 300, 500, and 800 condensed time series on the Weather and Traffic datasets. The results, shown in Figure~\ref{trade}, indicate that the performance initially drops and then 
either stabilizes (Figure~\ref{weather_trade}) or increase slightly (Figure~\ref{traffic_trade}). In addition, as the number of condensed time series increases, 
the running time decreases considerably. Generally, the results show that model performance improves with an increase in condensed time series data, but at the cost of an increased training time.
When the number of condensed time series is set to 500, TimeDC achieves outstanding performance with an acceptable training time, making 500 an appropriate setting for balancing performance and efficiency.

\begin{table}[h]
\caption{Comparison between Autoformer$_r$ and TimeDC on Weather ($PL=96$)}
\vspace{-0.34cm}
\label{streaming_wea}
\setlength{\tabcolsep}{1.5mm}{
\begin{tabular}{ccccc}
\hline
 Metric & \multicolumn{2}{c}{MAE} & \multicolumn{2}{c}{RMSE} \\\cline{1-5} 
                              Method & Autoformer${_r}$   & TimeDC   & Autoformer${_r}$    & TimeDC   \\ \hline
 $\mathcal{B}_0$     & 0.603        & 0.456    & 0.625         & 0.477    \\ \cline{1-5}
                              $\mathcal{B}_1$     & 0.835        & 0.512    & 1.055         & 0.533    \\
                              $\mathcal{I}$      & 0.638        & 0.656    & 0.687         & 0.512    \\ \hline
\end{tabular}}
\end{table}

\begin{figure*}[!htbp] \centering
\subfigure[Traffic] {
    \includegraphics[scale=0.205]{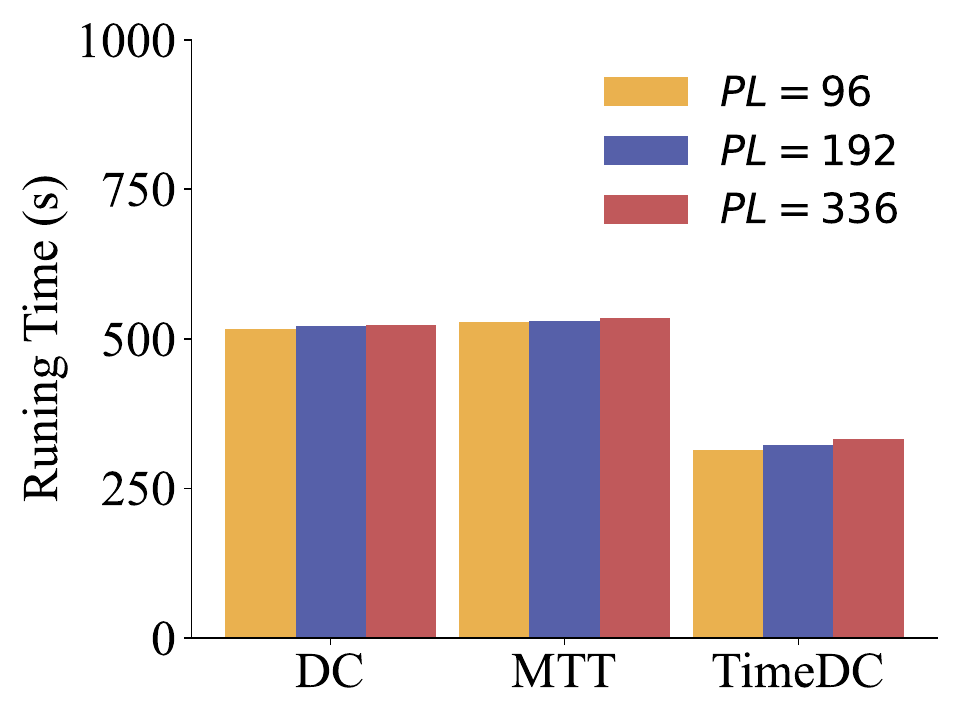} 
    \label{a}
}
\subfigure[Electricity] {
    \includegraphics[scale=0.205]{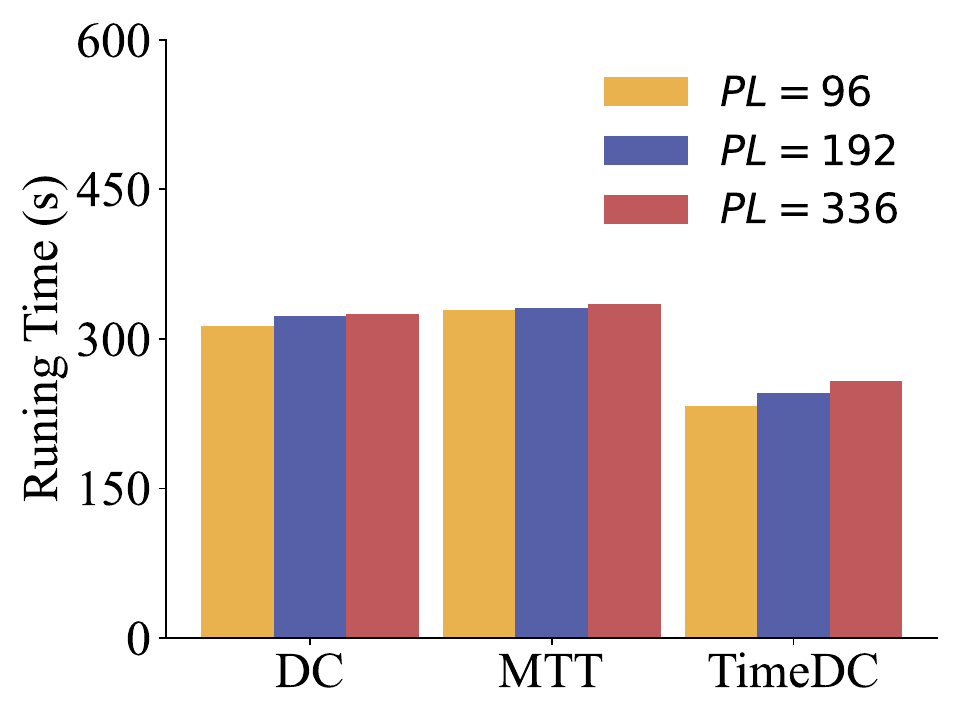} 
    \label{}
}   
\subfigure[ETTh2] {
    \includegraphics[scale=0.205]{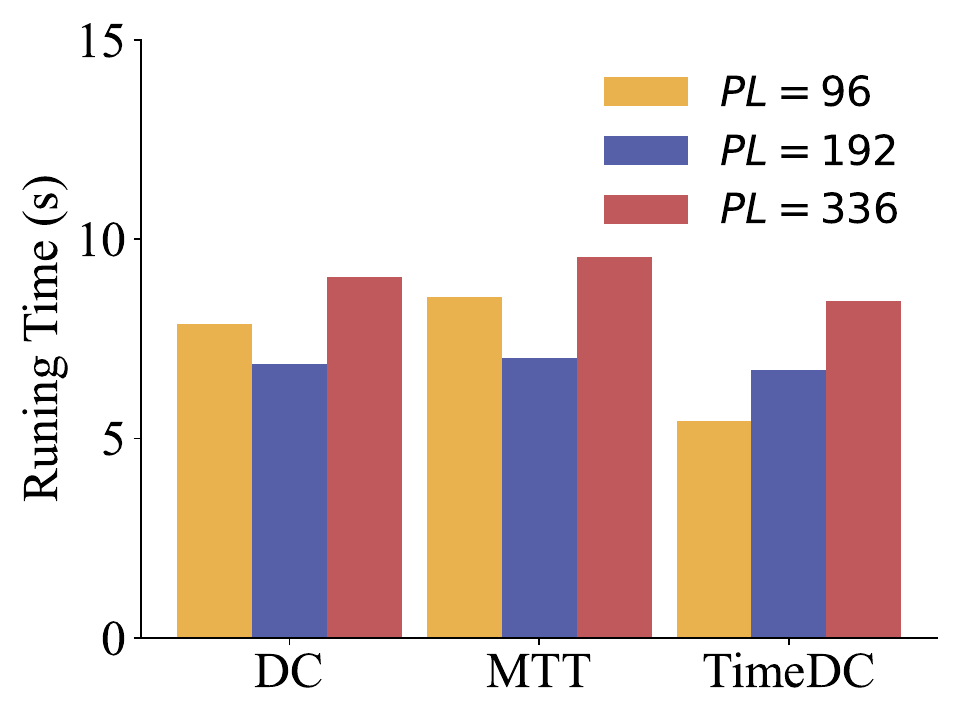} 
    \label{}
} 
\subfigure[ETTm1] { 
    \includegraphics[scale=0.205]{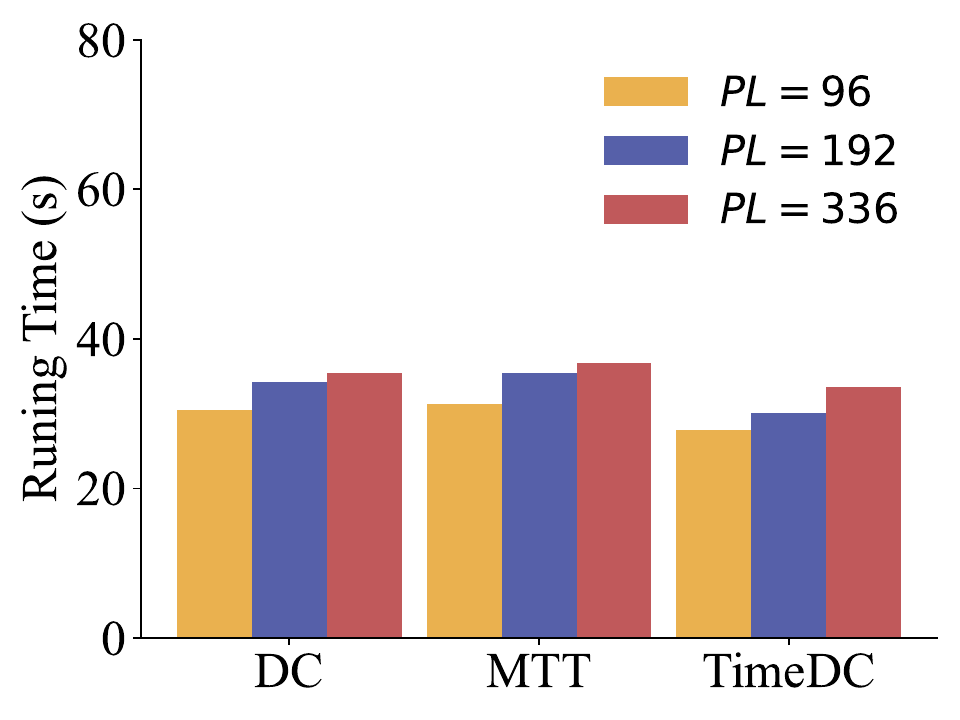}   
    \label{}  
} 
    \subfigure[ETTm2] {
    \includegraphics[scale=0.205]{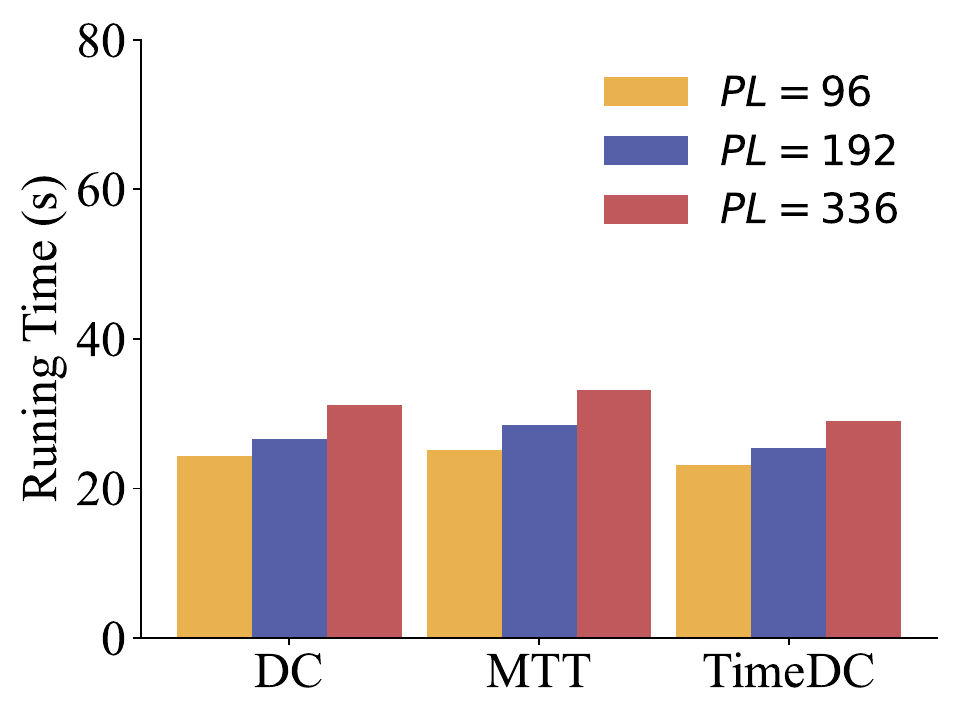} 
    \label{e}
}  
    \subfigure[Traffic] {
    \includegraphics[scale=0.205]{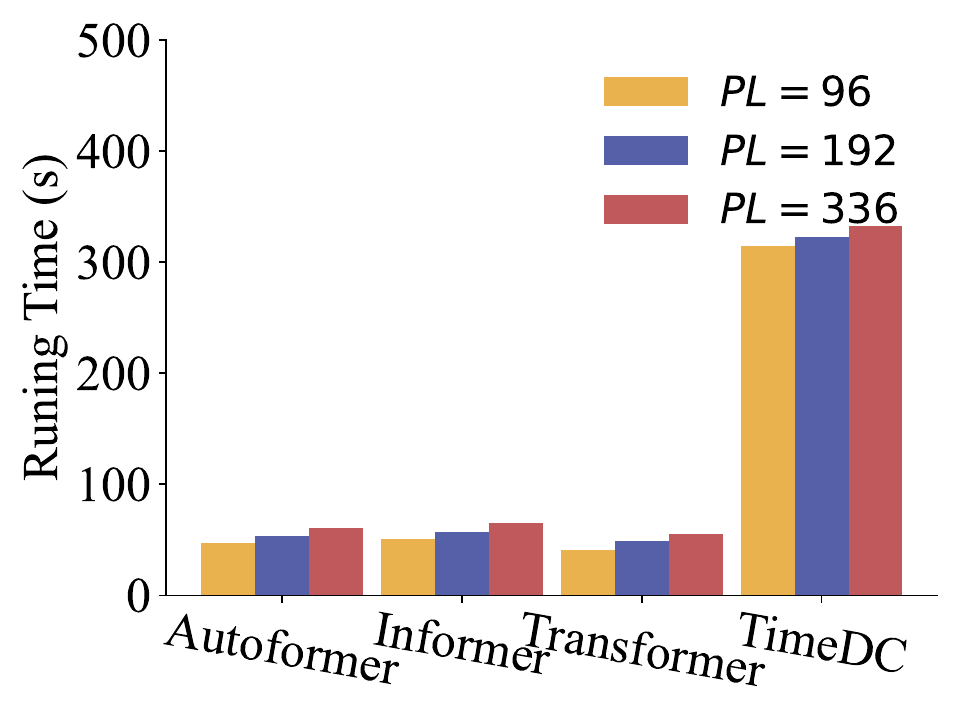} 
    \label{f}
}   
    \subfigure[Electricity] {
    \includegraphics[scale=0.205]{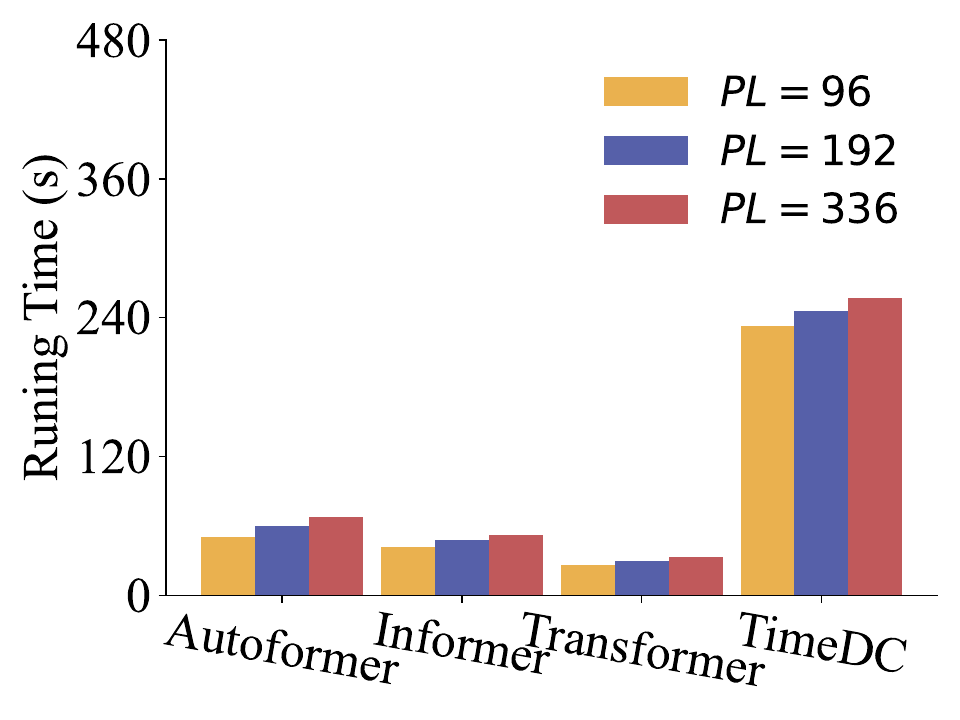} 
    \label{}
}            
\subfigure[ETTh2] {
    \includegraphics[scale=0.205]{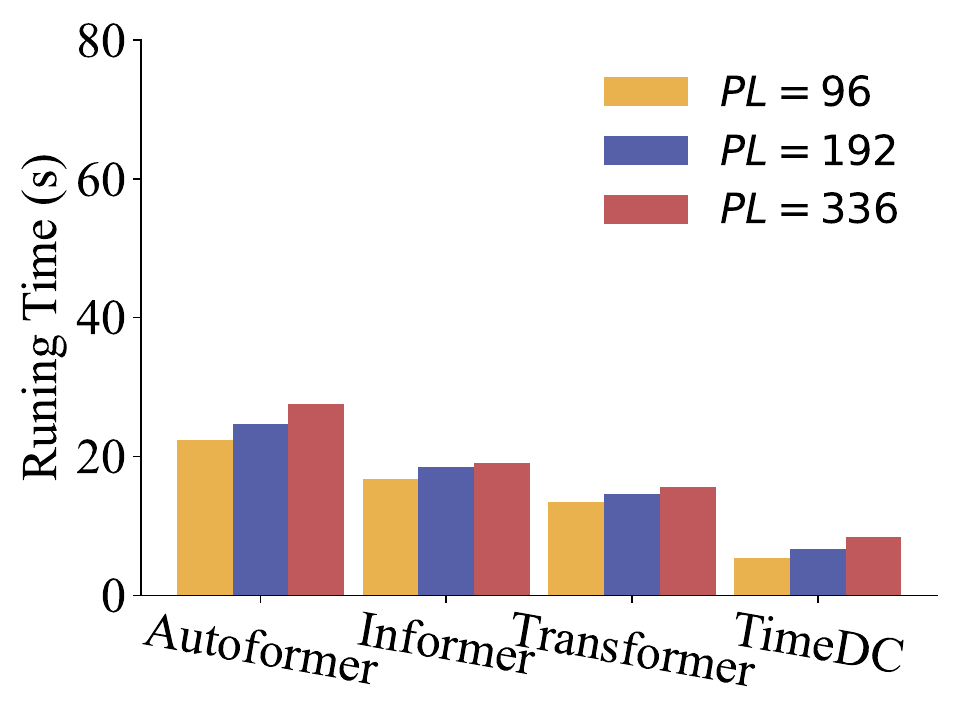} 
    \label{}
}         
\subfigure[ETTm1] { 
    \includegraphics[scale=0.205]{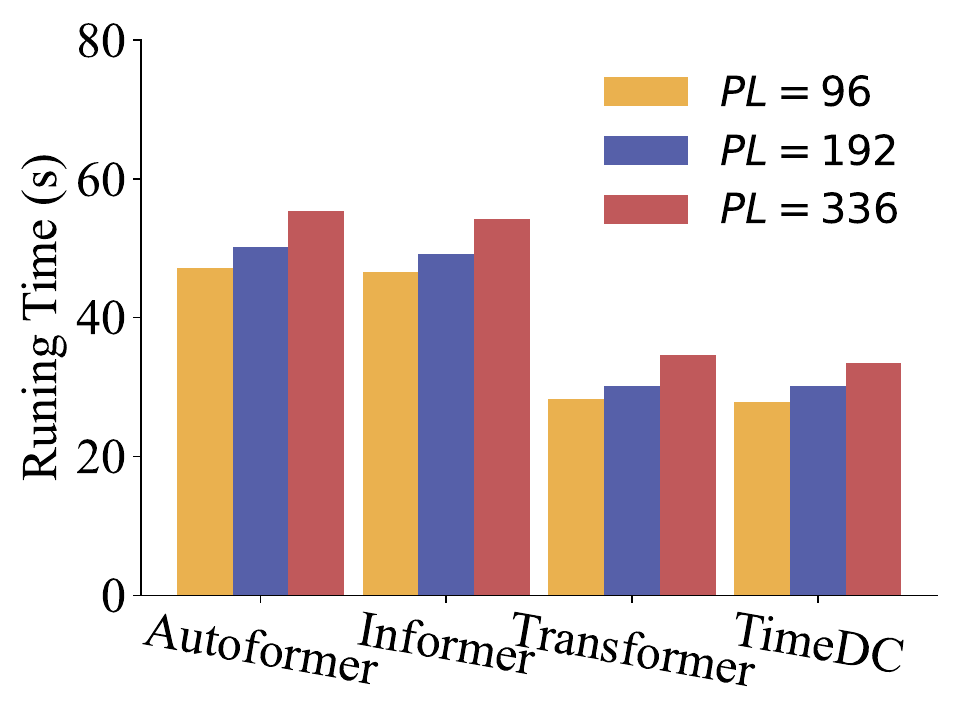}   
    \label{}  
}  
    \subfigure[ETTm2] {
    \includegraphics[scale=0.205]{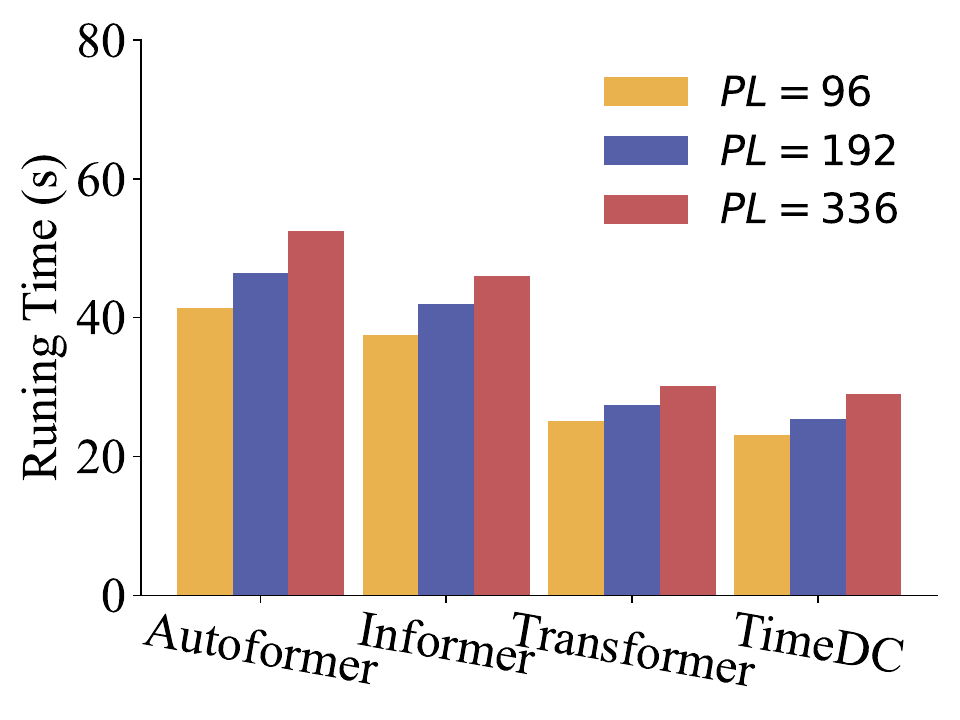} 
    \label{j}
}      
\caption{Running Time Comparison}
\label{efficiency_} 
\end{figure*}

\subsubsection{Performance Comparison on Streaming Data}
To assess more comprehensively the effectiveness of TimeDC, we compare it with the replay-based method Autoformer$_r$ on Weather, which adopts an explicit buffer to store a random subset of the base set $\mathcal{B}$. When the incremental set $\mathcal{I}$ arrives, we randomly select 500 samples from the buffer and then fuse them with the new data to update the model parameters. The results are shown in Table~\ref{streaming_wea}, where TimeDC outperforms Autoformer$_r$, which shows a simple replay-based method is insufficient for time series streaming learning due to the concept drift problems. The results show that TimeDC achieves relatively stable performance as well as notably better performance on the base and incremental sets in five out of six cases.

\begin{table}[t]
    \centering
    \caption{Training Time of TimeDC and Training Time on Original Datasets (s/epoch)}
    \vspace{-0.35cm}
    \setlength{\tabcolsep}{3.6mm}
    \begin{tabular}{ccc}
    \hline
       Dataset  & Condensed Dataset & Original Dataset\\ \hline
       Weather & 22.39 & 35.26\\
       Traffic & 232.34&346.76\\
       Electricity &314.56&522.85\\ 
       ETTh1 & 14.38&20.43\\ 
       ETTh2 & 5.43&8.95\\
       ETTm1 & 27.83&35.75\\
       ETTm2 & 23.15&30.46 \\ \hline
    \end{tabular}
    \vspace{-0.3cm}
    \label{costreduction}
\end{table}

\subsubsection{Training Time on Dataset Condensation and Original Datasets} 
We compare the training time of TimeDC and training time on original datasets based on the stacked \emph{TSOperators} in terms of an epoch, as shown in Table~\ref{costreduction}. The results indicate that the training time of TimeDC is significantly lower than training on the original dataset. 
For example, the training time of TimeDC for dataset condensation is reduced by 39.84\% compared to that of training a model on the original dataset on Electricity. 
Additionally, Table~1 shows that the model trained on the condensed dataset 
performs comparably to the model trained on the original dataset. Thus, TimeDC not only reduces 
the training time to achieve convergence for dataset condensation but also 
maintains good performance, showing the efficiency and practicality of time series dataset condensation, especially for resource-intensive tasks.

\subsubsection{Running Time Efficiecy} As efficiency is important in dataset condensation to enable scalability, especially on resource-constrained edge computing devices, we study the training time (of an epoch) for the condensation methods and different architectures. We conduct experiments on five datasets, i.e., Traffic, Electricity, ETTh2, ETTm1, and ETTm2, as shown in Figure~\ref{efficiency_}. Overall, TimeDC consumes the least training time in most cases. 


We study the training time (of an epoch) for the condensation methods. Figures~\ref{a}--\ref{e} report the training time on Traffic, Electricity, ETTh2, ETTm1, and ETTm2. We see that the training time of TimeDC is below those of DC and MTT, which is largely because of the expert buffer in the CT2M module that stores pre-computed trajectories. This indicates the feasibility of TimeDC for model deployment in large time series dataset reduction scenarios.

\begin{figure*}[t]
    \centering
    \subfigure[Random] {
    \includegraphics[scale=0.3]{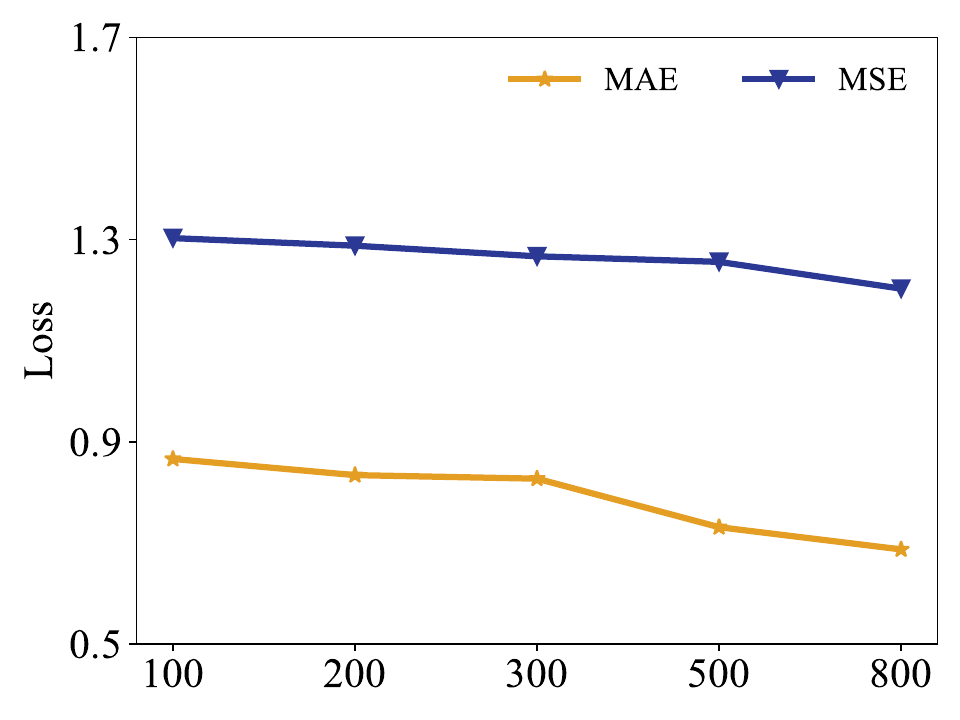} 
    \label{}
    }
    \subfigure[K-Center] {
    \includegraphics[scale=0.3]{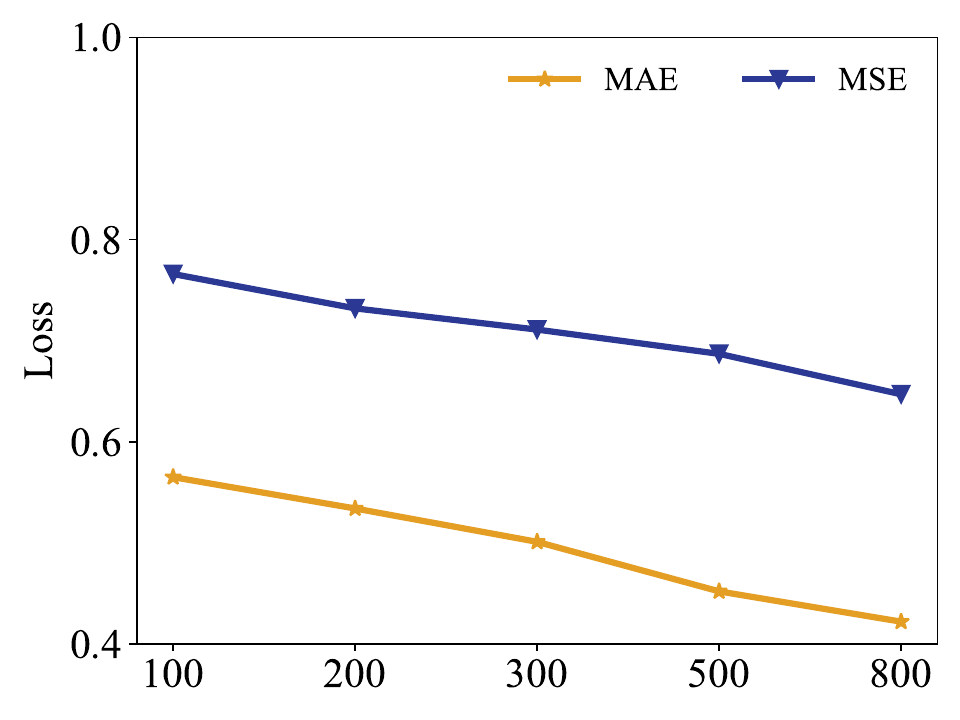} 
    \label{}
    }
    \subfigure[Herding] {
    \includegraphics[scale=0.3]{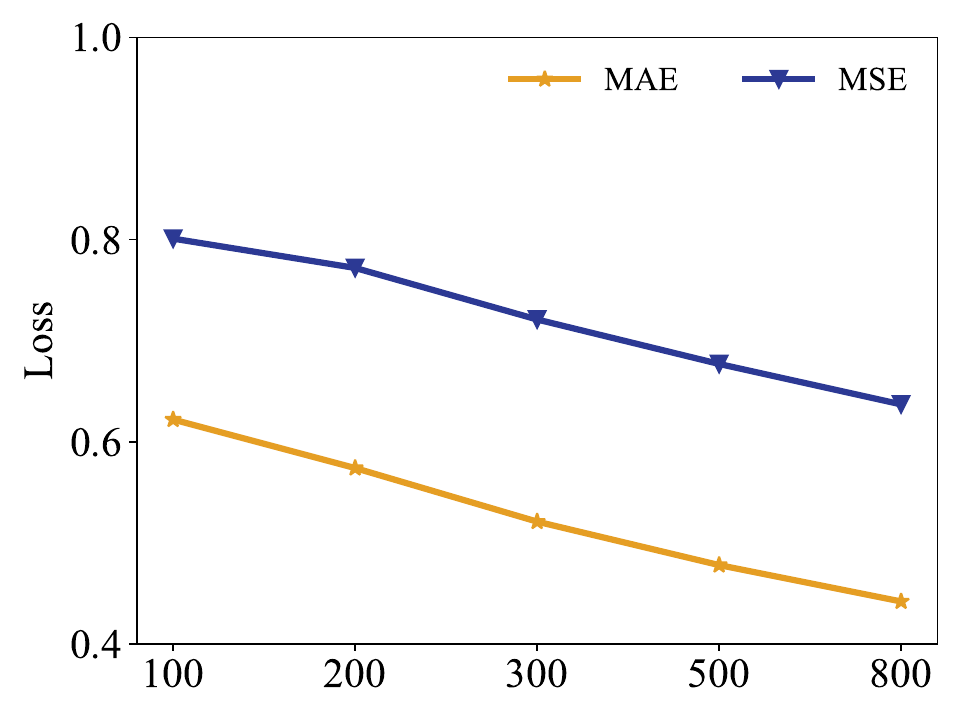} 
    \label{}
    }
    \subfigure[DC] {
    \includegraphics[scale=0.3]{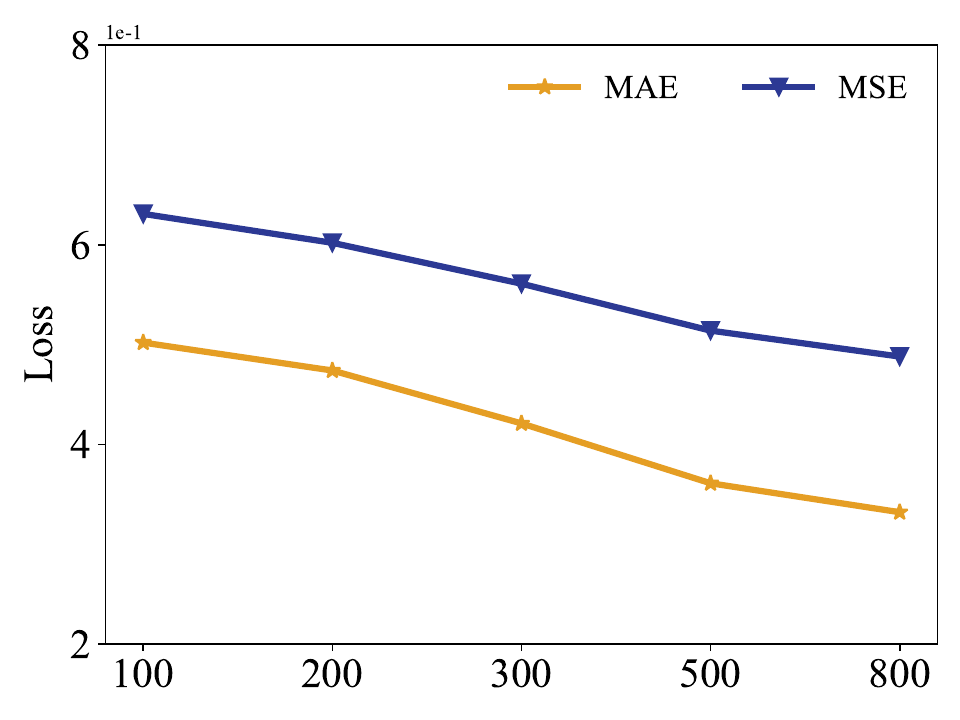} 
    \label{}
    }
    \subfigure[MTT] {
    \includegraphics[scale=0.3]{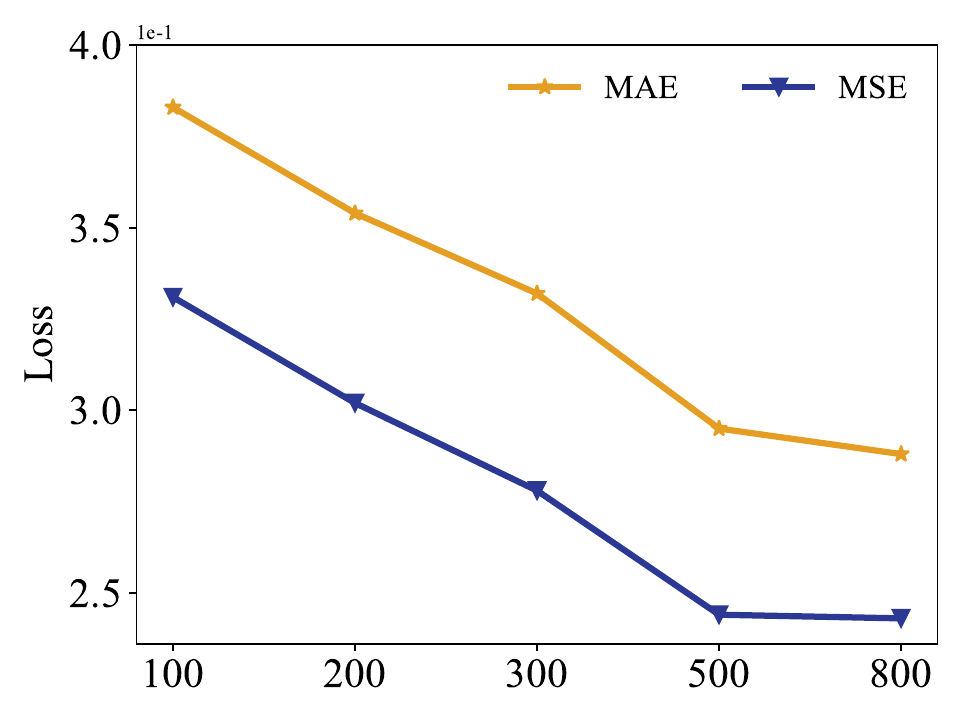} 
    \label{}
    }
    \subfigure[TimeDC] {
    \includegraphics[scale=0.3]{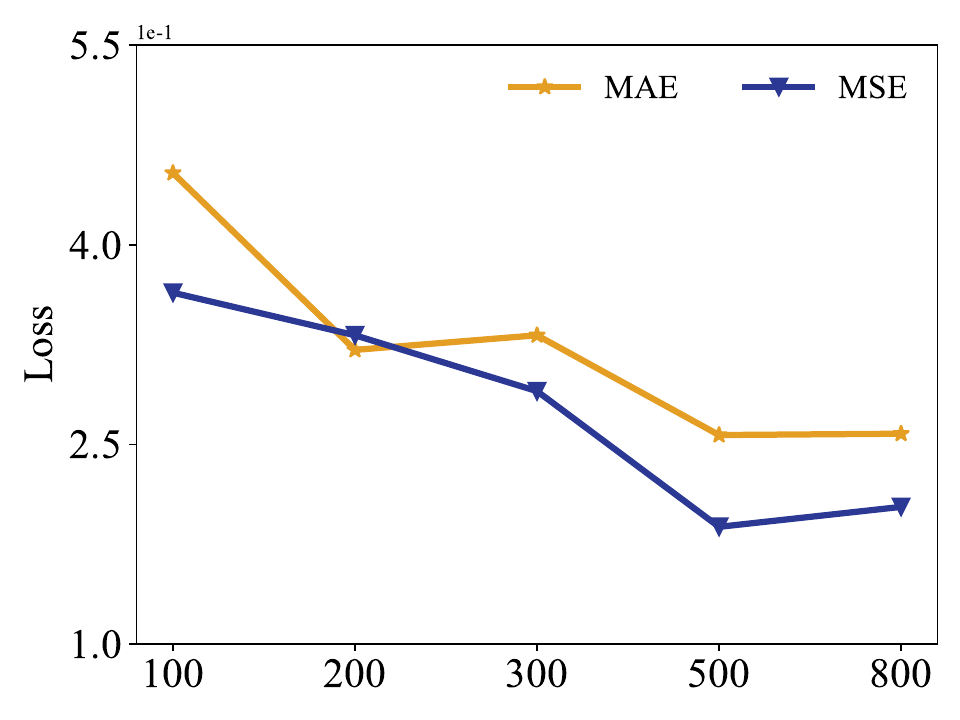} 
    \label{}
    }
    \caption{Effect of the Size of Condensed TS Datasets across Different Methods on Weather}
    \label{size_efff}
\end{figure*}

We also study training time across different network architectures on Traffic, Electricity, ETTh2, ETTm1, and ETTm2---see Figures~\ref{f}--\ref{j}. Here, TimeDC consumes the least training time in most cases. TimeDC is faster than the other methods due to its patching mechanism, which reduces the complexity of the self-attention mechanism through input data simplification. However, TimeDC consumes more training time on Traffic and Electricity. This is because Traffic and Electricity have many more features than the other datasets---862 versus 321 features. The proposed \emph{TSOperators} require more training time to learn these features simultaneously because of the channel-independent mechanism. Nonetheless, TimeDC achieves much better performance on Traffic and Electricity, indicating its ability to learn correlations across different channels.


\subsubsection{Training Time of TimeDC and Its Variants}We report the training time of TimeDC and its variants in Table~\ref{trainingtime}. 
The training time of \emph{w/o\_patch} is much lower than those of the other variants, primarily due to the patch construction and storage in the patching mechanism. However, the patching mechanism enables the modeling of local semantics, which enables the model to process longer historical sequences to improve the feature extraction, thus enabling better model performance.

\begin{table}[h]
    \centering
    \caption{Training Time of TimeDC and Its Variants (s/epoch)}
    \vspace{-0.35cm}
    \begin{tabular}{ccccc}
    \hline
       Dataset  & \emph{w/o\_Patch} & \emph{w/o\_DDFM} & \emph{w/o\_CT$^2$M} & TimeDC \\ \hline
     Weather & 20.34 & 22.44 & 33.4 & 22.84\\ 
     Traffic &50.45 & 278.74 & 510.4 & 314.56\\ 
     Electricity & 60.43 & 201.45 &331.32 &232.34\\ 
     ETTh1 & 15.68& 16.45& 25.64 & 18.75\\ \hline
    \end{tabular}
    \label{trainingtime}
\end{table}

\subsubsection{Effect of the Size of Condensed TS Datasets across Different Methods} 

We study the effect of the size of the condensed time series (TS) dataset across different methods in Figure~\ref{size_efff}. We observe that the performance curves drop significantly in most cases. This shows that the model performance improves with a larger condensed time series dataset.
This increase in performance is attributed to the availability of more training data containing valuable knowledge.
It is noteworthy that TimeDC shows a slight decrease in performance when using 800 condensed time series compared to 500 condensed time series on the Weather dataset.
This may be because the patterns in these time series are relatively straightforward. 
Additional condensed time series data might introduce recurring patterns, thereby making the model overfit to these patterns and degrading performance on other data. 
For detailed results for other datasets, please refer to the repository at \url{https://github.com/uestc-liuzq/STdistillation}.

\subsubsection{Dynamic Tensor Memory Cost} We conduct experiments to compare the memory used by the dynamic (online) tensor across DC, MTT, and TimeDC on all datasets in Table~\ref{memory_all}. TimeDC is able to significantly alleviate heavy online memory and computation costs thanks to the training trajectories precomputed offline. 
\begin{table}[h]
\caption{Dynamic Tensor Memory Cost on All Datasets}
\vspace{-0.35cm}
\label{memory_all}
\setlength{\tabcolsep}{4.5mm}{
\begin{tabular}{c|cccc}
\hline
Dataset       & DC & MTT & TimeDC\\ \hline
Weather & 10.0 GB   & 8.9 GB & 3.3 GB\\
Traffic &17.8 GB&13.7 GB&10.9 GB\\
Electricity &8.5 GB&932.5 MB&516.0 MB\\
ETTh1  & 1.9 GB  & 845.5 MB   & 280.9 MB\\
ETTh2 &1.8 GB&932.5 MB&516.1 MB\\
ETTm1 &1.8 GB&932.2 MB&515.9 MB\\
ETTm2 &1.7 GB&932.5 MB&516.1 MB\\
\hline
\end{tabular}}
\end{table}

\begin{table*}[t]
\caption{Condensed Dataset Size for Seven Datasets}
\label{size_condense}
\vspace{-0.3cm}
\setlength{\tabcolsep}{6mm}{
\begin{tabular}{c|cccccc}
\hline
Dataset     & Random  & K-Center & Herding & DC      & MTT     & TimeDC  \\ \hline
Weather     & 40.3 MB  & 40.3 MB   & 40.3 MB  & 39.8 MB  & 38.6 MB  & 38.5 MB  \\
Traffic     & 1.7 GB  & 1.7 GB   & 1.7 GB  & 1.7 MB  & 1.7 GB  & 1.7 GB  \\
Electricity & 316.3 MB & 308.2 MB  & 316.3 MB & 316.3 MB & 310.8 MB & 308.2 MB \\
ETTh1       & 13.4 MB  & 13.4 MB   & 13.4 MB  & 13.3 MB  & 13.4 MB  & 12.8 MB  \\
ETTh2       & 13.3 MB  & 13.4 MB   & 13.4 MB  & 13.4 MB  & 13.4 MB  & 13.4 MB  \\
ETTm1       & 13.4 MB  & 13.4 MB   & 13.4 MB  & 13.4 MB  & 13.4 MB  & 13.4 MB  \\
ETTm2       & 13.4 MB  & 13.4 MB   & 13.4 MB  & 13.4 MB  & 13.4 MB  & 13.4 MB \\ \hline
\end{tabular}}
\end{table*}

\subsubsection{The Size of Condensed Datasets on Seven Datasets} We report the condensed dataset size for each baseline and TimeDC in Table~\ref{size_condense}. 
We observe that the dataset sizes are similar across the different baselines because we fix the number of condensed time series at 500 for each method, resulting in minimal variation in the dataset size. 

\begin{figure*}[!htbp] \centering
\subfigure[Traffic] {
    \includegraphics[scale=0.205]{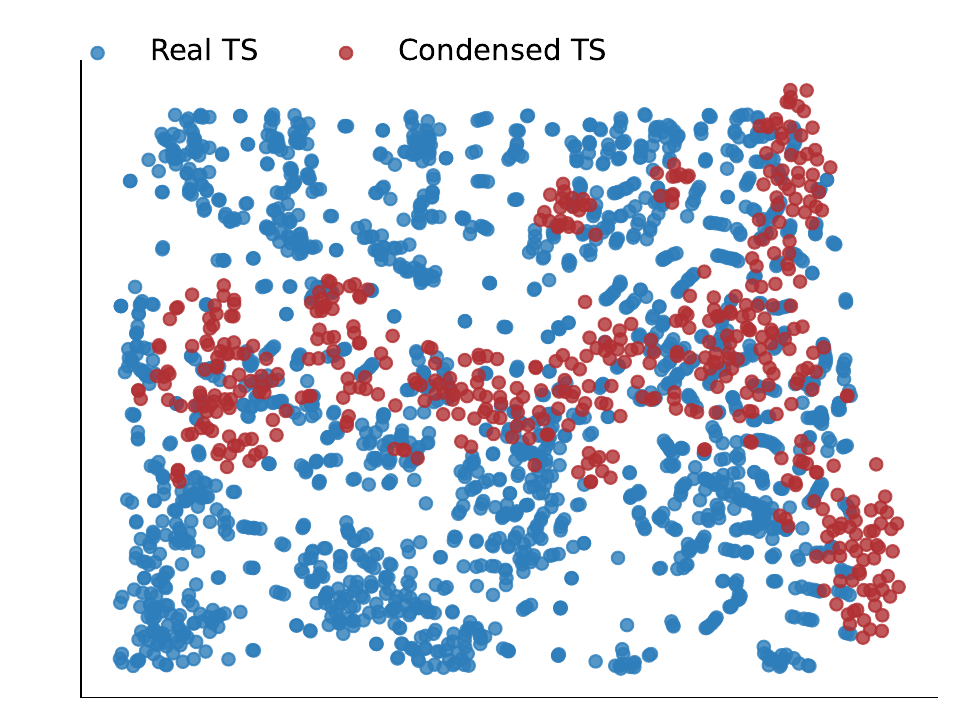} 
    \label{a_}
}
\subfigure[Electricity] {
    \includegraphics[scale=0.205]{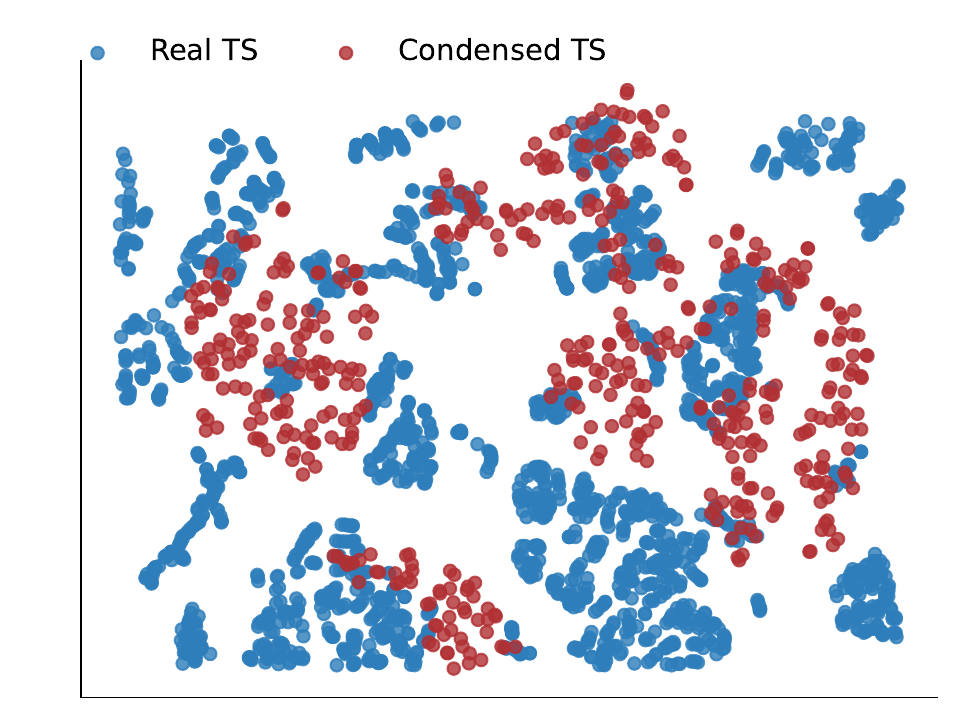} 
    \label{}
}   
\subfigure[ETTh2] {
    \includegraphics[scale=0.205]{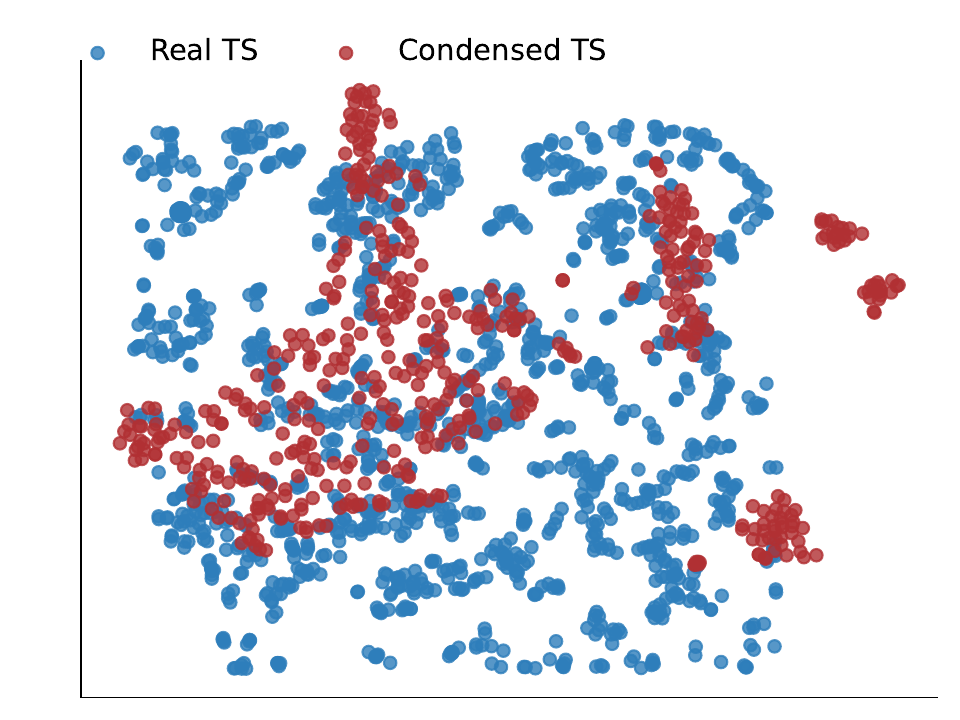} 
    \label{}
} 
\subfigure[ETTm1] { 
    \includegraphics[scale=0.205]{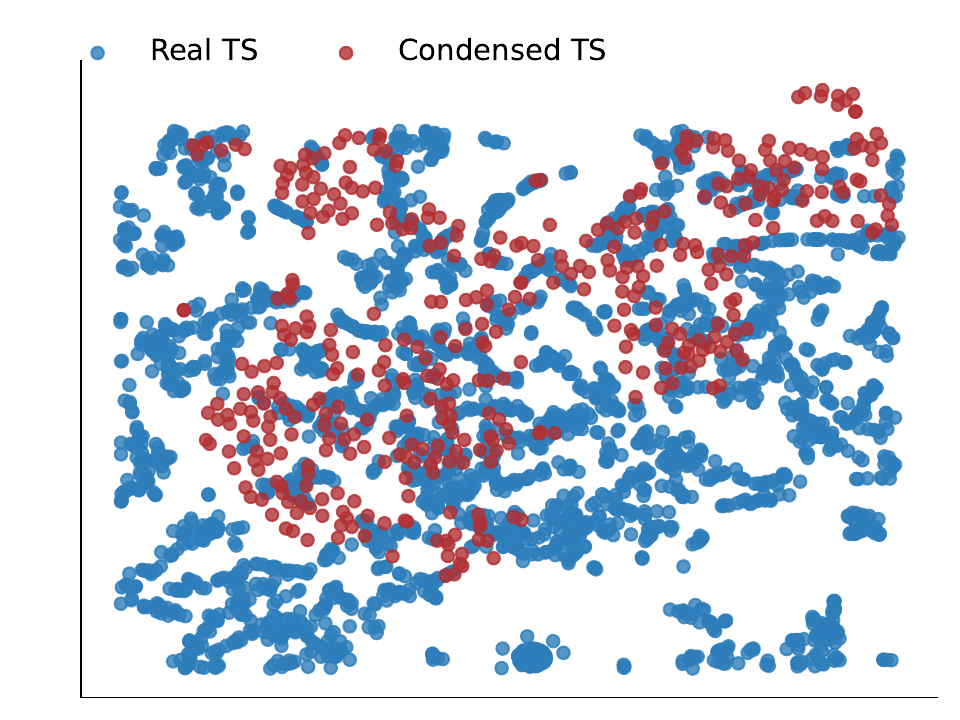}   
    \label{}  
} 
    \subfigure[ETTm2] {
    \includegraphics[scale=0.205]{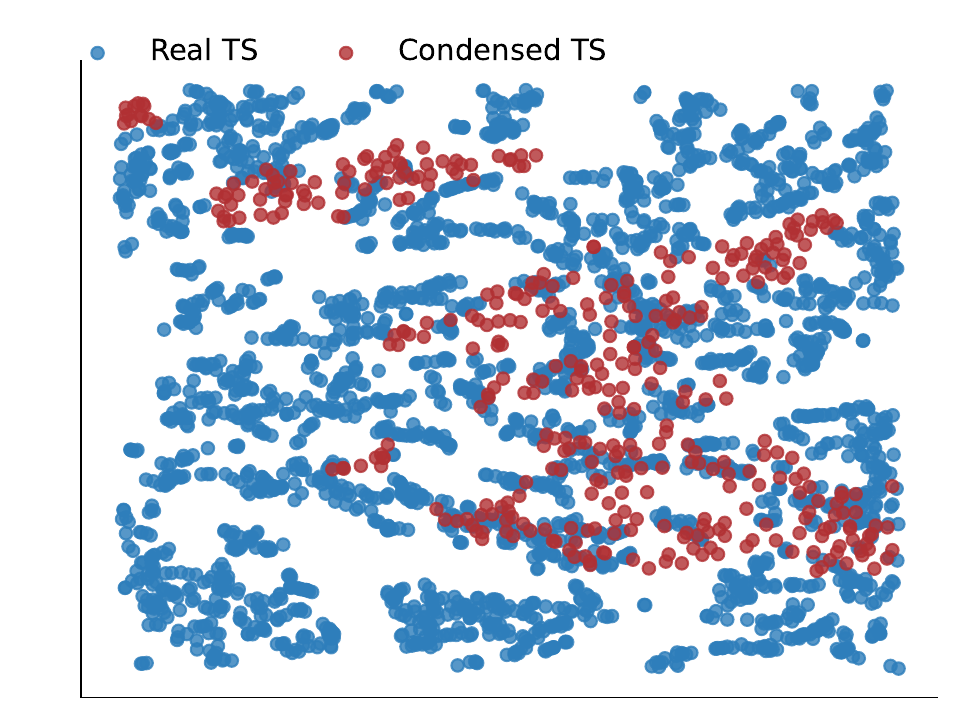} 
    \label{e_}
}       
\caption{Dataset Distribution Comparison on Five Datasets}
\label{tnse__} 
\end{figure*}

\subsubsection{Case Study on Dataset Condensation} To further assess the effectiveness of TimeDC on synthesizing condensed time series datasets that cover the original time series distribution well, we show t-SNE graphs of the original time series dataset and the condensed time series dataset for Traffic, Electricity, ETTh2, ETTm1, and ETTm2. 
Figure~\ref{tnse__} compares the dataset distributions, where blue and red dots represent the original (i.e., real) and condensed dataset, respectively. We randomly sample 500 time series from the original time series dataset for the visualization of the original time series. We observe that the red dots are well integrated with the blue dots, indicating similarity in distribution between the original and condensed datasets. 
This indicates that the condensed dataset is of high quality and that the condensation method is effective. 

\subsubsection{Scalability}
To assess the scalability of TimeDC, we conduct experiments to study the effect of different prediction lengths on Weather. The results are shown in Table~\ref{predictionlength}. With an increase in the prediction length, TimeDC achieves similar dynamic tensor costs and training times, demonstrating its efficiency and scalability in terms of time series prediction length.

In addition, TimeDC scales with the input dimensionality (i.e., number of features). In particular, the increase in the dynamic tensor and storage costs of TimeDC is significantly smaller than the difference in the number of input features of various datasets in most cases (see Tables~3 and~6). For example, the number of features in Traffic (i.e., 862) is approximately 123 times that of ETTh1 (i.e., 7), but the dynamic tensor and storage costs of TimeDC on Traffic are 12 and 64 times those of ETTh1, respectively.


Further, TimeDC scales with the number of condensed time series. According to Figures~\ref{weather_efff} and~\ref{traffic_efff}, the running time of TimeDC increases approximately linearly with the number of time series.

\begin{table}[t]
    \centering
    \caption{Effect of Prediction Length on Weather}
    \vspace{-0.35cm}
    \setlength{\tabcolsep}{3mm}
    \begin{tabular}{ccc}
    \hline
\makecell{Prediction Length}& \makecell{Dynamic Tensor} & \makecell{Training Time}\\ \hline
      96 & 3.34 GB & 20.82 s \\
    192 & 3.41 GB & 21.73 s\\
       336 & 3.48 GB & 22.39 s\\ \hline
    \end{tabular}
    \label{predictionlength}
\end{table}


%% file: main.bbl

\begin{thebibliography}{55}


\ifx \showCODEN    \undefined \def \showCODEN     #1{\unskip}     \fi
\ifx \showDOI      \undefined \def \showDOI       #1{#1}\fi
\ifx \showISBNx    \undefined \def \showISBNx     #1{\unskip}     \fi
\ifx \showISBNxiii \undefined \def \showISBNxiii  #1{\unskip}     \fi
\ifx \showISSN     \undefined \def \showISSN      #1{\unskip}     \fi
\ifx \showLCCN     \undefined \def \showLCCN      #1{\unskip}     \fi
\ifx \shownote     \undefined \def \shownote      #1{#1}          \fi
\ifx \showarticletitle \undefined \def \showarticletitle #1{#1}   \fi
\ifx \showURL      \undefined \def \showURL       {\relax}        \fi
\providecommand\bibfield[2]{#2}
\providecommand\bibinfo[2]{#2}
\providecommand\natexlab[1]{#1}
\providecommand\showeprint[2][]{arXiv:#2}

\bibitem[\protect\citeauthoryear{Agarwal, Har-Peled, and Varadarajan}{Agarwal et~al\mbox{.}}{2004}]%
        {agarwal2004approximating}
\bibfield{author}{\bibinfo{person}{Pankaj~K Agarwal}, \bibinfo{person}{Sariel Har-Peled}, {and} \bibinfo{person}{Kasturi~R Varadarajan}.} \bibinfo{year}{2004}\natexlab{}.
\newblock \showarticletitle{Approximating extent measures of points}.
\newblock \bibinfo{journal}{\emph{JACM}} \bibinfo{volume}{51}, \bibinfo{number}{4} (\bibinfo{year}{2004}), \bibinfo{pages}{606--635}.
\newblock


\bibitem[\protect\citeauthoryear{Aggarwal and Yu}{Aggarwal and Yu}{2008}]%
        {aggarwal2008static}
\bibfield{author}{\bibinfo{person}{Charu~C Aggarwal} {and} \bibinfo{person}{Philip~S Yu}.} \bibinfo{year}{2008}\natexlab{}.
\newblock \showarticletitle{On static and dynamic methods for condensation-based privacy-preserving data mining}.
\newblock \bibinfo{journal}{\emph{TODS}} \bibinfo{volume}{33}, \bibinfo{number}{1} (\bibinfo{year}{2008}), \bibinfo{pages}{1--39}.
\newblock


\bibitem[\protect\citeauthoryear{Angiulli}{Angiulli}{2007}]%
        {angiulli2007fast}
\bibfield{author}{\bibinfo{person}{Fabrizio Angiulli}.} \bibinfo{year}{2007}\natexlab{}.
\newblock \showarticletitle{Fast nearest neighbor condensation for large data sets classification}.
\newblock \bibinfo{journal}{\emph{TKDE}} \bibinfo{volume}{19}, \bibinfo{number}{11} (\bibinfo{year}{2007}), \bibinfo{pages}{1450--1464}.
\newblock


\bibitem[\protect\citeauthoryear{Bonifati, Buono, Guerra, and Tiano}{Bonifati et~al\mbox{.}}{2022}]%
        {bonifati2022time2feat}
\bibfield{author}{\bibinfo{person}{Angela Bonifati}, \bibinfo{person}{Francesco~Del Buono}, \bibinfo{person}{Francesco Guerra}, {and} \bibinfo{person}{Donato Tiano}.} \bibinfo{year}{2022}\natexlab{}.
\newblock \showarticletitle{Time2Feat: learning interpretable representations for multivariate time series clustering}.
\newblock \bibinfo{journal}{\emph{PVLDB}} \bibinfo{volume}{16}, \bibinfo{number}{2} (\bibinfo{year}{2022}), \bibinfo{pages}{193--201}.
\newblock


\bibitem[\protect\citeauthoryear{Campos, Yang, Kieu, Zhang, Guo, and Jensen}{Campos et~al\mbox{.}}{2024}]%
        {David2024Qcore}
\bibfield{author}{\bibinfo{person}{David Campos}, \bibinfo{person}{Bin Yang}, \bibinfo{person}{Tung Kieu}, \bibinfo{person}{Miao Zhang}, \bibinfo{person}{Chenjuan Guo}, {and} \bibinfo{person}{Christian~S. Jensen}.} \bibinfo{year}{2024}\natexlab{}.
\newblock \showarticletitle{QCore: Data-Efficient, On-Device Continual Calibration for Quantized Models}.
\newblock \bibinfo{journal}{\emph{{PVLDB}}} \bibinfo{volume}{17}, \bibinfo{number}{11} (\bibinfo{year}{2024}), \bibinfo{pages}{2708--2721}.
\newblock


\bibitem[\protect\citeauthoryear{Cazenavette, Wang, Torralba, Efros, and Zhu}{Cazenavette et~al\mbox{.}}{2022}]%
        {cazenavette2022dataset}
\bibfield{author}{\bibinfo{person}{George Cazenavette}, \bibinfo{person}{Tongzhou Wang}, \bibinfo{person}{Antonio Torralba}, \bibinfo{person}{Alexei~A Efros}, {and} \bibinfo{person}{Jun-Yan Zhu}.} \bibinfo{year}{2022}\natexlab{}.
\newblock \showarticletitle{Dataset distillation by matching training trajectories}. In \bibinfo{booktitle}{\emph{CVPR}}. \bibinfo{pages}{4750--4759}.
\newblock


\bibitem[\protect\citeauthoryear{Chai, Liu, Tang, Fan, Miao, Wang, Luo, and Li}{Chai et~al\mbox{.}}{2023}]%
        {chai2023goodcore}
\bibfield{author}{\bibinfo{person}{Chengliang Chai}, \bibinfo{person}{Jiabin Liu}, \bibinfo{person}{Nan Tang}, \bibinfo{person}{Ju Fan}, \bibinfo{person}{Dongjing Miao}, \bibinfo{person}{Jiayi Wang}, \bibinfo{person}{Yuyu Luo}, {and} \bibinfo{person}{Guoliang Li}.} \bibinfo{year}{2023}\natexlab{}.
\newblock \showarticletitle{GoodCore: Data-effective and Data-efficient Machine Learning through Coreset Selection over Incomplete Data}.
\newblock \bibinfo{journal}{\emph{SIGMOD}} \bibinfo{volume}{1}, \bibinfo{number}{2} (\bibinfo{year}{2023}), \bibinfo{pages}{1--27}.
\newblock


\bibitem[\protect\citeauthoryear{Chen, Zhang, Cheng, Shu, Wang, Wen, Yang, and Guo}{Chen et~al\mbox{.}}{2024}]%
        {DBLP:conf/iclr/ChenZ0SWW0G24}
\bibfield{author}{\bibinfo{person}{Peng Chen}, \bibinfo{person}{Yingying Zhang}, \bibinfo{person}{Yunyao Cheng}, \bibinfo{person}{Yang Shu}, \bibinfo{person}{Yihang Wang}, \bibinfo{person}{Qingsong Wen}, \bibinfo{person}{Bin Yang}, {and} \bibinfo{person}{Chenjuan Guo}.} \bibinfo{year}{2024}\natexlab{}.
\newblock \showarticletitle{Pathformer: Multi-scale Transformers with Adaptive Pathways for Time Series Forecasting}. In \bibinfo{booktitle}{\emph{{ICLR}}}.
\newblock


\bibitem[\protect\citeauthoryear{Cheng, Chen, Guo, Zhao, Wen, Yang, and Jensen}{Cheng et~al\mbox{.}}{2024a}]%
        {cheng2023weakly}
\bibfield{author}{\bibinfo{person}{Yunyao Cheng}, \bibinfo{person}{Peng Chen}, \bibinfo{person}{Chenjuan Guo}, \bibinfo{person}{Kai Zhao}, \bibinfo{person}{Qingsong Wen}, \bibinfo{person}{Bin Yang}, {and} \bibinfo{person}{Christian~S Jensen}.} \bibinfo{year}{2024}\natexlab{a}.
\newblock \showarticletitle{Weakly guided adaptation for robust time series forecasting}.
\newblock \bibinfo{journal}{\emph{PVLDB}} \bibinfo{volume}{17}, \bibinfo{number}{4} (\bibinfo{year}{2024}), \bibinfo{pages}{766--779}.
\newblock


\bibitem[\protect\citeauthoryear{Cheng, Guo, Yang, Yu, Zhao, and Jensen}{Cheng et~al\mbox{.}}{2024b}]%
        {cheng2024memfromer}
\bibfield{author}{\bibinfo{person}{Yunyao Cheng}, \bibinfo{person}{Chenjuan Guo}, \bibinfo{person}{Bin Yang}, \bibinfo{person}{Haomin Yu}, \bibinfo{person}{Kai Zhao}, {and} \bibinfo{person}{Christian~S. Jensen}.} \bibinfo{year}{2024}\natexlab{b}.
\newblock \showarticletitle{A Memory Guided Transformer for Time Series Forecasting}.
\newblock \bibinfo{journal}{\emph{PVLDB}}  \bibinfo{volume}{18} (\bibinfo{year}{2024}).
\newblock


\bibitem[\protect\citeauthoryear{Chu, Li, Rathbun, and Li}{Chu et~al\mbox{.}}{2023}]%
        {chu2023continual}
\bibfield{author}{\bibinfo{person}{Zhixuan Chu}, \bibinfo{person}{Ruopeng Li}, \bibinfo{person}{Stephen Rathbun}, {and} \bibinfo{person}{Sheng Li}.} \bibinfo{year}{2023}\natexlab{}.
\newblock \showarticletitle{Continual causal inference with incremental observational data}. In \bibinfo{booktitle}{\emph{ICDE}}. \bibinfo{pages}{3430--3439}.
\newblock


\bibitem[\protect\citeauthoryear{Cleveland, Cleveland, McRae, and Terpenning}{Cleveland et~al\mbox{.}}{1990}]%
        {cleveland1990stl}
\bibfield{author}{\bibinfo{person}{Robert~B Cleveland}, \bibinfo{person}{William~S Cleveland}, \bibinfo{person}{Jean~E McRae}, {and} \bibinfo{person}{Irma Terpenning}.} \bibinfo{year}{1990}\natexlab{}.
\newblock \showarticletitle{STL: A seasonal-trend decomposition}.
\newblock \bibinfo{journal}{\emph{J. Off. Stat}} \bibinfo{volume}{6}, \bibinfo{number}{1} (\bibinfo{year}{1990}), \bibinfo{pages}{3--73}.
\newblock


\bibitem[\protect\citeauthoryear{Fang, Xie, Zhao, Chen, Gao, and Zheng}{Fang et~al\mbox{.}}{2024}]%
        {fang2024temporal}
\bibfield{author}{\bibinfo{person}{Yuchen Fang}, \bibinfo{person}{Jiandong Xie}, \bibinfo{person}{Yan Zhao}, \bibinfo{person}{Lu Chen}, \bibinfo{person}{Yunjun Gao}, {and} \bibinfo{person}{Kai Zheng}.} \bibinfo{year}{2024}\natexlab{}.
\newblock \showarticletitle{Temporal-Frequency Masked Autoencoders for Time Series Anomaly Detection}. In \bibinfo{booktitle}{\emph{ICDE}}. \bibinfo{pages}{1228--1241}.
\newblock


\bibitem[\protect\citeauthoryear{Farahani and Hekmatfar}{Farahani and Hekmatfar}{2009}]%
        {farahani2009facility}
\bibfield{author}{\bibinfo{person}{Reza~Zanjirani Farahani} {and} \bibinfo{person}{Masoud Hekmatfar}.} \bibinfo{year}{2009}\natexlab{}.
\newblock \bibinfo{booktitle}{\emph{Facility location: concepts, models, algorithms and case studies}}.
\newblock \bibinfo{publisher}{Springer Science \& Business Media}.
\newblock


\bibitem[\protect\citeauthoryear{Feldman and Langberg}{Feldman and Langberg}{2011}]%
        {feldman2011unified}
\bibfield{author}{\bibinfo{person}{Dan Feldman} {and} \bibinfo{person}{Michael Langberg}.} \bibinfo{year}{2011}\natexlab{}.
\newblock \showarticletitle{A unified framework for approximating and clustering data}. In \bibinfo{booktitle}{\emph{STOC}}. \bibinfo{pages}{569--578}.
\newblock


\bibitem[\protect\citeauthoryear{Feldman, Schmidt, and Sohler}{Feldman et~al\mbox{.}}{2020}]%
        {feldman2020turning}
\bibfield{author}{\bibinfo{person}{Dan Feldman}, \bibinfo{person}{Melanie Schmidt}, {and} \bibinfo{person}{Christian Sohler}.} \bibinfo{year}{2020}\natexlab{}.
\newblock \showarticletitle{Turning big data into tiny data: Constant-size coresets for k-means, PCA, and projective clustering}.
\newblock \bibinfo{journal}{\emph{SICOMP}} \bibinfo{volume}{49}, \bibinfo{number}{3} (\bibinfo{year}{2020}), \bibinfo{pages}{601--657}.
\newblock


\bibitem[\protect\citeauthoryear{Hasani, Thirumuruganathan, Asudeh, Koudas, and Das}{Hasani et~al\mbox{.}}{2018}]%
        {hasani2018efficient}
\bibfield{author}{\bibinfo{person}{Sona Hasani}, \bibinfo{person}{Saravanan Thirumuruganathan}, \bibinfo{person}{Abolfazl Asudeh}, \bibinfo{person}{Nick Koudas}, {and} \bibinfo{person}{Gautam Das}.} \bibinfo{year}{2018}\natexlab{}.
\newblock \showarticletitle{Efficient construction of approximate ad-hoc ML models through materialization and reuse}.
\newblock \bibinfo{journal}{\emph{PVLDB}} \bibinfo{volume}{11}, \bibinfo{number}{11} (\bibinfo{year}{2018}), \bibinfo{pages}{1468--1481}.
\newblock


\bibitem[\protect\citeauthoryear{Jensen, Pedersen, and Thomsen}{Jensen et~al\mbox{.}}{2018}]%
        {jensen2018modelardb}
\bibfield{author}{\bibinfo{person}{S{\o}ren~Kejser Jensen}, \bibinfo{person}{Torben~Bach Pedersen}, {and} \bibinfo{person}{Christian Thomsen}.} \bibinfo{year}{2018}\natexlab{}.
\newblock \showarticletitle{Modelardb: Modular model-based time series management with spark and cassandra}.
\newblock \bibinfo{journal}{\emph{PVLDB}} \bibinfo{volume}{11}, \bibinfo{number}{11} (\bibinfo{year}{2018}), \bibinfo{pages}{1688--1701}.
\newblock


\bibitem[\protect\citeauthoryear{Kieu, Kieu, Han, Yang, Jensen, and Le}{Kieu et~al\mbox{.}}{2024}]%
        {kieu2024Team}
\bibfield{author}{\bibinfo{person}{Duc Kieu}, \bibinfo{person}{Tung Kieu}, \bibinfo{person}{Peng Han}, \bibinfo{person}{Bin Yang}, \bibinfo{person}{Christian~S. Jensen}, {and} \bibinfo{person}{Bac Le}.} \bibinfo{year}{2024}\natexlab{}.
\newblock \showarticletitle{TEAM: Topological Evolution-aware Framework for Traffic Forecasting}.
\newblock \bibinfo{journal}{\emph{PVLDB}}  \bibinfo{volume}{18} (\bibinfo{year}{2024}).
\newblock


\bibitem[\protect\citeauthoryear{Kieu, Yang, Guo, Jensen, Zhao, Huang, and Zheng}{Kieu et~al\mbox{.}}{2022}]%
        {DBLP:conf/icde/KieuYGJZHZ22}
\bibfield{author}{\bibinfo{person}{Tung Kieu}, \bibinfo{person}{Bin Yang}, \bibinfo{person}{Chenjuan Guo}, \bibinfo{person}{Christian~S. Jensen}, \bibinfo{person}{Yan Zhao}, \bibinfo{person}{Feiteng Huang}, {and} \bibinfo{person}{Kai Zheng}.} \bibinfo{year}{2022}\natexlab{}.
\newblock \showarticletitle{Robust and Explainable Autoencoders for Unsupervised Time Series Outlier Detection}. In \bibinfo{booktitle}{\emph{{ICDE}}}. \bibinfo{pages}{3038--3050}.
\newblock


\bibitem[\protect\citeauthoryear{Lai, Zhang, Li, Jensen, Lu, and Zhao}{Lai et~al\mbox{.}}{2023}]%
        {lai2023lightcts}
\bibfield{author}{\bibinfo{person}{Zhichen Lai}, \bibinfo{person}{Dalin Zhang}, \bibinfo{person}{Huan Li}, \bibinfo{person}{Christian~S Jensen}, \bibinfo{person}{Hua Lu}, {and} \bibinfo{person}{Yan Zhao}.} \bibinfo{year}{2023}\natexlab{}.
\newblock \showarticletitle{LightCTS: A Lightweight Framework for Correlated Time Series Forecasting}.
\newblock \bibinfo{journal}{\emph{SIGMOD}} \bibinfo{volume}{1}, \bibinfo{number}{2} (\bibinfo{year}{2023}), \bibinfo{pages}{1--26}.
\newblock


\bibitem[\protect\citeauthoryear{Li, Shen, and Chen}{Li et~al\mbox{.}}{2022}]%
        {li2022camel}
\bibfield{author}{\bibinfo{person}{Yiming Li}, \bibinfo{person}{Yanyan Shen}, {and} \bibinfo{person}{Lei Chen}.} \bibinfo{year}{2022}\natexlab{}.
\newblock \showarticletitle{Camel: Managing Data for Efficient Stream Learning}. In \bibinfo{booktitle}{\emph{SIGMOD}}. \bibinfo{pages}{1271--1285}.
\newblock


\bibitem[\protect\citeauthoryear{Liu, Piao, Ma, Yuan, Tang, Wang, and Leung}{Liu et~al\mbox{.}}{2021}]%
        {liu2021modeling}
\bibfield{author}{\bibinfo{person}{Chi~Harold Liu}, \bibinfo{person}{Chengzhe Piao}, \bibinfo{person}{Xiaoxin Ma}, \bibinfo{person}{Ye Yuan}, \bibinfo{person}{Jian Tang}, \bibinfo{person}{Guoren Wang}, {and} \bibinfo{person}{Kin~K Leung}.} \bibinfo{year}{2021}\natexlab{}.
\newblock \showarticletitle{Modeling citywide crowd flows using attentive convolutional LSTM}. In \bibinfo{booktitle}{\emph{ICDE}}. \bibinfo{pages}{217--228}.
\newblock


\bibitem[\protect\citeauthoryear{Liu, Miao, Zhao, Liu, Zheng, and Li}{Liu et~al\mbox{.}}{2024}]%
        {DBLP:conf/icde/Liu00L0024}
\bibfield{author}{\bibinfo{person}{Ziqiao Liu}, \bibinfo{person}{Hao Miao}, \bibinfo{person}{Yan Zhao}, \bibinfo{person}{Chenxi Liu}, \bibinfo{person}{Kai Zheng}, {and} \bibinfo{person}{Huan Li}.} \bibinfo{year}{2024}\natexlab{}.
\newblock \showarticletitle{LightTR: {A} Lightweight Framework for Federated Trajectory Recovery}. In \bibinfo{booktitle}{\emph{{ICDE}}}. \bibinfo{pages}{4422--4434}.
\newblock


\bibitem[\protect\citeauthoryear{Lucic, Faulkner, Krause, and Feldman}{Lucic et~al\mbox{.}}{2018}]%
        {lucic2018training}
\bibfield{author}{\bibinfo{person}{Mario Lucic}, \bibinfo{person}{Matthew Faulkner}, \bibinfo{person}{Andreas Krause}, {and} \bibinfo{person}{Dan Feldman}.} \bibinfo{year}{2018}\natexlab{}.
\newblock \showarticletitle{Training gaussian mixture models at scale via coresets}.
\newblock \bibinfo{journal}{\emph{J. Mach. Learn. Res}} \bibinfo{volume}{18}, \bibinfo{number}{160} (\bibinfo{year}{2018}), \bibinfo{pages}{1--25}.
\newblock


\bibitem[\protect\citeauthoryear{Miao, Shen, Cao, Xia, and Wang}{Miao et~al\mbox{.}}{2023}]%
        {MiaoSCXW23}
\bibfield{author}{\bibinfo{person}{Hao Miao}, \bibinfo{person}{Jiaxing Shen}, \bibinfo{person}{Jiannong Cao}, \bibinfo{person}{Jiangnan Xia}, {and} \bibinfo{person}{Senzhang Wang}.} \bibinfo{year}{2023}\natexlab{}.
\newblock \showarticletitle{MBA-STNet: Bayes-Enhanced Discriminative Multi-Task Learning for Flow Prediction}.
\newblock \bibinfo{journal}{\emph{TKDE}} \bibinfo{volume}{35}, \bibinfo{number}{7} (\bibinfo{year}{2023}), \bibinfo{pages}{7164--7177}.
\newblock


\bibitem[\protect\citeauthoryear{Miao, Zhao, Guo, Yang, Zheng, Huang, Xie, and Jensen}{Miao et~al\mbox{.}}{2024}]%
        {DBLP:conf/icde/00010GY0HXJ24}
\bibfield{author}{\bibinfo{person}{Hao Miao}, \bibinfo{person}{Yan Zhao}, \bibinfo{person}{Chenjuan Guo}, \bibinfo{person}{Bin Yang}, \bibinfo{person}{Kai Zheng}, \bibinfo{person}{Feiteng Huang}, \bibinfo{person}{Jiandong Xie}, {and} \bibinfo{person}{Christian~S. Jensen}.} \bibinfo{year}{2024}\natexlab{}.
\newblock \showarticletitle{A Unified Replay-Based Continuous Learning Framework for Spatio-Temporal Prediction on Streaming Data}. In \bibinfo{booktitle}{\emph{{ICDE}}}. \bibinfo{pages}{1050--1062}.
\newblock


\bibitem[\protect\citeauthoryear{Nguyen, Chen, and Lee}{Nguyen et~al\mbox{.}}{2020}]%
        {nguyen2020dataset}
\bibfield{author}{\bibinfo{person}{Timothy Nguyen}, \bibinfo{person}{Zhourong Chen}, {and} \bibinfo{person}{Jaehoon Lee}.} \bibinfo{year}{2020}\natexlab{}.
\newblock \showarticletitle{Dataset Meta-Learning from Kernel Ridge-Regression}. In \bibinfo{booktitle}{\emph{ICLR}}.
\newblock


\bibitem[\protect\citeauthoryear{Nie, H.~Nguyen, Sinthong, and Kalagnanam}{Nie et~al\mbox{.}}{2023}]%
        {Yuqietal-2023-PatchTST}
\bibfield{author}{\bibinfo{person}{Yuqi Nie}, \bibinfo{person}{Nam H.~Nguyen}, \bibinfo{person}{Phanwadee Sinthong}, {and} \bibinfo{person}{Jayant Kalagnanam}.} \bibinfo{year}{2023}\natexlab{}.
\newblock \showarticletitle{A Time Series is Worth 64 Words: Long-term Forecasting with Transformers}. In \bibinfo{booktitle}{\emph{ICLR}}.
\newblock


\bibitem[\protect\citeauthoryear{Salinas, Flunkert, Gasthaus, and Januschowski}{Salinas et~al\mbox{.}}{2020}]%
        {salinas2020deepar}
\bibfield{author}{\bibinfo{person}{David Salinas}, \bibinfo{person}{Valentin Flunkert}, \bibinfo{person}{Jan Gasthaus}, {and} \bibinfo{person}{Tim Januschowski}.} \bibinfo{year}{2020}\natexlab{}.
\newblock \showarticletitle{DeepAR: Probabilistic forecasting with autoregressive recurrent networks}.
\newblock \bibinfo{journal}{\emph{Int J Forecast}} \bibinfo{volume}{36}, \bibinfo{number}{3} (\bibinfo{year}{2020}), \bibinfo{pages}{1181--1191}.
\newblock


\bibitem[\protect\citeauthoryear{Shekhar and Williams}{Shekhar and Williams}{2007}]%
        {shekhar2007adaptive}
\bibfield{author}{\bibinfo{person}{Shashank Shekhar} {and} \bibinfo{person}{Billy~M Williams}.} \bibinfo{year}{2007}\natexlab{}.
\newblock \showarticletitle{Adaptive seasonal time series models for forecasting short-term traffic flow}.
\newblock \bibinfo{journal}{\emph{TRR}} \bibinfo{volume}{2024}, \bibinfo{number}{1} (\bibinfo{year}{2007}), \bibinfo{pages}{116--125}.
\newblock


\bibitem[\protect\citeauthoryear{Sylligardos, Boniol, Paparrizos, Trahanias, and Palpanas}{Sylligardos et~al\mbox{.}}{2023}]%
        {sylligardos2023choose}
\bibfield{author}{\bibinfo{person}{Emmanouil Sylligardos}, \bibinfo{person}{Paul Boniol}, \bibinfo{person}{John Paparrizos}, \bibinfo{person}{Panos Trahanias}, {and} \bibinfo{person}{Themis Palpanas}.} \bibinfo{year}{2023}\natexlab{}.
\newblock \showarticletitle{Choose wisely: An extensive evaluation of model selection for anomaly detection in time series}.
\newblock \bibinfo{journal}{\emph{PVLDB}} \bibinfo{volume}{16}, \bibinfo{number}{11} (\bibinfo{year}{2023}), \bibinfo{pages}{3418--3432}.
\newblock


\bibitem[\protect\citeauthoryear{Tai, Sharan, Bailis, and Valiant}{Tai et~al\mbox{.}}{2018}]%
        {tai2018sketching}
\bibfield{author}{\bibinfo{person}{Kai~Sheng Tai}, \bibinfo{person}{Vatsal Sharan}, \bibinfo{person}{Peter Bailis}, {and} \bibinfo{person}{Gregory Valiant}.} \bibinfo{year}{2018}\natexlab{}.
\newblock \showarticletitle{Sketching linear classifiers over data streams}. In \bibinfo{booktitle}{\emph{SIGMOD}}. \bibinfo{pages}{757--772}.
\newblock


\bibitem[\protect\citeauthoryear{Van~der Maaten and Hinton}{Van~der Maaten and Hinton}{2008}]%
        {van2008visualizing}
\bibfield{author}{\bibinfo{person}{Laurens Van~der Maaten} {and} \bibinfo{person}{Geoffrey Hinton}.} \bibinfo{year}{2008}\natexlab{}.
\newblock \showarticletitle{Visualizing data using t-SNE.}
\newblock \bibinfo{journal}{\emph{J Mach Learn Res}} \bibinfo{volume}{9}, \bibinfo{number}{11} (\bibinfo{year}{2008}).
\newblock


\bibitem[\protect\citeauthoryear{Vaswani, Shazeer, Parmar, Uszkoreit, Jones, Gomez, Kaiser, and Polosukhin}{Vaswani et~al\mbox{.}}{2017}]%
        {vaswani2017attention}
\bibfield{author}{\bibinfo{person}{Ashish Vaswani}, \bibinfo{person}{Noam Shazeer}, \bibinfo{person}{Niki Parmar}, \bibinfo{person}{Jakob Uszkoreit}, \bibinfo{person}{Llion Jones}, \bibinfo{person}{Aidan~N Gomez}, \bibinfo{person}{{\L}ukasz Kaiser}, {and} \bibinfo{person}{Illia Polosukhin}.} \bibinfo{year}{2017}\natexlab{}.
\newblock \showarticletitle{Attention is all you need}.
\newblock \bibinfo{journal}{\emph{NeurIPS}}  \bibinfo{volume}{30} (\bibinfo{year}{2017}).
\newblock


\bibitem[\protect\citeauthoryear{Wang, Huang, Qiao, et~al\mbox{.}}{Wang et~al\mbox{.}}{2020}]%
        {wang2020apache}
\bibfield{author}{\bibinfo{person}{Chen Wang}, \bibinfo{person}{Xiangdong Huang}, \bibinfo{person}{Jialin Qiao}, {et~al\mbox{.}}} \bibinfo{year}{2020}\natexlab{}.
\newblock \showarticletitle{Apache IoTDB: Time-series database for internet of things}.
\newblock \bibinfo{journal}{\emph{PVLDB}} \bibinfo{volume}{13}, \bibinfo{number}{12} (\bibinfo{year}{2020}), \bibinfo{pages}{2901--2904}.
\newblock


\bibitem[\protect\citeauthoryear{Wang, Whitmarsh, Navarro, and Palpanas}{Wang et~al\mbox{.}}{2022a}]%
        {wang2022iedeal}
\bibfield{author}{\bibinfo{person}{Qitong Wang}, \bibinfo{person}{Stephen Whitmarsh}, \bibinfo{person}{Vincent Navarro}, {and} \bibinfo{person}{Themis Palpanas}.} \bibinfo{year}{2022}\natexlab{a}.
\newblock \showarticletitle{iEDeaL: A Deep Learning Framework for Detecting Highly Imbalanced Interictal Epileptiform Discharges}.
\newblock \bibinfo{journal}{\emph{PVLDB}} \bibinfo{volume}{16}, \bibinfo{number}{3} (\bibinfo{year}{2022}), \bibinfo{pages}{480--490}.
\newblock


\bibitem[\protect\citeauthoryear{Wang, Zhang, Miao, Peng, and Yu}{Wang et~al\mbox{.}}{2022b}]%
        {wang2022multivariate}
\bibfield{author}{\bibinfo{person}{Senzhang Wang}, \bibinfo{person}{Meiyue Zhang}, \bibinfo{person}{Hao Miao}, \bibinfo{person}{Zhaohui Peng}, {and} \bibinfo{person}{Philip~S Yu}.} \bibinfo{year}{2022}\natexlab{b}.
\newblock \showarticletitle{Multivariate correlation-aware spatio-temporal graph convolutional networks for multi-scale traffic prediction}.
\newblock \bibinfo{journal}{\emph{TIST}} \bibinfo{volume}{13}, \bibinfo{number}{3} (\bibinfo{year}{2022}), \bibinfo{pages}{1--22}.
\newblock


\bibitem[\protect\citeauthoryear{Welling}{Welling}{2009}]%
        {welling2009herding}
\bibfield{author}{\bibinfo{person}{Max Welling}.} \bibinfo{year}{2009}\natexlab{}.
\newblock \showarticletitle{Herding dynamical weights to learn}. In \bibinfo{booktitle}{\emph{ICML}}. \bibinfo{pages}{1121--1128}.
\newblock


\bibitem[\protect\citeauthoryear{Wen, He, Sun, Zhang, Ke, and Xu}{Wen et~al\mbox{.}}{2021}]%
        {wen2021robustperiod}
\bibfield{author}{\bibinfo{person}{Qingsong Wen}, \bibinfo{person}{Kai He}, \bibinfo{person}{Liang Sun}, \bibinfo{person}{Yingying Zhang}, \bibinfo{person}{Min Ke}, {and} \bibinfo{person}{Huan Xu}.} \bibinfo{year}{2021}\natexlab{}.
\newblock \showarticletitle{RobustPeriod: Robust time-frequency mining for multiple periodicity detection}. In \bibinfo{booktitle}{\emph{SIGMOD}}. \bibinfo{pages}{2328--2337}.
\newblock


\bibitem[\protect\citeauthoryear{Wu, Xu, Wang, and Long}{Wu et~al\mbox{.}}{2021}]%
        {wu2021autoformer}
\bibfield{author}{\bibinfo{person}{Haixu Wu}, \bibinfo{person}{Jiehui Xu}, \bibinfo{person}{Jianmin Wang}, {and} \bibinfo{person}{Mingsheng Long}.} \bibinfo{year}{2021}\natexlab{}.
\newblock \showarticletitle{Autoformer: Decomposition transformers with auto-correlation for long-term series forecasting}.
\newblock \bibinfo{journal}{\emph{NeurIPS}}  \bibinfo{volume}{34} (\bibinfo{year}{2021}), \bibinfo{pages}{22419--22430}.
\newblock


\bibitem[\protect\citeauthoryear{Wu, Wu, Yang, Zhou, Guo, Qiu, Hu, Sheng, and Jensen}{Wu et~al\mbox{.}}{2024a}]%
        {DBLP:journals/vldb/WuWYZGQHSJ24}
\bibfield{author}{\bibinfo{person}{Xinle Wu}, \bibinfo{person}{Xingjian Wu}, \bibinfo{person}{Bin Yang}, \bibinfo{person}{Lekui Zhou}, \bibinfo{person}{Chenjuan Guo}, \bibinfo{person}{Xiangfei Qiu}, \bibinfo{person}{Jilin Hu}, \bibinfo{person}{Zhenli Sheng}, {and} \bibinfo{person}{Christian~S. Jensen}.} \bibinfo{year}{2024}\natexlab{a}.
\newblock \showarticletitle{AutoCTS++: zero-shot joint neural architecture and hyperparameter search for correlated time series forecasting}.
\newblock \bibinfo{journal}{\emph{{VLDBJ}}} \bibinfo{volume}{33}, \bibinfo{number}{5} (\bibinfo{year}{2024}), \bibinfo{pages}{1743--1770}.
\newblock


\bibitem[\protect\citeauthoryear{Wu, Wu, Zhang, Zhang, Guo, Yang, and Jensen}{Wu et~al\mbox{.}}{2024b}]%
        {xinle2024FACTS}
\bibfield{author}{\bibinfo{person}{Xinle Wu}, \bibinfo{person}{Xingjian Wu}, \bibinfo{person}{Dalin Zhang}, \bibinfo{person}{Miao Zhang}, \bibinfo{person}{Chenjuan Guo}, \bibinfo{person}{Bin Yang}, {and} \bibinfo{person}{Christian~S. Jensen}.} \bibinfo{year}{2024}\natexlab{b}.
\newblock \showarticletitle{Fully Automated Correlated Time Series Forecasting in Minutes}.
\newblock \bibinfo{journal}{\emph{PVLDB.}}  \bibinfo{volume}{18} (\bibinfo{year}{2024}).
\newblock


\bibitem[\protect\citeauthoryear{Wu, Zhang, Zhang, Guo, Yang, and Jensen}{Wu et~al\mbox{.}}{2023}]%
        {wu2023autocts+}
\bibfield{author}{\bibinfo{person}{Xinle Wu}, \bibinfo{person}{Dalin Zhang}, \bibinfo{person}{Miao Zhang}, \bibinfo{person}{Chenjuan Guo}, \bibinfo{person}{Bin Yang}, {and} \bibinfo{person}{Christian~S Jensen}.} \bibinfo{year}{2023}\natexlab{}.
\newblock \showarticletitle{AutoCTS+: Joint Neural Architecture and Hyperparameter Search for Correlated Time Series Forecasting}.
\newblock \bibinfo{journal}{\emph{SIGMOD}} \bibinfo{volume}{1}, \bibinfo{number}{1} (\bibinfo{year}{2023}), \bibinfo{pages}{1--26}.
\newblock


\bibitem[\protect\citeauthoryear{Xiao, Huang, Hu, Song, Huang, and Wang}{Xiao et~al\mbox{.}}{2022}]%
        {xiao2022time}
\bibfield{author}{\bibinfo{person}{Jinzhao Xiao}, \bibinfo{person}{Yuxiang Huang}, \bibinfo{person}{Changyu Hu}, \bibinfo{person}{Shaoxu Song}, \bibinfo{person}{Xiangdong Huang}, {and} \bibinfo{person}{Jianmin Wang}.} \bibinfo{year}{2022}\natexlab{}.
\newblock \showarticletitle{Time series data encoding for efficient storage: A comparative analysis in apache IoTDB}.
\newblock \bibinfo{journal}{\emph{PVLDB}} \bibinfo{volume}{15}, \bibinfo{number}{10} (\bibinfo{year}{2022}), \bibinfo{pages}{2148--2160}.
\newblock


\bibitem[\protect\citeauthoryear{Xu, Miao, Wang, Yu, and Wang}{Xu et~al\mbox{.}}{2024}]%
        {xu2024pefad}
\bibfield{author}{\bibinfo{person}{Ronghui Xu}, \bibinfo{person}{Hao Miao}, \bibinfo{person}{Senzhang Wang}, \bibinfo{person}{Philip~S Yu}, {and} \bibinfo{person}{Jianxin Wang}.} \bibinfo{year}{2024}\natexlab{}.
\newblock \showarticletitle{PeFAD: A Parameter-Efficient Federated Framework for Time Series Anomaly Detection}. In \bibinfo{booktitle}{\emph{SIGKDD}}. \bibinfo{pages}{3621--3632}.
\newblock


\bibitem[\protect\citeauthoryear{Yu, Peng, Li, Wang, Shen, Mai, and Xie}{Yu et~al\mbox{.}}{2020}]%
        {yu2020two}
\bibfield{author}{\bibinfo{person}{Xinyang Yu}, \bibinfo{person}{Yanqing Peng}, \bibinfo{person}{Feifei Li}, \bibinfo{person}{Sheng Wang}, \bibinfo{person}{Xiaowei Shen}, \bibinfo{person}{Huijun Mai}, {and} \bibinfo{person}{Yue Xie}.} \bibinfo{year}{2020}\natexlab{}.
\newblock \showarticletitle{Two-level data compression using machine learning in time series database}. In \bibinfo{booktitle}{\emph{ICDE}}. \bibinfo{pages}{1333--1344}.
\newblock


\bibitem[\protect\citeauthoryear{Zhao, Mopuri, and Bilen}{Zhao et~al\mbox{.}}{2021}]%
        {zhao2021DC}
\bibfield{author}{\bibinfo{person}{Bo Zhao}, \bibinfo{person}{Konda~Reddy Mopuri}, {and} \bibinfo{person}{Hakan Bilen}.} \bibinfo{year}{2021}\natexlab{}.
\newblock \showarticletitle{Dataset Condensation with Gradient Matching}. In \bibinfo{booktitle}{\emph{ICLR}}.
\newblock


\bibitem[\protect\citeauthoryear{Zhao, Li, Qin, and Yu}{Zhao et~al\mbox{.}}{2023b}]%
        {zhao2023improved}
\bibfield{author}{\bibinfo{person}{Ganlong Zhao}, \bibinfo{person}{Guanbin Li}, \bibinfo{person}{Yipeng Qin}, {and} \bibinfo{person}{Yizhou Yu}.} \bibinfo{year}{2023}\natexlab{b}.
\newblock \showarticletitle{Improved distribution matching for dataset condensation}. In \bibinfo{booktitle}{\emph{CVPR}}. \bibinfo{pages}{7856--7865}.
\newblock


\bibitem[\protect\citeauthoryear{Zhao, Guo, Cheng, Han, Zhang, and Yang}{Zhao et~al\mbox{.}}{2023a}]%
        {DBLP:journals/pvldb/ZhaoGCHZY23}
\bibfield{author}{\bibinfo{person}{Kai Zhao}, \bibinfo{person}{Chenjuan Guo}, \bibinfo{person}{Yunyao Cheng}, \bibinfo{person}{Peng Han}, \bibinfo{person}{Miao Zhang}, {and} \bibinfo{person}{Bin Yang}.} \bibinfo{year}{2023}\natexlab{a}.
\newblock \showarticletitle{Multiple Time Series Forecasting with Dynamic Graph Modeling}.
\newblock \bibinfo{journal}{\emph{PVLDB}} \bibinfo{volume}{17}, \bibinfo{number}{4} (\bibinfo{year}{2023}), \bibinfo{pages}{753--765}.
\newblock


\bibitem[\protect\citeauthoryear{Zhao, Chen, Deng, Kieu, Guo, Yang, Zheng, and Jensen}{Zhao et~al\mbox{.}}{2022}]%
        {yanwww2022}
\bibfield{author}{\bibinfo{person}{Yan Zhao}, \bibinfo{person}{Xuanhao Chen}, \bibinfo{person}{Liwei Deng}, \bibinfo{person}{Tung Kieu}, \bibinfo{person}{Chenjuan Guo}, \bibinfo{person}{Bin Yang}, \bibinfo{person}{Kai Zheng}, {and} \bibinfo{person}{Christian~S. Jensen}.} \bibinfo{year}{2022}\natexlab{}.
\newblock \showarticletitle{Outlier Detection for Streaming Task Assignment in Crowdsourcing}. In \bibinfo{booktitle}{\emph{{WWW}}}. \bibinfo{pages}{1933--1943}.
\newblock


\bibitem[\protect\citeauthoryear{Zheng, Chen, Herschel, Ngiam, Ooi, and Gao}{Zheng et~al\mbox{.}}{2021}]%
        {zheng2021pace}
\bibfield{author}{\bibinfo{person}{Kaiping Zheng}, \bibinfo{person}{Gang Chen}, \bibinfo{person}{Melanie Herschel}, \bibinfo{person}{Kee~Yuan Ngiam}, \bibinfo{person}{Beng~Chin Ooi}, {and} \bibinfo{person}{Jinyang Gao}.} \bibinfo{year}{2021}\natexlab{}.
\newblock \showarticletitle{PACE: learning effective task decomposition for human-in-the-loop healthcare delivery}. In \bibinfo{booktitle}{\emph{SIGMOD}}. \bibinfo{pages}{2156--2168}.
\newblock


\bibitem[\protect\citeauthoryear{Zhou, Zhang, Peng, Zhang, Li, Xiong, and Zhang}{Zhou et~al\mbox{.}}{2021}]%
        {zhou2021informer}
\bibfield{author}{\bibinfo{person}{Haoyi Zhou}, \bibinfo{person}{Shanghang Zhang}, \bibinfo{person}{Jieqi Peng}, \bibinfo{person}{Shuai Zhang}, \bibinfo{person}{Jianxin Li}, \bibinfo{person}{Hui Xiong}, {and} \bibinfo{person}{Wancai Zhang}.} \bibinfo{year}{2021}\natexlab{}.
\newblock \showarticletitle{Informer: Beyond efficient transformer for long sequence time-series forecasting}. In \bibinfo{booktitle}{\emph{AAAI}}, Vol.~\bibinfo{volume}{35}. \bibinfo{pages}{11106--11115}.
\newblock


\bibitem[\protect\citeauthoryear{Zhou, Ma, Wen, Wang, Sun, and Jin}{Zhou et~al\mbox{.}}{2022a}]%
        {zhou2022fedformer}
\bibfield{author}{\bibinfo{person}{Tian Zhou}, \bibinfo{person}{Ziqing Ma}, \bibinfo{person}{Qingsong Wen}, \bibinfo{person}{Xue Wang}, \bibinfo{person}{Liang Sun}, {and} \bibinfo{person}{Rong Jin}.} \bibinfo{year}{2022}\natexlab{a}.
\newblock \showarticletitle{Fedformer: Frequency enhanced decomposed transformer for long-term series forecasting}. In \bibinfo{booktitle}{\emph{ICML}}. \bibinfo{pages}{27268--27286}.
\newblock


\bibitem[\protect\citeauthoryear{Zhou, Nezhadarya, and Ba}{Zhou et~al\mbox{.}}{2022b}]%
        {zhou2022dataset}
\bibfield{author}{\bibinfo{person}{Yongchao Zhou}, \bibinfo{person}{Ehsan Nezhadarya}, {and} \bibinfo{person}{Jimmy Ba}.} \bibinfo{year}{2022}\natexlab{b}.
\newblock \showarticletitle{Dataset distillation using neural feature regression}.
\newblock \bibinfo{journal}{\emph{NeurIPS}}  \bibinfo{volume}{35} (\bibinfo{year}{2022}), \bibinfo{pages}{9813--9827}.
\newblock


\end{thebibliography}
